\documentclass[showpacs,showkeys,amssymb,aps]{revtex4}

\usepackage{amsmath}

\begin{document}

\title[Zernike Basis to Cartesian Transforms]{Zernike Basis to Cartesian Transformations}

\author{Richard J. Mathar}
\homepage{http://www.strw.leidenuniv.nl/~mathar}
\email{mathar@strw.leidenuniv.nl}
\affiliation{Leiden Observatory, P.O. Box 9513, 2300 RA Leiden, The Netherlands}

\date{\today}

\begin{abstract}
The radial polynomials of the 2D (circular) and 3D (spherical) Zernike
functions are tabulated as powers of the radial distance. The reciprocal
tabulation of powers of the radial distance in series of radial polynomials is
also given,
based on projections that take advantage of the orthogonality of the polynomials
over the unit interval. They may play a role in the
expansion of products of the polynomials into sums,
which is demonstrated by some examples.

Multiplication of the polynomials by the angular bases (azimuth, polar angle)
defines the Zernike
functions, for which we derive and tabulate transformations
to and from the Cartesian coordinate system centered at the middle of the circle
or sphere.
\end{abstract}

\maketitle

\section{Scope} 
The incentive of the following exercises in standard analysis
is to model paths of optical rays through a turbulent atmosphere based
on Snell's law, if the dielectric function is created
on a computer with modes founded on a Zernike basis inside
a sphere \cite{MatharArxiv0805}. Since calculation of gradients is more tedious in
curvilinear local spherical coordinates than in Cartesian coordinates,
a transformation of the basis is useful.

There is sufficient  similarity between the 3D and 2D Zernike functions
to present both cases conjointly. The
more familiar 2D Zernike functions in Noll's nomenclature are covered in
Section \ref{sec.2D}---although no new aspects beyond
earlier work emerge \cite{DaiJOSAA23}---and similar
methodology is independently applied to the 3D case in Section \ref{sec.3D}.

\section{Zernike Circle Functions} \label{sec.2D} 
\subsection{Zernike Radial Polynomials} \label{sec.rofR}
\subsubsection{Definition}
We define Zernike radial polynomials in Noll's nomenclature
\cite{NollJOSA66,PrataAO28,KintnerJModOpt23,TysonOL7,ConfortiOL8,TangoApplPhys13,MatharArxiv0705a}
for $0\le m\le n$, even $n-m$ as
\begin{eqnarray}
R_n^m(r)
&=&
\sum_{s=0}^{(n-m)/2}
(-1)^s\binom{n-s}{s}\binom{n-2s}{(n-m)/2-s} r^{n-2s}
\label{eq.Rpolyxn}
\\
&=&
(-1)^{(n-m)/2}
\sum_{s=0}^{(n-m)/2}
(-1)^s\binom{(n+m)/2+s}{(n-m)/2-s}\binom{m+2s}{ s} r^{m+2s}
\label{eq.Rpolyxm}
\\
&=&
(-1)^a
\binom{-b}{ -a}
r^m\,_2F_1\left(a,1-b;1+m;r^2\right)
=
\binom{n}{ -a}
r^n\,_2F_1\left(a,b;-n;\frac{1}{r^2}\right)
,
\end{eqnarray}
where an equivalent set of two negative parameters
\begin{equation}
a\equiv -(n-m)/2;\quad b\equiv -(n+m)/2
\label{eq.R2Dhyp}
\end{equation}
abbreviates the notation in the context of hypergeometric series.
The formulas show that each expansion coefficient can be quickly
generated by multiplying two binomial integers.
Following the original notation, we will not put the upper index $m$ in $R_n^m$---which
is not a power---into parentheses.
\small \begin{verbatim}
R_0^0(r) = 1.
R_1^1(r) = r.
R_2^0(r) = -1 +2*r^2.
R_2^2(r) = r^2.
R_3^1(r) = -2*r +3*r^3.
R_3^3(r) = r^3.
R_4^0(r) = 1 -6*r^2 +6*r^4.
R_4^2(r) = -3*r^2 +4*r^4.
R_4^4(r) = r^4.
R_5^1(r) = 3*r -12*r^3 +10*r^5.
R_5^3(r) = -4*r^3 +5*r^5.
R_5^5(r) = r^5.
R_6^0(r) = -1 +12*r^2 -30*r^4 +20*r^6.
R_6^2(r) = 6*r^2 -20*r^4 +15*r^6.
R_6^4(r) = -5*r^4 +6*r^6.
R_6^6(r) = r^6.
R_7^1(r) = -4*r +30*r^3 -60*r^5 +35*r^7.
R_7^3(r) = 10*r^3 -30*r^5 +21*r^7.
R_7^5(r) = -6*r^5 +7*r^7.
R_7^7(r) = r^7.
R_8^0(r) = 1 -20*r^2 +90*r^4 -140*r^6 +70*r^8.
R_8^2(r) = -10*r^2 +60*r^4 -105*r^6 +56*r^8.
R_8^4(r) = 15*r^4 -42*r^6 +28*r^8.
R_8^6(r) = -7*r^6 +8*r^8.
R_8^8(r) = r^8.
R_9^1(r) = 5*r -60*r^3 +210*r^5 -280*r^7 +126*r^9.
R_9^3(r) = -20*r^3 +105*r^5 -168*r^7 +84*r^9.
R_9^5(r) = 21*r^5 -56*r^7 +36*r^9.
R_9^7(r) = -8*r^7 +9*r^9.
R_9^9(r) = r^9.
R_10^0(r) = -1 +30*r^2 -210*r^4 +560*r^6 -630*r^8 +252*r^10.
R_10^2(r) = 15*r^2 -140*r^4 +420*r^6 -504*r^8 +210*r^10.
R_10^4(r) = -35*r^4 +168*r^6 -252*r^8 +120*r^10.
R_10^6(r) = 28*r^6 -72*r^8 +45*r^10.
R_10^8(r) = -9*r^8 +10*r^10.
R_10^10(r) = r^10.
R_11^1(r) = -6*r +105*r^3 -560*r^5 +1260*r^7 -1260*r^9 +462*r^11.
R_11^3(r) = 35*r^3 -280*r^5 +756*r^7 -840*r^9 +330*r^11.
R_11^5(r) = -56*r^5 +252*r^7 -360*r^9 +165*r^11.
R_11^7(r) = 36*r^7 -90*r^9 +55*r^11.
R_11^9(r) = -10*r^9 +11*r^11.
R_11^11(r) = r^11.
R_12^0(r) = 1 -42*r^2 +420*r^4 -1680*r^6 +3150*r^8 -2772*r^10 +924*r^12.
R_12^2(r) = -21*r^2 +280*r^4 -1260*r^6 +2520*r^8 -2310*r^10 +792*r^12.
R_12^4(r) = 70*r^4 -504*r^6 +1260*r^8 -1320*r^10 +495*r^12.
R_12^6(r) = -84*r^6 +360*r^8 -495*r^10 +220*r^12.
R_12^8(r) = 45*r^8 -110*r^10 +66*r^12.
R_12^10(r) = -11*r^10 +12*r^12.
R_12^12(r) = r^12.
R_13^1(r) = 7*r -168*r^3 +1260*r^5 -4200*r^7 +6930*r^9 -5544*r^11 +1716*r^13.
R_13^3(r) = -56*r^3 +630*r^5 -2520*r^7 +4620*r^9 -3960*r^11 +1287*r^13.
R_13^5(r) = 126*r^5 -840*r^7 +1980*r^9 -1980*r^11 +715*r^13.
R_13^7(r) = -120*r^7 +495*r^9 -660*r^11 +286*r^13.
R_13^9(r) = 55*r^9 -132*r^11 +78*r^13.
R_13^11(r) = -12*r^11 +13*r^13.
R_13^13(r) = r^13.
\end{verbatim}\normalsize

\subsubsection{Inverse Representation}
The inversion of the representations (\ref{eq.Rpolyxm}) decomposes powers $r^j$ into sums
of $R_{n}^m(r)$,
\begin{equation}
r^j \equiv \sum_{n=m \mod 2}^j h_{j,n,m} R_n^m(r);\quad
j-m =0,2,4,6,\ldots
\label{eq.rofR}
\end{equation}
Multiplying this equation by $R_{n'}^m(r)$, contraction
on the right hand side with the normalization integral
\begin{equation}
\int_0^1 r\, R_n^m(r) R_{n'}^m(r) dr=\frac{1}{2(n+1)}\delta_{n,n'}
\label{eq.ortho}
\end{equation}
to isolate one $h$-coefficient, and using the explicit
polynomial expression (\ref{eq.Rpolyxm}) on the left hand side yields
 \cite{MatharArxiv0705a,RuizJCompApplMath89}:
\begin{equation}
h_{j,n,m}=2(n+1)\int_0^1 r^{j+1}(-1)^a \binom{ -b}{ -a}r^m\,_2F_1(a,1-b;1+m;r^2)dr
=(n+1)(-1)^a
\frac{ \left(\frac{m-j}{2}\right)_{-a} } {\left(1+\frac{m+j}{2}\right)_{1-a}}.
\label{eq.h}
\end{equation}
(See App.\ \ref{sec.nots} for notational standards like Pochhammer's symbol.)
From
\begin{equation}
R_n^m(1)=1,
\label{eq.Rat1}
\end{equation}
and (\ref{eq.rofR}) at $r=1$ we obtain the sum rule
\begin{equation}
\sum_{n=m \mod 2}^j h_{j,n,m} =1.
\end{equation}
Recurrences follow from basic properties of the Gamma Functions
implicitly contained in (\ref{eq.h}):
\begin{eqnarray}
h_{j+2,n,m} &=& \frac{(j+2+m)(j+2-m)}{(j+2-n)(j+4+n)} h_{j,n,m} ;
\\
h_{j,n+2,m} &=& \frac{(n+3)(j-n)}{(j+4+n)(n+1)} h_{j,n,m} ;
\\
h_{j,n,m+2} &=& \frac{j+2+m}{j-m} h_{j,n,m}.
\end{eqnarray}
If we write \texttt{r\symbol{94}j} for $r^j$ and \texttt{R\_n\symbol{94}m} for $R_n^m$,
expansions of (\ref{eq.rofR}) are
\small \begin{verbatim}
r^0 = R_0^0(r).
r^1 = R_1^1(r).
r^2 = 1/2*R_0^0(r) +1/2*R_2^0(r)
   = R_2^2(r).
r^3 = 2/3*R_1^1(r) +1/3*R_3^1(r)
   = R_3^3(r).
r^4 = 1/3*R_0^0(r) +1/2*R_2^0(r) +1/6*R_4^0(r)
   = 3/4*R_2^2(r) +1/4*R_4^2(r)
   = R_4^4(r).
r^5 = 1/2*R_1^1(r) +2/5*R_3^1(r) +1/10*R_5^1(r)
   = 4/5*R_3^3(r) +1/5*R_5^3(r)
   = R_5^5(r).
r^6 = 1/4*R_0^0(r) +9/20*R_2^0(r) +1/4*R_4^0(r) +1/20*R_6^0(r)
   = 3/5*R_2^2(r) +1/3*R_4^2(r) +1/15*R_6^2(r)
   = 5/6*R_4^4(r) +1/6*R_6^4(r)
   = R_6^6(r).
r^7 = 2/5*R_1^1(r) +2/5*R_3^1(r) +6/35*R_5^1(r) +1/35*R_7^1(r)
   = 2/3*R_3^3(r) +2/7*R_5^3(r) +1/21*R_7^3(r)
   = 6/7*R_5^5(r) +1/7*R_7^5(r)
   = R_7^7(r).
r^8 = 1/5*R_0^0(r) +2/5*R_2^0(r) +2/7*R_4^0(r) +1/10*R_6^0(r) +1/70*R_8^0(r)
   = 1/2*R_2^2(r) +5/14*R_4^2(r) +1/8*R_6^2(r) +1/56*R_8^2(r)
   = 5/7*R_4^4(r) +1/4*R_6^4(r) +1/28*R_8^4(r)
   = 7/8*R_6^6(r) +1/8*R_8^6(r)
   = R_8^8(r).
r^9 = 1/3*R_1^1(r) +8/21*R_3^1(r) +3/14*R_5^1(r) +4/63*R_7^1(r) +1/126*R_9^1(r)
   = 4/7*R_3^3(r) +9/28*R_5^3(r) +2/21*R_7^3(r) +1/84*R_9^3(r)
   = 3/4*R_5^5(r) +2/9*R_7^5(r) +1/36*R_9^5(r)
   = 8/9*R_7^7(r) +1/9*R_9^7(r)
   = R_9^9(r).
r^10 = 1/6*R_0^0(r) +5/14*R_2^0(r) +25/84*R_4^0(r) +5/36*R_6^0(r) +1/28*R_8^0(r) +1/252*R_10^0(r)
   = 3/7*R_2^2(r) +5/14*R_4^2(r) +1/6*R_6^2(r) +3/70*R_8^2(r) +1/210*R_10^2(r)
   = 5/8*R_4^4(r) +7/24*R_6^4(r) +3/40*R_8^4(r) +1/120*R_10^4(r)
   = 7/9*R_6^6(r) +1/5*R_8^6(r) +1/45*R_10^6(r)
   = 9/10*R_8^8(r) +1/10*R_10^8(r)
   = R_10^10(r).
r^11 = 2/7*R_1^1(r) +5/14*R_3^1(r) +5/21*R_5^1(r) +2/21*R_7^1(r) +5/231*R_9^1(r) +1/462*R_11^1(r)
   = 1/2*R_3^3(r) +1/3*R_5^3(r) +2/15*R_7^3(r) +1/33*R_9^3(r) +1/330*R_11^3(r)
   = 2/3*R_5^5(r) +4/15*R_7^5(r) +2/33*R_9^5(r) +1/165*R_11^5(r)
   = 4/5*R_7^7(r) +2/11*R_9^7(r) +1/55*R_11^7(r)
   = 10/11*R_9^9(r) +1/11*R_11^9(r)
   = R_11^11(r).
r^12 = 1/7*R_0^0(r) +9/28*R_2^0(r) +25/84*R_4^0(r) +1/6*R_6^0(r) +9/154*R_8^0(r) +1/84*R_10^0(r) 
   +1/924*R_12^0(r)
   = 3/8*R_2^2(r) +25/72*R_4^2(r) +7/36*R_6^2(r) +3/44*R_8^2(r) +1/72*R_10^2(r) +1/792*R_12^2(r)
   = 5/9*R_4^4(r) +14/45*R_6^4(r) +6/55*R_8^4(r) +1/45*R_10^4(r) +1/495*R_12^4(r)
   = 7/10*R_6^6(r) +27/110*R_8^6(r) +1/20*R_10^6(r) +1/220*R_12^6(r)
   = 9/11*R_8^8(r) +1/6*R_10^8(r) +1/66*R_12^8(r)
   = 11/12*R_10^10(r) +1/12*R_12^10(r)
   = R_12^12(r).
r^13 = 1/4*R_1^1(r) +1/3*R_3^1(r) +1/4*R_5^1(r) +4/33*R_7^1(r) +5/132*R_9^1(r) +1/143*R_11^1(r) 
   +1/1716*R_13^1(r)
   = 4/9*R_3^3(r) +1/3*R_5^3(r) +16/99*R_7^3(r) +5/99*R_9^3(r) +4/429*R_11^3(r) +1/1287*R_13^3(r)
   = 3/5*R_5^5(r) +16/55*R_7^5(r) +1/11*R_9^5(r) +12/715*R_11^5(r) +1/715*R_13^5(r)
   = 8/11*R_7^7(r) +5/22*R_9^7(r) +6/143*R_11^7(r) +1/286*R_13^7(r)
   = 5/6*R_9^9(r) +2/13*R_11^9(r) +1/78*R_13^9(r)
   = 12/13*R_11^11(r) +1/13*R_13^11(r)
   = R_13^13(r).
r^14 = 1/8*R_0^0(r) +7/24*R_2^0(r) +7/24*R_4^0(r) +49/264*R_6^0(r) +7/88*R_8^0(r) +7/312*R_10^0(r) 
   +1/264*R_12^0(r) +1/3432*R_14^0(r)
   = 1/3*R_2^2(r) +1/3*R_4^2(r) +7/33*R_6^2(r) +1/11*R_8^2(r) +1/39*R_10^2(r) +1/231*R_12^2(r) 
   +1/3003*R_14^2(r)
   = 1/2*R_4^4(r) +7/22*R_6^4(r) +3/22*R_8^4(r) +1/26*R_10^4(r) +1/154*R_12^4(r) +1/2002*R_14^4(r)
   = 7/11*R_6^6(r) +3/11*R_8^6(r) +1/13*R_10^6(r) +1/77*R_12^6(r) +1/1001*R_14^6(r)
   = 3/4*R_8^8(r) +11/52*R_10^8(r) +1/28*R_12^8(r) +1/364*R_14^8(r)
   = 11/13*R_10^10(r) +1/7*R_12^10(r) +1/91*R_14^10(r)
   = 13/14*R_12^12(r) +1/14*R_14^12(r)
   = R_14^14(r).
\end{verbatim}\normalsize

\subsection{Zernike Functions} 
The Zernike functions $Z_j$ are products of the radial polynomials by azimuth
functions, $\cos(m\varphi)$ or $\sin(m\varphi)$, related to Cartesian
coordinates
\begin{equation}
x=r\cos\varphi;\quad y=r\sin\varphi.
\label{eq.xy}
\end{equation}
Starting at $Z_1=1$, the index $j$ is increased by increasing $m$, increasing
$n$, if this is exhausted as $n=m$ has been reached, and keeping $j$ even
for functions $\propto \cos(m\varphi)$, $m>0$ and keeping $j$ odd for functions
$\propto \sin(m\varphi)$ \cite{NollJOSA66},
\begin{equation}
Z_j=\left\{
\begin{array}{ll}
\sqrt{2n+2} R_n^m(r)\cos(m\varphi),& m>0, j\, \text{even} ;\\
\sqrt{2n+2} R_n^m(r)\sin(m\varphi),& m>0, j\, \text{odd} ;\\
\sqrt{n+1} R_n^m(r),& m=0. \\
\end{array}
\right.
\end{equation}
\begin{equation}
\int_{r<1} Z_j(r,\varphi) Z_k(r,\varphi) d^2r = \pi \delta_{j,k};
\quad
d^2r =r\,drd\varphi.
\label{eq.Zortho}
\end{equation}
\small \begin{verbatim}
Z_1 = 1*R_0^0(r).
Z_2 = 2*R_1^1(r)*cos(phi).    Z_3 = 2*R_1^1(r)*sin(phi).
Z_4 = 3^(1/2)*R_2^0(r).
Z_5 = 6^(1/2)*R_2^2(r)*sin(2*phi).    Z_6 = 6^(1/2)*R_2^2(r)*cos(2*phi).
Z_7 = 2*2^(1/2)*R_3^1(r)*sin(phi).    Z_8 = 2*2^(1/2)*R_3^1(r)*cos(phi).
Z_9 = 2*2^(1/2)*R_3^3(r)*sin(3*phi).    Z_10 = 2*2^(1/2)*R_3^3(r)*cos(3*phi).
Z_11 = 5^(1/2)*R_4^0(r).
Z_12 = 10^(1/2)*R_4^2(r)*cos(2*phi).    Z_13 = 10^(1/2)*R_4^2(r)*sin(2*phi).
Z_14 = 10^(1/2)*R_4^4(r)*cos(4*phi).    Z_15 = 10^(1/2)*R_4^4(r)*sin(4*phi).
Z_16 = 2*3^(1/2)*R_5^1(r)*cos(phi).    Z_17 = 2*3^(1/2)*R_5^1(r)*sin(phi).
Z_18 = 2*3^(1/2)*R_5^3(r)*cos(3*phi).    Z_19 = 2*3^(1/2)*R_5^3(r)*sin(3*phi).
Z_20 = 2*3^(1/2)*R_5^5(r)*cos(5*phi).    Z_21 = 2*3^(1/2)*R_5^5(r)*sin(5*phi).
Z_22 = 7^(1/2)*R_6^0(r).
Z_23 = 14^(1/2)*R_6^2(r)*sin(2*phi).    Z_24 = 14^(1/2)*R_6^2(r)*cos(2*phi).
Z_25 = 14^(1/2)*R_6^4(r)*sin(4*phi).    Z_26 = 14^(1/2)*R_6^4(r)*cos(4*phi).
Z_27 = 14^(1/2)*R_6^6(r)*sin(6*phi).    Z_28 = 14^(1/2)*R_6^6(r)*cos(6*phi).
Z_29 = 4*R_7^1(r)*sin(phi).    Z_30 = 4*R_7^1(r)*cos(phi).
Z_31 = 4*R_7^3(r)*sin(3*phi).    Z_32 = 4*R_7^3(r)*cos(3*phi).
Z_33 = 4*R_7^5(r)*sin(5*phi).    Z_34 = 4*R_7^5(r)*cos(5*phi).
Z_35 = 4*R_7^7(r)*sin(7*phi).    Z_36 = 4*R_7^7(r)*cos(7*phi).
Z_37 = 3*R_8^0(r).
Z_38 = 3*2^(1/2)*R_8^2(r)*cos(2*phi).    Z_39 = 3*2^(1/2)*R_8^2(r)*sin(2*phi).
Z_40 = 3*2^(1/2)*R_8^4(r)*cos(4*phi).    Z_41 = 3*2^(1/2)*R_8^4(r)*sin(4*phi).
Z_42 = 3*2^(1/2)*R_8^6(r)*cos(6*phi).    Z_43 = 3*2^(1/2)*R_8^6(r)*sin(6*phi).
Z_44 = 3*2^(1/2)*R_8^8(r)*cos(8*phi).    Z_45 = 3*2^(1/2)*R_8^8(r)*sin(8*phi).
Z_46 = 2*5^(1/2)*R_9^1(r)*cos(phi).    Z_47 = 2*5^(1/2)*R_9^1(r)*sin(phi).
Z_48 = 2*5^(1/2)*R_9^3(r)*cos(3*phi).    Z_49 = 2*5^(1/2)*R_9^3(r)*sin(3*phi).
Z_50 = 2*5^(1/2)*R_9^5(r)*cos(5*phi).    Z_51 = 2*5^(1/2)*R_9^5(r)*sin(5*phi).
Z_52 = 2*5^(1/2)*R_9^7(r)*cos(7*phi).    Z_53 = 2*5^(1/2)*R_9^7(r)*sin(7*phi).
Z_54 = 2*5^(1/2)*R_9^9(r)*cos(9*phi).    Z_55 = 2*5^(1/2)*R_9^9(r)*sin(9*phi).
Z_56 = 11^(1/2)*R_10^0(r).
Z_57 = 22^(1/2)*R_10^2(r)*sin(2*phi).    Z_58 = 22^(1/2)*R_10^2(r)*cos(2*phi).
Z_59 = 22^(1/2)*R_10^4(r)*sin(4*phi).    Z_60 = 22^(1/2)*R_10^4(r)*cos(4*phi).
Z_61 = 22^(1/2)*R_10^6(r)*sin(6*phi).    Z_62 = 22^(1/2)*R_10^6(r)*cos(6*phi).
Z_63 = 22^(1/2)*R_10^8(r)*sin(8*phi).    Z_64 = 22^(1/2)*R_10^8(r)*cos(8*phi).
Z_65 = 22^(1/2)*R_10^10(r)*sin(10*phi).    Z_66 = 22^(1/2)*R_10^10(r)*cos(10*phi).
\end{verbatim}\normalsize
Square roots $\surd k$ are written in the style \texttt{k\symbol{94}(1/2)}.

\subsection{Cartesian to Zernike} 
With (\ref{eq.xy}), each multinomial $x^py^q$ can be decomposed into terms
proportional to $R_n^m(r)\cos(m\varphi)$ if $q$ is even, or
proportional to $R_n^m(r)\sin(m\varphi)$ if $q$ is odd,
\begin{equation}
x^py^q = r^j\cos^p\varphi \sin^q\varphi
,\quad j\equiv p+q
.
\label{eq.c2z2D}
\end{equation}
\begin{eqnarray}
\cos^p\varphi \sin^q\varphi
&=&
\frac{(e^{i\varphi}+e^{-i\varphi})^p}{2^p}
\frac{(e^{i\varphi}-e^{-i\varphi})^q}{(2i)^q}
\\
&=&
\frac{(-1)^{\lfloor q/2\rfloor}}{2^{p+q}}
\sum_{s=0}^p\sum_{l=0}^q
\binom{p}{ s}
\binom{q}{ l}
(-1)^{q-l}
\times
\left\{
\begin{array}{ll}
\cos[(2s-p+2l-q)\varphi], & q\, \mathrm{even};
\\
\sin[(2s-p+2l-q)\varphi], & q\, \mathrm{odd};
\end{array}
\right.
\\
&=&
\frac{(-1)^{\lfloor q/2\rfloor}}{2^j}
\times
\left\{
\begin{array}{ll}
\left[
2\sum_{m=0}^{\lfloor(j-1)/2\rfloor }\sum_{l=\max(0,m-p)}^{\min(q,m)}
\binom{p}{ m-l}
\binom{q}{ l}
(-1)^l
\cos[(j-2m)\varphi]+
C(p,q)\right]
, & q\, \mathrm{even};
\\
2\sum_{m=0}^{\lfloor(j-1)/2\rfloor }\sum_{l=\max(0,m-p)}^{\min(q,m)}
\binom{p}{ m-l}
\binom{q}{ l}
(-1)^l
\sin[(j-2m)\varphi], & q\, \mathrm{odd};
\end{array}
\right.
\label{eq.phiProd}
\end{eqnarray}
where $j\equiv p+q$,
and where
\begin{equation}
C(p,q)\equiv \left\{
\begin{array}{ll}
\sum_{l=0}^q\binom{p}{ j/2-l}\binom{q}{ l}(-1)^l
=
2^j\frac{\Gamma((p+1)/2)}{\Gamma(j/2+1)\Gamma((1-q)/2)}
=
(-1)^{q/2}
2^{j/2}
\frac{(q-1)!!(p-1)!!}{(j/2)!}
, & j\,\mathrm{even};
\\
0, & j\,\mathrm{odd}
\end{array}
\right.
\end{equation}
is a ``dangling'' component representing the contribution at $j=2m$ if $q$ and $j$ are both even \cite[(B2b)]{DaiJOSAA12}.

This extends Section 1.32 of the Gradstein-Ryshik tables \cite{GR}
and distributes
$\cos^p\varphi \sin^q\varphi$ into sums over $\cos(m\varphi)$
or $\sin(m\varphi)$:
\small \begin{verbatim}
cos^2 phi = [ 1 +cos(2*phi) ]/2.
cos phi sin phi = [ sin(2*phi) ]/2.
sin^2 phi = [ 1 -cos(2*phi) ]/2.
cos^3 phi = [ 3*cos(phi) +cos(3*phi) ]/4.
cos^2 phi sin phi = [ sin(phi) +sin(3*phi) ]/4.
cos phi sin^2 phi = [ cos(phi) -cos(3*phi) ]/4.
sin^3 phi = [ 3*sin(phi) -sin(3*phi) ]/4.
cos^4 phi = [ 3 +4*cos(2*phi) +cos(4*phi) ]/8.
cos^3 phi sin phi = [ 2*sin(2*phi) +sin(4*phi) ]/8.
cos^2 phi sin^2 phi = [ 1 -cos(4*phi) ]/8.
cos phi sin^3 phi = [ 2*sin(2*phi) -sin(4*phi) ]/8.
sin^4 phi = [ 3 -4*cos(2*phi) +cos(4*phi) ]/8.
\end{verbatim}\normalsize

Back to (\ref{eq.c2z2D}), these are multiplied by $r^j=r^{p+q}$ for an intermediate table
of Cartesian to spherical transformations:
\small \begin{verbatim}
x = r*cos(phi).
y = r*sin(phi).
x^2 = 1/2*r^2*cos(2*phi) +1/2*r^2.
x y = 1/2*r^2*sin(2*phi).
y^2 =  -1/2*r^2*cos(2*phi) +1/2*r^2.
x^3 = 1/4*r^3*cos(3*phi) +3/4*r^3*cos(phi).
x^2 y = 1/4*r^3*sin(3*phi) +1/4*r^3*sin(phi).
x y^2 =  -1/4*r^3*cos(3*phi) +1/4*r^3*cos(phi).
y^3 =  -1/4*r^3*sin(3*phi) +3/4*r^3*sin(phi).
x^4 = 1/8*r^4*cos(4*phi) +1/2*r^4*cos(2*phi) +3/8*r^4.
x^3 y = 1/8*r^4*sin(4*phi) +1/4*r^4*sin(2*phi).
x^2 y^2 =  -1/8*r^4*cos(4*phi) +1/8*r^4.
x y^3 =  -1/8*r^4*sin(4*phi) +1/4*r^4*sin(2*phi).
y^4 = 1/8*r^4*cos(4*phi) -1/2*r^4*cos(2*phi) +3/8*r^4.
x^5 = 1/16*r^5*cos(5*phi) +5/16*r^5*cos(3*phi) +5/8*r^5*cos(phi).
x^4 y = 1/16*r^5*sin(5*phi) +3/16*r^5*sin(3*phi) +1/8*r^5*sin(phi).
x^3 y^2 =  -1/16*r^5*cos(5*phi) -1/16*r^5*cos(3*phi) +1/8*r^5*cos(phi).
x^2 y^3 =  -1/16*r^5*sin(5*phi) +1/16*r^5*sin(3*phi) +1/8*r^5*sin(phi).
x y^4 = 1/16*r^5*cos(5*phi) -3/16*r^5*cos(3*phi) +1/8*r^5*cos(phi).
y^5 = 1/16*r^5*sin(5*phi) -5/16*r^5*sin(3*phi) +5/8*r^5*sin(phi).
\end{verbatim}\normalsize
Each factor $r^j$ is expanded with
(\ref{eq.rofR})
in a sum over $R$,
selecting the line of the table of section \ref{sec.rofR} associated with $m$:
\small \begin{verbatim}
x = R_1^1(r)*cos(phi).
y = R_1^1(r)*sin(phi).
x^2 = 1/2*R_2^2(r)*cos(2*phi) +1/4*R_0^0(r) +1/4*R_2^0(r).
x y = 1/2*R_2^2(r)*sin(2*phi).
y^2 = -1/2*R_2^2(r)*cos(2*phi) +1/4*R_0^0(r) +1/4*R_2^0(r).
x^3 = 1/4*R_3^3(r)*cos(3*phi) +1/2*cos(phi)*R_1^1(r) +1/4*cos(phi)*R_3^1(r).
x^2 y = 1/4*R_3^3(r)*sin(3*phi) +1/6*sin(phi)*R_1^1(r) +1/12*sin(phi)*R_3^1(r).
x y^2 = -1/4*R_3^3(r)*cos(3*phi) +1/6*cos(phi)*R_1^1(r) +1/12*cos(phi)*R_3^1(r).
y^3 = -1/4*R_3^3(r)*sin(3*phi) +1/2*sin(phi)*R_1^1(r) +1/4*sin(phi)*R_3^1(r).
x^4 = 1/8*R_4^4(r)*cos(4*phi) +3/8*cos(2*phi)*R_2^2(r) +1/8*cos(2*phi)*R_4^2(r) +1/8*R_0^0(r) 
   +3/16*R_2^0(r) +1/16*R_4^0(r).
x^3 y = 1/8*R_4^4(r)*sin(4*phi) +3/16*sin(2*phi)*R_2^2(r) +1/16*sin(2*phi)*R_4^2(r).
x^2 y^2 = -1/8*R_4^4(r)*cos(4*phi) +1/24*R_0^0(r) +1/16*R_2^0(r) +1/48*R_4^0(r).
x y^3 = -1/8*R_4^4(r)*sin(4*phi) +3/16*sin(2*phi)*R_2^2(r) +1/16*sin(2*phi)*R_4^2(r).
y^4 = 1/8*R_4^4(r)*cos(4*phi) -3/8*cos(2*phi)*R_2^2(r) -1/8*cos(2*phi)*R_4^2(r) +1/8*R_0^0(r) 
   +3/16*R_2^0(r) +1/16*R_4^0(r).
x^5 = 1/16*R_5^5(r)*cos(5*phi) +1/4*cos(3*phi)*R_3^3(r) +1/16*cos(3*phi)*R_5^3(r) 
   +5/16*cos(phi)*R_1^1(r) +1/4*cos(phi)*R_3^1(r) +1/16*cos(phi)*R_5^1(r).
x^4 y = 1/16*R_5^5(r)*sin(5*phi) +3/20*sin(3*phi)*R_3^3(r) +3/80*sin(3*phi)*R_5^3(r) 
   +1/16*sin(phi)*R_1^1(r) +1/20*sin(phi)*R_3^1(r) +1/80*sin(phi)*R_5^1(r).
x^3 y^2 = -1/16*R_5^5(r)*cos(5*phi) -1/20*cos(3*phi)*R_3^3(r) -1/80*cos(3*phi)*R_5^3(r) 
   +1/16*cos(phi)*R_1^1(r) +1/20*cos(phi)*R_3^1(r) +1/80*cos(phi)*R_5^1(r).
x^2 y^3 = -1/16*R_5^5(r)*sin(5*phi) +1/20*sin(3*phi)*R_3^3(r) +1/80*sin(3*phi)*R_5^3(r) 
   +1/16*sin(phi)*R_1^1(r) +1/20*sin(phi)*R_3^1(r) +1/80*sin(phi)*R_5^1(r).
x y^4 = 1/16*R_5^5(r)*cos(5*phi) -3/20*cos(3*phi)*R_3^3(r) -3/80*cos(3*phi)*R_5^3(r) 
   +1/16*cos(phi)*R_1^1(r) +1/20*cos(phi)*R_3^1(r) +1/80*cos(phi)*R_5^1(r).
y^5 = 1/16*R_5^5(r)*sin(5*phi) -1/4*sin(3*phi)*R_3^3(r) -1/16*sin(3*phi)*R_5^3(r) 
   +5/16*sin(phi)*R_1^1(r) +1/4*sin(phi)*R_3^1(r) +1/16*sin(phi)*R_5^1(r).
x^6 = 1/32*R_6^6(r)*cos(6*phi) +5/32*cos(4*phi)*R_4^4(r) +1/32*cos(4*phi)*R_6^4(r) 
   +9/32*cos(2*phi)*R_2^2(r) +5/32*cos(2*phi)*R_4^2(r) +1/32*cos(2*phi)*R_6^2(r) +5/64*R_0^0(r) 
   +9/64*R_2^0(r) +5/64*R_4^0(r) +1/64*R_6^0(r).
x^5 y = 1/32*R_6^6(r)*sin(6*phi) +5/48*sin(4*phi)*R_4^4(r) +1/48*sin(4*phi)*R_6^4(r) 
   +3/32*sin(2*phi)*R_2^2(r) +5/96*sin(2*phi)*R_4^2(r) +1/96*sin(2*phi)*R_6^2(r).
x^4 y^2 = -1/32*R_6^6(r)*cos(6*phi) -5/96*cos(4*phi)*R_4^4(r) -1/96*cos(4*phi)*R_6^4(r) 
   +3/160*cos(2*phi)*R_2^2(r) +1/96*cos(2*phi)*R_4^2(r) +1/480*cos(2*phi)*R_6^2(r) +1/64*R_0^0(r) 
   +9/320*R_2^0(r) +1/64*R_4^0(r) +1/320*R_6^0(r).
x^3 y^3 = -1/32*R_6^6(r)*sin(6*phi) +9/160*sin(2*phi)*R_2^2(r) +1/32*sin(2*phi)*R_4^2(r) 
   +1/160*sin(2*phi)*R_6^2(r).
x^2 y^4 = 1/32*R_6^6(r)*cos(6*phi) -5/96*cos(4*phi)*R_4^4(r) -1/96*cos(4*phi)*R_6^4(r) 
   -3/160*cos(2*phi)*R_2^2(r) -1/96*cos(2*phi)*R_4^2(r) -1/480*cos(2*phi)*R_6^2(r) +1/64*R_0^0(r) 
   +9/320*R_2^0(r) +1/64*R_4^0(r) +1/320*R_6^0(r).
x y^5 = 1/32*R_6^6(r)*sin(6*phi) -5/48*sin(4*phi)*R_4^4(r) -1/48*sin(4*phi)*R_6^4(r) 
   +3/32*sin(2*phi)*R_2^2(r) +5/96*sin(2*phi)*R_4^2(r) +1/96*sin(2*phi)*R_6^2(r).
y^6 = -1/32*R_6^6(r)*cos(6*phi) +5/32*cos(4*phi)*R_4^4(r) +1/32*cos(4*phi)*R_6^4(r) 
   -9/32*cos(2*phi)*R_2^2(r) -5/32*cos(2*phi)*R_4^2(r) -1/32*cos(2*phi)*R_6^2(r) +5/64*R_0^0(r) 
   +9/64*R_2^0(r) +5/64*R_4^0(r) +1/64*R_6^0(r).
x^7 = 1/64*R_7^7(r)*cos(7*phi) +3/32*cos(5*phi)*R_5^5(r) +1/64*cos(5*phi)*R_7^5(r) 
   +7/32*cos(3*phi)*R_3^3(r) +3/32*cos(3*phi)*R_5^3(r) +1/64*cos(3*phi)*R_7^3(r) +7/32*cos(phi)*R_1^1(r) 
   +7/32*cos(phi)*R_3^1(r) +3/32*cos(phi)*R_5^1(r) +1/64*cos(phi)*R_7^1(r).
x^6 y = 1/64*R_7^7(r)*sin(7*phi) +15/224*sin(5*phi)*R_5^5(r) +5/448*sin(5*phi)*R_7^5(r) 
   +3/32*sin(3*phi)*R_3^3(r) +9/224*sin(3*phi)*R_5^3(r) +3/448*sin(3*phi)*R_7^3(r) 
   +1/32*sin(phi)*R_1^1(r) +1/32*sin(phi)*R_3^1(r) +3/224*sin(phi)*R_5^1(r) +1/448*sin(phi)*R_7^1(r).
x^5 y^2 = -1/64*R_7^7(r)*cos(7*phi) -9/224*cos(5*phi)*R_5^5(r) -3/448*cos(5*phi)*R_7^5(r) 
   -1/96*cos(3*phi)*R_3^3(r) -1/224*cos(3*phi)*R_5^3(r) -1/1344*cos(3*phi)*R_7^3(r) 
   +1/32*cos(phi)*R_1^1(r) +1/32*cos(phi)*R_3^1(r) +3/224*cos(phi)*R_5^1(r) +1/448*cos(phi)*R_7^1(r).
x^4 y^3 = -1/64*R_7^7(r)*sin(7*phi) -3/224*sin(5*phi)*R_5^5(r) -1/448*sin(5*phi)*R_7^5(r) 
   +1/32*sin(3*phi)*R_3^3(r) +3/224*sin(3*phi)*R_5^3(r) +1/448*sin(3*phi)*R_7^3(r) 
   +3/160*sin(phi)*R_1^1(r) +3/160*sin(phi)*R_3^1(r) +9/1120*sin(phi)*R_5^1(r) +3/2240*sin(phi)*R_7^1(r).
x^3 y^4 = 1/64*R_7^7(r)*cos(7*phi) -3/224*cos(5*phi)*R_5^5(r) -1/448*cos(5*phi)*R_7^5(r) 
   -1/32*cos(3*phi)*R_3^3(r) -3/224*cos(3*phi)*R_5^3(r) -1/448*cos(3*phi)*R_7^3(r) 
   +3/160*cos(phi)*R_1^1(r) +3/160*cos(phi)*R_3^1(r) +9/1120*cos(phi)*R_5^1(r) +3/2240*cos(phi)*R_7^1(r).
x^2 y^5 = 1/64*R_7^7(r)*sin(7*phi) -9/224*sin(5*phi)*R_5^5(r) -3/448*sin(5*phi)*R_7^5(r) 
   +1/96*sin(3*phi)*R_3^3(r) +1/224*sin(3*phi)*R_5^3(r) +1/1344*sin(3*phi)*R_7^3(r) 
   +1/32*sin(phi)*R_1^1(r) +1/32*sin(phi)*R_3^1(r) +3/224*sin(phi)*R_5^1(r) +1/448*sin(phi)*R_7^1(r).
x y^6 = -1/64*R_7^7(r)*cos(7*phi) +15/224*cos(5*phi)*R_5^5(r) +5/448*cos(5*phi)*R_7^5(r) 
   -3/32*cos(3*phi)*R_3^3(r) -9/224*cos(3*phi)*R_5^3(r) -3/448*cos(3*phi)*R_7^3(r) 
   +1/32*cos(phi)*R_1^1(r) +1/32*cos(phi)*R_3^1(r) +3/224*cos(phi)*R_5^1(r) +1/448*cos(phi)*R_7^1(r).
y^7 = -1/64*R_7^7(r)*sin(7*phi) +3/32*sin(5*phi)*R_5^5(r) +1/64*sin(5*phi)*R_7^5(r) 
   -7/32*sin(3*phi)*R_3^3(r) -3/32*sin(3*phi)*R_5^3(r) -1/64*sin(3*phi)*R_7^3(r) +7/32*sin(phi)*R_1^1(r) 
   +7/32*sin(phi)*R_3^1(r) +3/32*sin(phi)*R_5^1(r) +1/64*sin(phi)*R_7^1(r).
x^8 = 1/128*R_8^8(r)*cos(8*phi) +7/128*cos(6*phi)*R_6^6(r) +1/128*cos(6*phi)*R_8^6(r) 
   +5/32*cos(4*phi)*R_4^4(r) +7/128*cos(4*phi)*R_6^4(r) +1/128*cos(4*phi)*R_8^4(r) 
   +7/32*cos(2*phi)*R_2^2(r) +5/32*cos(2*phi)*R_4^2(r) +7/128*cos(2*phi)*R_6^2(r) 
   +1/128*cos(2*phi)*R_8^2(r) +7/128*R_0^0(r) +7/64*R_2^0(r) +5/64*R_4^0(r) +7/256*R_6^0(r) 
   +1/256*R_8^0(r).
x^7 y = 1/128*R_8^8(r)*sin(8*phi) +21/512*sin(6*phi)*R_6^6(r) +3/512*sin(6*phi)*R_8^6(r) 
   +5/64*sin(4*phi)*R_4^4(r) +7/256*sin(4*phi)*R_6^4(r) +1/256*sin(4*phi)*R_8^4(r) 
   +7/128*sin(2*phi)*R_2^2(r) +5/128*sin(2*phi)*R_4^2(r) +7/512*sin(2*phi)*R_6^2(r) 
   +1/512*sin(2*phi)*R_8^2(r).
x^6 y^2 = -1/128*R_8^8(r)*cos(8*phi) -7/256*cos(6*phi)*R_6^6(r) -1/256*cos(6*phi)*R_8^6(r) 
   -5/224*cos(4*phi)*R_4^4(r) -1/128*cos(4*phi)*R_6^4(r) -1/896*cos(4*phi)*R_8^4(r) 
   +1/64*cos(2*phi)*R_2^2(r) +5/448*cos(2*phi)*R_4^2(r) +1/256*cos(2*phi)*R_6^2(r) 
   +1/1792*cos(2*phi)*R_8^2(r) +1/128*R_0^0(r) +1/64*R_2^0(r) +5/448*R_4^0(r) +1/256*R_6^0(r) 
   +1/1792*R_8^0(r).
x^5 y^3 = -1/128*R_8^8(r)*sin(8*phi) -7/512*sin(6*phi)*R_6^6(r) -1/512*sin(6*phi)*R_8^6(r) 
   +5/448*sin(4*phi)*R_4^4(r) +1/256*sin(4*phi)*R_6^4(r) +1/1792*sin(4*phi)*R_8^4(r) 
   +3/128*sin(2*phi)*R_2^2(r) +15/896*sin(2*phi)*R_4^2(r) +3/512*sin(2*phi)*R_6^2(r) 
   +3/3584*sin(2*phi)*R_8^2(r).
x^4 y^4 = 1/128*R_8^8(r)*cos(8*phi) -5/224*cos(4*phi)*R_4^4(r) -1/128*cos(4*phi)*R_6^4(r) 
   -1/896*cos(4*phi)*R_8^4(r) +3/640*R_0^0(r) +3/320*R_2^0(r) +3/448*R_4^0(r) +3/1280*R_6^0(r) 
   +3/8960*R_8^0(r).
x^3 y^5 = 1/128*R_8^8(r)*sin(8*phi) -7/512*sin(6*phi)*R_6^6(r) -1/512*sin(6*phi)*R_8^6(r) 
   -5/448*sin(4*phi)*R_4^4(r) -1/256*sin(4*phi)*R_6^4(r) -1/1792*sin(4*phi)*R_8^4(r) 
   +3/128*sin(2*phi)*R_2^2(r) +15/896*sin(2*phi)*R_4^2(r) +3/512*sin(2*phi)*R_6^2(r) 
   +3/3584*sin(2*phi)*R_8^2(r).
x^2 y^6 = -1/128*R_8^8(r)*cos(8*phi) +7/256*cos(6*phi)*R_6^6(r) +1/256*cos(6*phi)*R_8^6(r) 
   -5/224*cos(4*phi)*R_4^4(r) -1/128*cos(4*phi)*R_6^4(r) -1/896*cos(4*phi)*R_8^4(r) 
   -1/64*cos(2*phi)*R_2^2(r) -5/448*cos(2*phi)*R_4^2(r) -1/256*cos(2*phi)*R_6^2(r) 
   -1/1792*cos(2*phi)*R_8^2(r) +1/128*R_0^0(r) +1/64*R_2^0(r) +5/448*R_4^0(r) +1/256*R_6^0(r) 
   +1/1792*R_8^0(r).
x y^7 = -1/128*R_8^8(r)*sin(8*phi) +21/512*sin(6*phi)*R_6^6(r) +3/512*sin(6*phi)*R_8^6(r) 
   -5/64*sin(4*phi)*R_4^4(r) -7/256*sin(4*phi)*R_6^4(r) -1/256*sin(4*phi)*R_8^4(r) 
   +7/128*sin(2*phi)*R_2^2(r) +5/128*sin(2*phi)*R_4^2(r) +7/512*sin(2*phi)*R_6^2(r) 
   +1/512*sin(2*phi)*R_8^2(r).
y^8 = 1/128*R_8^8(r)*cos(8*phi) -7/128*cos(6*phi)*R_6^6(r) -1/128*cos(6*phi)*R_8^6(r) 
   +5/32*cos(4*phi)*R_4^4(r) +7/128*cos(4*phi)*R_6^4(r) +1/128*cos(4*phi)*R_8^4(r) 
   -7/32*cos(2*phi)*R_2^2(r) -5/32*cos(2*phi)*R_4^2(r) -7/128*cos(2*phi)*R_6^2(r) 
   -1/128*cos(2*phi)*R_8^2(r) +7/128*R_0^0(r) +7/64*R_2^0(r) +5/64*R_4^0(r) +7/256*R_6^0(r) 
   +1/256*R_8^0(r).
\end{verbatim}\normalsize
If the factor $y$ is absent,
ie, for pure powers \texttt{x\symbol{94}p}, the coefficients could also 
be taken from
the column headed with two \texttt{p}
in Conforti's Table 1 \cite{ConfortiOL8}.

\subsection{Zernike to Cartesian} 
The reversal of the transformation of the previous section
splits
each $r^j\cos(m\varphi)$ or $r^j\sin(m\varphi)$ in the expansions
of Section \ref{sec.rofR} into
\begin{eqnarray}
r^{j-m}r^m\cos(m\varphi)&=&(x^2+y^2)^{(j-m)/2}\sum_{k=0,2,4,\ldots}^m (-1)^{\lfloor k/2\rfloor}\binom{m }{ k} x^{m-k}y^k,
\\
r^{j-m}r^m\sin(m\varphi)&=&(x^2+y^2)^{(j-m)/2}\sum_{k=1,3,5,\ldots}^m (-1)^{\lfloor k/2\rfloor}\binom{m }{ k} x^{m-k}y^k,
\end{eqnarray}
which yields a sum over the bivariate Cartesian products after binomial expansion
of
$(x^2+y^2)^{(j-m)/2}
$.
The two cases are
\begin{eqnarray}
r^j\cos(m\varphi)
&=&
\sum_{t=0}^{(j-m)/2}\binom{(j-m)/2 }{ t} x^{2t}y^{j-m-2t}\sum_{k=0,2,4,\ldots}^m (-1)^{k/2}
\binom{m }{ k}x^{m-k}y^k
\\
&=&
\sum_{t=-\lfloor m/2\rfloor}^{(j-m)/2}
\sum_{k=\max(0,-t)}^{\min((j-m)/2-t,\lfloor m/2\rfloor)}
(-1)^k
\binom{m }{ 2k}
\binom{(j-m)/2 }{ t+k}
x^{2t+m}y^{j-m-2t}
,
\label{eq.rjcos}
\end{eqnarray}
and
\begin{eqnarray}
r^j\sin(m\varphi)
&=&
\sum_{t=0}^{(j-m)/2}\binom{(j-m)/2 }{ t} x^{2t}y^{j-m-2t}\sum_{k=1,3,5,\ldots}^m (-1)^{(k-1)/2}
\binom{m }{ k}x^{m-k}y^k
\\
&=&
\sum_{t=-\lfloor (m-1)/2\rfloor}^{(j-m)/2}
\sum_{k=\max(0,-t)}^{\min((j-m)/2-t,\lfloor (m-1)/2\rfloor)}
(-1)^k
\binom{m }{ 2k+1}
\binom{(j-m)/2 }{ t+k}
x^{2t+m-1}y^{j-m-2t+1}
\label{eq.rjsin}
.
\end{eqnarray}
The explicit tabulation for small $j$ and small $m$ starts as follows:
\small \begin{verbatim}
r cos(phi) = x.
r sin(phi) = y.
r^2 = x^2 +y^2.
r^2 cos(2*phi) = x^2 -y^2.
r^2 sin(2*phi) = 2*x*y.
r^3 cos(phi) = x^3 +x*y^2.
r^3 sin(phi) = y*x^2 +y^3.
r^3 cos(3*phi) = x^3 -3*x*y^2.
r^3 sin(3*phi) = 3*y*x^2 -y^3.
r^4 = x^4 +2*x^2*y^2 +y^4.
r^4 cos(2*phi) = x^4 -y^4.
r^4 sin(2*phi) = 2*x^3*y +2*x*y^3.
r^4 cos(4*phi) = x^4 -6*x^2*y^2 +y^4.
r^4 sin(4*phi) = 4*x^3*y -4*x*y^3.
r^5 cos(phi) = x^5 +2*x^3*y^2 +x*y^4.
r^5 sin(phi) = y*x^4 +2*x^2*y^3 +y^5.
r^5 cos(3*phi) = x^5 -2*x^3*y^2 -3*x*y^4.
r^5 sin(3*phi) = 3*y*x^4 +2*x^2*y^3 -y^5.
r^5 cos(5*phi) = x^5 -10*x^3*y^2 +5*x*y^4.
r^5 sin(5*phi) = 5*y*x^4 -10*x^2*y^3 +y^5.
\end{verbatim}\normalsize

Linear superposition by inserting these into the right hand sides of (\ref{eq.Rpolyxm})
as tabulated in Section \ref{sec.rofR} yields
\small \begin{verbatim}
R_0^0(r) = 1 Z_1 = 1.
R_1^1(r) cos(phi) = 1/2 Z_2 = x.
R_1^1(r) sin(phi) = 1/2 Z_3 = y.
R_2^0(r) = 1/3*3^(1/2) Z_4 = 2*x^2 +2*y^2 -1.
R_2^2(r) cos(2*phi) = 1/6*6^(1/2) Z_6 = x^2 -y^2.
R_2^2(r) sin(2*phi) = 1/6*6^(1/2) Z_5 = 2*x*y.
R_3^1(r) cos(phi) = 1/4*2^(1/2) Z_8 = 3*x^3 +3*x*y^2 -2*x.
R_3^1(r) sin(phi) = 1/4*2^(1/2) Z_7 = 3*x^2*y +3*y^3 -2*y.
R_3^3(r) cos(3*phi) = 1/4*2^(1/2) Z_10 = x^3 -3*x*y^2.
R_3^3(r) sin(3*phi) = 1/4*2^(1/2) Z_9 = 3*x^2*y -y^3.
R_4^0(r) = 1/5*5^(1/2) Z_11 = 6*x^4 +12*x^2*y^2 +6*y^4 -6*x^2 -6*y^2 +1.
R_4^2(r) cos(2*phi) = 1/10*10^(1/2) Z_12 = 4*x^4 -4*y^4 -3*x^2 +3*y^2.
R_4^2(r) sin(2*phi) = 1/10*10^(1/2) Z_13 = 8*x^3*y +8*x*y^3 -6*x*y.
R_4^4(r) cos(4*phi) = 1/10*10^(1/2) Z_14 = x^4 -6*x^2*y^2 +y^4.
R_4^4(r) sin(4*phi) = 1/10*10^(1/2) Z_15 = 4*x^3*y -4*x*y^3.
R_5^1(r) cos(phi) = 1/6*3^(1/2) Z_16 = 10*x^5 +20*x^3*y^2 +10*x*y^4 -12*x^3 -12*x*y^2 +3*x.
R_5^1(r) sin(phi) = 1/6*3^(1/2) Z_17 = 10*x^4*y +20*x^2*y^3 +10*y^5 -12*x^2*y -12*y^3 +3*y.
R_5^3(r) cos(3*phi) = 1/6*3^(1/2) Z_18 = 5*x^5 -10*x^3*y^2 -15*x*y^4 -4*x^3 +12*x*y^2.
R_5^3(r) sin(3*phi) = 1/6*3^(1/2) Z_19 = 15*x^4*y +10*x^2*y^3 -5*y^5 -12*x^2*y +4*y^3.
R_5^5(r) cos(5*phi) = 1/6*3^(1/2) Z_20 = x^5 -10*x^3*y^2 +5*x*y^4.
R_5^5(r) sin(5*phi) = 1/6*3^(1/2) Z_21 = 5*x^4*y -10*x^2*y^3 +y^5.
R_6^0(r) = 1/7*7^(1/2) Z_22 = 20*x^6 +60*x^4*y^2 +60*x^2*y^4 +20*y^6 -30*x^4 -60*x^2*y^2 -30*y^4 
   +12*x^2 +12*y^2 -1.
R_6^2(r) cos(2*phi) = 1/14*14^(1/2) Z_24 = 15*x^6 +15*x^4*y^2 -15*x^2*y^4 -15*y^6 -20*x^4 +20*y^4 
   +6*x^2 -6*y^2.
R_6^2(r) sin(2*phi) = 1/14*14^(1/2) Z_23 = 30*x^5*y +60*x^3*y^3 +30*x*y^5 -40*x^3*y -40*x*y^3 +12*x*y.
R_6^4(r) cos(4*phi) = 1/14*14^(1/2) Z_26 = 6*x^6 -30*x^4*y^2 -30*x^2*y^4 +6*y^6 -5*x^4 +30*x^2*y^2 
   -5*y^4.
R_6^4(r) sin(4*phi) = 1/14*14^(1/2) Z_25 = 24*x^5*y -24*x*y^5 -20*x^3*y +20*x*y^3.
R_6^6(r) cos(6*phi) = 1/14*14^(1/2) Z_28 = x^6 -15*x^4*y^2 +15*x^2*y^4 -y^6.
R_6^6(r) sin(6*phi) = 1/14*14^(1/2) Z_27 = 6*x^5*y -20*x^3*y^3 +6*x*y^5.
R_7^1(r) cos(phi) = 1/4 Z_30 = 35*x^7 +105*x^5*y^2 +105*x^3*y^4 +35*x*y^6 -60*x^5 -120*x^3*y^2 
   -60*x*y^4 +30*x^3 +30*x*y^2 -4*x.
R_7^1(r) sin(phi) = 1/4 Z_29 = 35*x^6*y +105*x^4*y^3 +105*x^2*y^5 +35*y^7 -60*x^4*y -120*x^2*y^3 
   -60*y^5 +30*x^2*y +30*y^3 -4*y.
R_7^3(r) cos(3*phi) = 1/4 Z_32 = 21*x^7 -21*x^5*y^2 -105*x^3*y^4 -63*x*y^6 -30*x^5 +60*x^3*y^2 
   +90*x*y^4 +10*x^3 -30*x*y^2.
R_7^3(r) sin(3*phi) = 1/4 Z_31 = 63*x^6*y +105*x^4*y^3 +21*x^2*y^5 -21*y^7 -90*x^4*y -60*x^2*y^3 
   +30*y^5 +30*x^2*y -10*y^3.
R_7^5(r) cos(5*phi) = 1/4 Z_34 = 7*x^7 -63*x^5*y^2 -35*x^3*y^4 +35*x*y^6 -6*x^5 +60*x^3*y^2 -30*x*y^4.
R_7^5(r) sin(5*phi) = 1/4 Z_33 = 35*x^6*y -35*x^4*y^3 -63*x^2*y^5 +7*y^7 -30*x^4*y +60*x^2*y^3 -6*y^5.
R_7^7(r) cos(7*phi) = 1/4 Z_36 = x^7 -21*x^5*y^2 +35*x^3*y^4 -7*x*y^6.
R_7^7(r) sin(7*phi) = 1/4 Z_35 = 7*x^6*y -35*x^4*y^3 +21*x^2*y^5 -y^7.
R_8^0(r) = 1/3 Z_37 = 70*x^8 +280*x^6*y^2 +420*x^4*y^4 +280*x^2*y^6 +70*y^8 -140*x^6 -420*x^4*y^2 
   -420*x^2*y^4 -140*y^6 +90*x^4 +180*x^2*y^2 +90*y^4 -20*x^2 -20*y^2 +1.
R_8^2(r) cos(2*phi) = 1/6*2^(1/2) Z_38 = 56*x^8 +112*x^6*y^2 -112*x^2*y^6 -56*y^8 -105*x^6 
   -105*x^4*y^2 +105*x^2*y^4 +105*y^6 +60*x^4 -60*y^4 -10*x^2 +10*y^2.
R_8^2(r) sin(2*phi) = 1/6*2^(1/2) Z_39 = 112*x^7*y +336*x^5*y^3 +336*x^3*y^5 +112*x*y^7 -210*x^5*y 
   -420*x^3*y^3 -210*x*y^5 +120*x^3*y +120*x*y^3 -20*x*y.
R_8^4(r) cos(4*phi) = 1/6*2^(1/2) Z_40 = 28*x^8 -112*x^6*y^2 -280*x^4*y^4 -112*x^2*y^6 +28*y^8 -42*x^6 
   +210*x^4*y^2 +210*x^2*y^4 -42*y^6 +15*x^4 -90*x^2*y^2 +15*y^4.
R_8^4(r) sin(4*phi) = 1/6*2^(1/2) Z_41 = 112*x^7*y +112*x^5*y^3 -112*x^3*y^5 -112*x*y^7 -168*x^5*y 
   +168*x*y^5 +60*x^3*y -60*x*y^3.
R_8^6(r) cos(6*phi) = 1/6*2^(1/2) Z_42 = 8*x^8 -112*x^6*y^2 +112*x^2*y^6 -8*y^8 -7*x^6 +105*x^4*y^2 
   -105*x^2*y^4 +7*y^6.
R_8^6(r) sin(6*phi) = 1/6*2^(1/2) Z_43 = 48*x^7*y -112*x^5*y^3 -112*x^3*y^5 +48*x*y^7 -42*x^5*y 
   +140*x^3*y^3 -42*x*y^5.
R_8^8(r) cos(8*phi) = 1/6*2^(1/2) Z_44 = x^8 -28*x^6*y^2 +70*x^4*y^4 -28*x^2*y^6 +y^8.
R_8^8(r) sin(8*phi) = 1/6*2^(1/2) Z_45 = 8*x^7*y -56*x^5*y^3 +56*x^3*y^5 -8*x*y^7.
R_9^1(r) cos(phi) = 1/10*5^(1/2) Z_46 = 126*x^9 +504*x^7*y^2 +756*x^5*y^4 +504*x^3*y^6 +126*x*y^8 
   -280*x^7 -840*x^5*y^2 -840*x^3*y^4 -280*x*y^6 +210*x^5 +420*x^3*y^2 +210*x*y^4 -60*x^3 -60*x*y^2 +5*x.
R_9^1(r) sin(phi) = 1/10*5^(1/2) Z_47 = 126*x^8*y +504*x^6*y^3 +756*x^4*y^5 +504*x^2*y^7 +126*y^9 
   -280*x^6*y -840*x^4*y^3 -840*x^2*y^5 -280*y^7 +210*x^4*y +420*x^2*y^3 +210*y^5 -60*x^2*y -60*y^3 +5*y.
R_9^3(r) cos(3*phi) = 1/10*5^(1/2) Z_48 = 84*x^9 -504*x^5*y^4 -672*x^3*y^6 -252*x*y^8 -168*x^7 
   +168*x^5*y^2 +840*x^3*y^4 +504*x*y^6 +105*x^5 -210*x^3*y^2 -315*x*y^4 -20*x^3 +60*x*y^2.
R_9^3(r) sin(3*phi) = 1/10*5^(1/2) Z_49 = 252*x^8*y +672*x^6*y^3 +504*x^4*y^5 -84*y^9 -504*x^6*y 
   -840*x^4*y^3 -168*x^2*y^5 +168*y^7 +315*x^4*y +210*x^2*y^3 -105*y^5 -60*x^2*y +20*y^3.
R_9^5(r) cos(5*phi) = 1/10*5^(1/2) Z_50 = 36*x^9 -288*x^7*y^2 -504*x^5*y^4 +180*x*y^8 -56*x^7 
   +504*x^5*y^2 +280*x^3*y^4 -280*x*y^6 +21*x^5 -210*x^3*y^2 +105*x*y^4.
R_9^5(r) sin(5*phi) = 1/10*5^(1/2) Z_51 = 180*x^8*y -504*x^4*y^5 -288*x^2*y^7 +36*y^9 -280*x^6*y 
   +280*x^4*y^3 +504*x^2*y^5 -56*y^7 +105*x^4*y -210*x^2*y^3 +21*y^5.
R_9^7(r) cos(7*phi) = 1/10*5^(1/2) Z_52 = 9*x^9 -180*x^7*y^2 +126*x^5*y^4 +252*x^3*y^6 -63*x*y^8 
   -8*x^7 +168*x^5*y^2 -280*x^3*y^4 +56*x*y^6.
R_9^7(r) sin(7*phi) = 1/10*5^(1/2) Z_53 = 63*x^8*y -252*x^6*y^3 -126*x^4*y^5 +180*x^2*y^7 -9*y^9 
   -56*x^6*y +280*x^4*y^3 -168*x^2*y^5 +8*y^7.
R_9^9(r) cos(9*phi) = 1/10*5^(1/2) Z_54 = x^9 -36*x^7*y^2 +126*x^5*y^4 -84*x^3*y^6 +9*x*y^8.
R_9^9(r) sin(9*phi) = 1/10*5^(1/2) Z_55 = 9*x^8*y -84*x^6*y^3 +126*x^4*y^5 -36*x^2*y^7 +y^9.
\end{verbatim}\normalsize
If $m$ is odd, some of these lines are redundant; then the transformation
from $\cos(m\varphi)$ to $\sin(m\varphi)$ and vice versa is a straight exchange
of $x$ and $y$ plus a multiplication by $(-1)^{\lfloor m/2\rfloor}$.

\subsection{Product Expansion (Linearization Coefficients) } 
Products of Zernike functions are  products
$R_{n_1}^{m_1}R_{n_2}^{m_2}$
of the radial polynomials
by products of azimuthal functions of essentially
three different types \cite[4.3]{AS},
\begin{eqnarray}
\cos(m_1\varphi)\cos(m_2\varphi)&=&\frac{1}{2}\cos[(m_1-m_2)\varphi]+\frac{1}{2}\cos[(m_1+m_2)\varphi];
\\
\sin(m_1\varphi)\cos(m_2\varphi)&=&\frac{1}{2}\sin[(m_1-m_2)\varphi]+\frac{1}{2}\sin[(m_1+m_2)\varphi];
\\
\sin(m_1\varphi)\sin(m_2\varphi)&=&\frac{1}{2}\cos[(m_1-m_2)\varphi]-\frac{1}{2}\cos[(m_1+m_2)\varphi].
\end{eqnarray}
Since the azimuthal terms couple to $m_3=m_1\pm m_2$, most applications seek an
expansion of the form
\begin{equation}
R_{n_1}^{m_1}(r)R_{n_2}^{m_2}(r)
=\sum_{n_3=m_3}^{n_1+n_2} g_{n_1,m_1,n_2,m_2,n_3,m_3} R_{n_3}^{m_3}(r)
\label{eq.gdef}
\end{equation}
given any of these two cases of $m_3$. The sum is only over even values of $n_1+n_2-n_3$.
Equations (\ref{eq.Rat1}) and (\ref{eq.gdef}) establish the sum rule
\begin{equation} 
\sum_{n_3 = (n_1+n_2)\mod 2}^{n_1+n_2} g_{n_1,m_1,n_2,m_2,n_3,m_3}
=1
\end{equation} 
for the linearization coefficients.

The orthogonality (\ref{eq.ortho}) rewrites (\ref{eq.gdef}) as
\begin{equation}
g_{n_1,m_1,n_2,m_2,n_3,m_3}=2(n_3+1)\int_0^1 r
R_{n_1}^{m_1}(r)R_{n_2}^{m_2}(r)R_{n_3}^{m_3}(r) dr.
\label{eq.gtripls}
\end{equation}
Threefold use of (\ref{eq.Rpolyxn}) reduces the right hand side
to a triple sum over elementary integrals
over polynomials in $r$,
\begin{equation}
g_{n_1,m_1,n_2,m_2,n_3,m_3}
=2(n_3+1)
\sum_{s_1=0}^{-a_1}
\sum_{s_2=0}^{-a_2}
\sum_{s_3=0}^{-a_3}
\frac{1}{n_1+n_2+n_3+2(1-s_1-s_2-s_3)}\prod_{j=1}^3 (-1)^{s_j}
\binom{ n_j-s_j}{ s_j }
\binom{ n_j-2s_j}{ -a_j-s_j }
.
\label{eq.gdirect}
\end{equation}

As an alternative, Bailey resummation \cite{SlaterHyp}
of the polynomial product on the left hand side of (\ref{eq.gdef})
with $\sigma=s_1+s_2$ 
and $1+m_j=1-b_j+a_j$ for $j=1,2$
gives
\begin{eqnarray}
R_{n_1}^{m_1}R_{n_2}^{m_2}
&=&
(-1)^{a_1+a_2}
\binom{-b_1 }{ -a_1}
\binom{-b_2 }{ -a_2}
r^{m_1+m_2}
\sum_{s_1=0}^{-a_1}
\sum_{s_2=0}^{-a_2}
\frac{(a_1)_{s_1}(1-b_1)_{s_1}}{(1+a_1-b_1)_{s_1}s_1!}
\frac{(a_2)_{s_2}(1-b_2)_{s_2}}{(1+a_2-b_2)_{s_2}s_2!}
r^{2(s_1+s_2)}
\\
&=&
(-1)^{a_1+a_2}
\binom{-b_1 }{ -a_1}
\binom{-b_2 }{ -a_2}
r^{m_1+m_2}
\nonumber \\ && \times
\sum_{\sigma=0}^{-a_1-a_2}
r^{2\sigma}
\frac{(a_2)_\sigma (1-b_2)_\sigma}{(1+a_2-b_2)_\sigma \sigma!}
\,_4F_3(a_1,-\sigma,1-b_1,-a_2+b_2-\sigma;b_2-\sigma,1+a_1-b_1,1-a_2-\sigma;1)
.
\label{eq.RBayl}
\end{eqnarray}
(The terminating Hypergeometric Function $_4F_3$
at the right hand side
is ``well poised'' and has various other representations \cite{WhipplePLMS25,SlaterHyp}.)
This points at two more ways besides (\ref{eq.gdirect}) to compute the $g$:
\begin{enumerate}
\item
Comparing coefficients of equal powers of $r$ on both sides of (\ref{eq.gdef}), we
obtain a linear system of equations. $g$ with row index $n_3$
is the column vector of unknowns, the matrix has entries of the form
\begin{equation}
(-1)^{a_3}\binom{-b_3}{ -a_3} \frac{(a_3)_{s_3}(1-b_3)_{s_3}}{(1+m_3)_{s_3}s_3!}
\end{equation}
with column index $n_3$ and row index $s_3$,
and the constant vector is given by the coefficients of (\ref{eq.RBayl})
with row index $s_3=\sigma_3+(m_1+m_2-m_3)/2$.
\item
Substitution of the terms $r^{m_1+m_2+2\sigma}$ in (\ref{eq.RBayl}) by
(\ref{eq.rofR}) also converts the product to the format demanded
by (\ref{eq.gdef}).
\end{enumerate}

Omitting the trivial case of either $n_1$ or $n_2$ being zero,
and showing only the cases $n_1\le n_2$ (the others follow by swapping
the two factors on the left hand sides) for $m_3=m_1\pm m_2$, examples
of (\ref{eq.gdef}) are:

\small \begin{verbatim}
R_1^1(r)*R_1^1(r) = R_2^2(r)
   = 1/2*R_0^0(r) +1/2*R_2^0(r).
R_1^1(r)*R_2^0(r) = 1/3*R_1^1(r) +2/3*R_3^1(r).
R_1^1(r)*R_2^2(r) = R_3^3(r)
   = 2/3*R_1^1(r) +1/3*R_3^1(r).
R_1^1(r)*R_3^1(r) = 1/4*R_2^2(r) +3/4*R_4^2(r)
   = 1/2*R_2^0(r) +1/2*R_4^0(r).
R_1^1(r)*R_3^3(r) = R_4^4(r)
   = 3/4*R_2^2(r) +1/4*R_4^2(r).
R_2^0(r)*R_2^0(r) = 1/3*R_0^0(r) +2/3*R_4^0(r).
R_2^0(r)*R_2^2(r) = 1/2*R_2^2(r) +1/2*R_4^2(r).
R_2^2(r)*R_2^2(r) = R_4^4(r)
   = 1/3*R_0^0(r) +1/2*R_2^0(r) +1/6*R_4^0(r).
R_1^1(r)*R_4^0(r) = 2/5*R_3^1(r) +3/5*R_5^1(r).
R_1^1(r)*R_4^2(r) = 1/5*R_3^3(r) +4/5*R_5^3(r)
   = 3/5*R_3^1(r) +2/5*R_5^1(r).
R_1^1(r)*R_4^4(r) = R_5^5(r)
   = 4/5*R_3^3(r) +1/5*R_5^3(r).
R_2^0(r)*R_3^1(r) = 1/3*R_1^1(r) +1/15*R_3^1(r) +3/5*R_5^1(r).
R_2^0(r)*R_3^3(r) = 3/5*R_3^3(r) +2/5*R_5^3(r).
R_2^2(r)*R_3^1(r) = 2/5*R_3^3(r) +3/5*R_5^3(r)
   = 1/6*R_1^1(r) +8/15*R_3^1(r) +3/10*R_5^1(r).
R_2^2(r)*R_3^3(r) = R_5^5(r)
   = 1/2*R_1^1(r) +2/5*R_3^1(r) +1/10*R_5^1(r).
R_1^1(r)*R_5^1(r) = 1/3*R_4^2(r) +2/3*R_6^2(r)
   = 1/2*R_4^0(r) +1/2*R_6^0(r).
R_1^1(r)*R_5^3(r) = 1/6*R_4^4(r) +5/6*R_6^4(r)
   = 2/3*R_4^2(r) +1/3*R_6^2(r).
R_1^1(r)*R_5^5(r) = R_6^6(r)
   = 5/6*R_4^4(r) +1/6*R_6^4(r).
R_2^0(r)*R_4^0(r) = 2/5*R_2^0(r) +3/5*R_6^0(r).
R_2^0(r)*R_4^2(r) = 3/10*R_2^2(r) +1/6*R_4^2(r) +8/15*R_6^2(r).
R_2^0(r)*R_4^4(r) = 2/3*R_4^4(r) +1/3*R_6^4(r).
R_2^2(r)*R_4^0(r) = 1/10*R_2^2(r) +1/2*R_4^2(r) +2/5*R_6^2(r).
R_2^2(r)*R_4^2(r) = 1/3*R_4^4(r) +2/3*R_6^4(r)
   = 3/10*R_2^0(r) +1/2*R_4^0(r) +1/5*R_6^0(r).
R_2^2(r)*R_4^4(r) = R_6^6(r)
   = 3/5*R_2^2(r) +1/3*R_4^2(r) +1/15*R_6^2(r).
R_3^1(r)*R_3^1(r) = 2/5*R_2^2(r) +3/5*R_6^2(r)
   = 1/4*R_0^0(r) +1/20*R_2^0(r) +1/4*R_4^0(r) +9/20*R_6^0(r).
R_3^1(r)*R_3^3(r) = 1/2*R_4^4(r) +1/2*R_6^4(r)
   = 3/10*R_2^2(r) +1/2*R_4^2(r) +1/5*R_6^2(r).
R_3^3(r)*R_3^3(r) = R_6^6(r)
   = 1/4*R_0^0(r) +9/20*R_2^0(r) +1/4*R_4^0(r) +1/20*R_6^0(r).
R_1^1(r)*R_6^0(r) = 3/7*R_5^1(r) +4/7*R_7^1(r).
R_1^1(r)*R_6^2(r) = 2/7*R_5^3(r) +5/7*R_7^3(r)
   = 4/7*R_5^1(r) +3/7*R_7^1(r).
R_1^1(r)*R_6^4(r) = 1/7*R_5^5(r) +6/7*R_7^5(r)
   = 5/7*R_5^3(r) +2/7*R_7^3(r).
R_1^1(r)*R_6^6(r) = R_7^7(r)
   = 6/7*R_5^5(r) +1/7*R_7^5(r).
R_2^0(r)*R_5^1(r) = 2/5*R_3^1(r) +1/35*R_5^1(r) +4/7*R_7^1(r).
R_2^0(r)*R_5^3(r) = 4/15*R_3^3(r) +9/35*R_5^3(r) +10/21*R_7^3(r).
R_2^0(r)*R_5^5(r) = 5/7*R_5^5(r) +2/7*R_7^5(r).
R_2^2(r)*R_5^1(r) = 1/15*R_3^3(r) +16/35*R_5^3(r) +10/21*R_7^3(r)
   = 1/5*R_3^1(r) +18/35*R_5^1(r) +2/7*R_7^1(r).
R_2^2(r)*R_5^3(r) = 2/7*R_5^5(r) +5/7*R_7^5(r)
   = 2/5*R_3^1(r) +16/35*R_5^1(r) +1/7*R_7^1(r).
R_2^2(r)*R_5^5(r) = R_7^7(r)
   = 2/3*R_3^3(r) +2/7*R_5^3(r) +1/21*R_7^3(r).
R_3^1(r)*R_4^0(r) = 1/5*R_1^1(r) +1/5*R_3^1(r) +3/35*R_5^1(r) +18/35*R_7^1(r).
R_3^1(r)*R_4^2(r) = 2/5*R_3^3(r) +1/35*R_5^3(r) +4/7*R_7^3(r)
   = 3/10*R_1^1(r) +5/14*R_5^1(r) +12/35*R_7^1(r).
R_3^1(r)*R_4^4(r) = 4/7*R_5^5(r) +3/7*R_7^5(r)
   = 2/5*R_3^3(r) +16/35*R_5^3(r) +1/7*R_7^3(r).
R_3^3(r)*R_4^0(r) = 1/5*R_3^3(r) +18/35*R_5^3(r) +2/7*R_7^3(r).
R_3^3(r)*R_4^2(r) = 3/7*R_5^5(r) +4/7*R_7^5(r)
   = 1/10*R_1^1(r) +2/5*R_3^1(r) +27/70*R_5^1(r) +4/35*R_7^1(r).
R_3^3(r)*R_4^4(r) = R_7^7(r)
   = 2/5*R_1^1(r) +2/5*R_3^1(r) +6/35*R_5^1(r) +1/35*R_7^1(r).
R_1^1(r)*R_7^1(r) = 3/8*R_6^2(r) +5/8*R_8^2(r)
   = 1/2*R_6^0(r) +1/2*R_8^0(r).
R_1^1(r)*R_7^3(r) = 1/4*R_6^4(r) +3/4*R_8^4(r)
   = 5/8*R_6^2(r) +3/8*R_8^2(r).
R_1^1(r)*R_7^5(r) = 1/8*R_6^6(r) +7/8*R_8^6(r)
   = 3/4*R_6^4(r) +1/4*R_8^4(r).
R_1^1(r)*R_7^7(r) = R_8^8(r)
   = 7/8*R_6^6(r) +1/8*R_8^6(r).
R_2^0(r)*R_6^0(r) = 3/7*R_4^0(r) +4/7*R_8^0(r).
R_2^0(r)*R_6^2(r) = 8/21*R_4^2(r) +1/12*R_6^2(r) +15/28*R_8^2(r).
R_2^0(r)*R_6^4(r) = 5/21*R_4^4(r) +1/3*R_6^4(r) +3/7*R_8^4(r).
R_2^0(r)*R_6^6(r) = 3/4*R_6^6(r) +1/4*R_8^6(r).
R_2^2(r)*R_6^0(r) = 1/7*R_4^2(r) +1/2*R_6^2(r) +5/14*R_8^2(r).
R_2^2(r)*R_6^2(r) = 1/21*R_4^4(r) +5/12*R_6^4(r) +15/28*R_8^4(r)
   = 2/7*R_4^0(r) +1/2*R_6^0(r) +3/14*R_8^0(r).
R_2^2(r)*R_6^4(r) = 1/4*R_6^6(r) +3/4*R_8^6(r)
   = 10/21*R_4^2(r) +5/12*R_6^2(r) +3/28*R_8^2(r).
R_2^2(r)*R_6^6(r) = R_8^8(r)
   = 5/7*R_4^4(r) +1/4*R_6^4(r) +1/28*R_8^4(r).
R_3^1(r)*R_5^1(r) = 3/20*R_2^2(r) +25/84*R_4^2(r) +1/60*R_6^2(r) +15/28*R_8^2(r)
   = 3/10*R_2^0(r) +1/14*R_4^0(r) +1/5*R_6^0(r) +3/7*R_8^0(r).
R_3^1(r)*R_5^3(r) = 8/21*R_4^4(r) +1/12*R_6^4(r) +15/28*R_8^4(r)
   = 3/10*R_2^2(r) +1/42*R_4^2(r) +49/120*R_6^2(r) +15/56*R_8^2(r).
R_3^1(r)*R_5^5(r) = 5/8*R_6^6(r) +3/8*R_8^6(r)
   = 10/21*R_4^4(r) +5/12*R_6^4(r) +3/28*R_8^4(r).
R_3^3(r)*R_5^1(r) = 1/7*R_4^4(r) +1/2*R_6^4(r) +5/14*R_8^4(r)
   = 1/20*R_2^2(r) +9/28*R_4^2(r) +9/20*R_6^2(r) +5/28*R_8^2(r).
R_3^3(r)*R_5^3(r) = 3/8*R_6^6(r) +5/8*R_8^6(r)
   = 1/5*R_2^0(r) +3/7*R_4^0(r) +3/10*R_6^0(r) +1/14*R_8^0(r).
R_3^3(r)*R_5^5(r) = R_8^8(r)
   = 1/2*R_2^2(r) +5/14*R_4^2(r) +1/8*R_6^2(r) +1/56*R_8^2(r).
R_4^0(r)*R_4^0(r) = 1/5*R_0^0(r) +2/7*R_4^0(r) +18/35*R_8^0(r).
R_4^0(r)*R_4^2(r) = 3/10*R_2^2(r) +1/14*R_4^2(r) +1/5*R_6^2(r) +3/7*R_8^2(r).
R_4^0(r)*R_4^4(r) = 2/7*R_4^4(r) +1/2*R_6^4(r) +3/14*R_8^4(r).
R_4^2(r)*R_4^2(r) = 3/7*R_4^4(r) +4/7*R_8^4(r)
   = 1/5*R_0^0(r) +1/10*R_2^0(r) +1/14*R_4^0(r) +2/5*R_6^0(r) +8/35*R_8^0(r).
R_4^2(r)*R_4^4(r) = 1/2*R_6^6(r) +1/2*R_8^6(r)
   = 1/5*R_2^2(r) +3/7*R_4^2(r) +3/10*R_6^2(r) +1/14*R_8^2(r).
R_4^4(r)*R_4^4(r) = R_8^8(r)
   = 1/5*R_0^0(r) +2/5*R_2^0(r) +2/7*R_4^0(r) +1/10*R_6^0(r) +1/70*R_8^0(r).
\end{verbatim}\normalsize
Recursive application expands triple, quadruple etc.\ products of Zernike functions.

\section{3D Zernike Functions} \label{sec.3D} 

\subsection{Radial Polynomials}
\subsubsection{Definition} \label{sec.3dR}
The radial polynomials of the 3D Zernike functions are
\cite{MatharArxiv0805,MakJMGM26,NovotniCAD36}
\begin{eqnarray}
R_n^{(l)}(r)
&=&
\frac{\sqrt{2n+3}}{2^{n-l}}
\frac{(-1)^{(n-l)/2}}{\binom{n }{ l}}
\sum_{s=0}^{(n-l)/2} (-1)^s
\binom{n}{ (n-l)/2-s}\binom{l+s }{ l}\binom{l+1+n+2s}{ n-l}r^{l+2s}
\\
&=&
\frac{\sqrt{2n+3}}{2^{n-l}}
\frac{1}{\binom{n }{ l}}
\sum_{s=0}^{(n-l)/2} (-1)^s
\binom{n}{ s} \binom{l+\frac{n-l}{2}-s }{ l}\binom{2n+1-2s}{ n-l}r^{n-2s}
,
\label{eq.Rofr3D}
\end{eqnarray}
defined for $0\le l\le n$,
even $n-l$,
and normalized by design as
\begin{equation}
\int_0^1 r^2
R_n^{(l)}(r)R_{n'}^{(l)}(r)dr
=
\delta_{n,n'}
.
\label{eq.ortho3D}
\end{equation}
A noteworthy special value is
\begin{equation}
R_n^{(l)}(1)= \sqrt{2n+3},
\label{eq.R3dat1}
\end{equation}
which is derived from representations as terminating Hypergeometric
Functions or Jacobi Polynomials \cite{MatharArxiv0805}\cite[(2.6)]{Fields}
\begin{eqnarray}
R_n^{(l)}(r)
&=& \frac{\sqrt{2n+3}}{2^{n-l}}\frac{\binom{(l+n)/2}{ l}\binom{1+2n}{ n-l}}{\binom{n }{ l}}
r^n \,_2F_1\left(-\alpha,\alpha-n-\frac{1}{2};-n-\frac{1}{2};\frac{1}{r^2}\right)
\\
&=& \sqrt{2n+3}\frac{(q)_\alpha}{\alpha!}(-1)^\alpha r^l
\,_2F_1\left(-\alpha,q+\alpha;q;r^2\right)
\label{eq.3dhyp}
\\
&=&
\sqrt{2n+3} r^l
P_\alpha^{(0,q-1)}(2r^2-1);
\quad \alpha\equiv (n-l)/2 ;
\quad q\equiv l+3/2,
\label{eq.Jac3D}
\end{eqnarray}
via equations (15.4.6) and (22.4.1) of \cite{AS}.
To set them apart from the radial polynomials of Section \ref{sec.2D},
the upper index is enclosed in parentheses. A stylistic difference to the
2D case is that Noll \cite{NollJOSA66} maintained integer coefficients with the
polynomials $R_n^m$ by removing a factor $\sqrt{2n+2}$ from the
$R_n^m$ and a factor $\sqrt\pi$ from the azimuths,
which resurface at places like (\ref{eq.Zortho}) and (\ref{eq.ortho}). Examples
of (\ref{eq.Rofr3D}) for small $n$ are:

\small \begin{verbatim}
R_0^(0)(r) = 3^(1/2).
R_1^(1)(r) = 5^(1/2)*r.
R_2^(0)(r) = 1/2*7^(1/2)*( -3 +5*r^2).
R_2^(2)(r) = 7^(1/2)*r^2.
R_3^(1)(r) = 3/2*r*( -5 +7*r^2).
R_3^(3)(r) = 3*r^3.
R_4^(0)(r) = 1/8*11^(1/2)*(15 -70*r^2 +63*r^4).
R_4^(2)(r) = 1/2*11^(1/2)*r^2*( -7 +9*r^2).
R_4^(4)(r) = 11^(1/2)*r^4.
R_5^(1)(r) = 1/8*13^(1/2)*r*(35 -126*r^2 +99*r^4).
R_5^(3)(r) = 1/2*13^(1/2)*r^3*( -9 +11*r^2).
R_5^(5)(r) = 13^(1/2)*r^5.
R_6^(0)(r) = 1/16*15^(1/2)*( -35 +315*r^2 -693*r^4 +429*r^6).
R_6^(2)(r) = 1/8*15^(1/2)*r^2*(63 -198*r^2 +143*r^4).
R_6^(4)(r) = 1/2*15^(1/2)*r^4*( -11 +13*r^2).
R_6^(6)(r) = 15^(1/2)*r^6.
R_7^(1)(r) = 1/16*17^(1/2)*r*( -105 +693*r^2 -1287*r^4 +715*r^6).
R_7^(3)(r) = 1/8*17^(1/2)*r^3*(99 -286*r^2 +195*r^4).
R_7^(5)(r) = 1/2*17^(1/2)*r^5*( -13 +15*r^2).
R_7^(7)(r) = 17^(1/2)*r^7.
R_8^(0)(r) = 1/128*19^(1/2)*(315 -4620*r^2 +18018*r^4 -25740*r^6 +12155*r^8).
R_8^(2)(r) = 1/16*19^(1/2)*r^2*( -231 +1287*r^2 -2145*r^4 +1105*r^6).
R_8^(4)(r) = 1/8*19^(1/2)*r^4*(143 -390*r^2 +255*r^4).
R_8^(6)(r) = 1/2*19^(1/2)*r^6*( -15 +17*r^2).
R_8^(8)(r) = 19^(1/2)*r^8.
R_9^(1)(r) = 1/128*21^(1/2)*r*(1155 -12012*r^2 +38610*r^4 -48620*r^6 +20995*r^8).
R_9^(3)(r) = 1/16*21^(1/2)*r^3*( -429 +2145*r^2 -3315*r^4 +1615*r^6).
R_9^(5)(r) = 1/8*21^(1/2)*r^5*(195 -510*r^2 +323*r^4).
R_9^(7)(r) = 1/2*21^(1/2)*r^7*( -17 +19*r^2).
R_9^(9)(r) = 21^(1/2)*r^9.
R_10^(0)(r) = 1/256*23^(1/2)*( -693 +15015*r^2 -90090*r^4 +218790*r^6 -230945*r^8 +88179*r^10).
R_10^(2)(r) = 1/128*23^(1/2)*r^2*(3003 -25740*r^2 +72930*r^4 -83980*r^6 +33915*r^8).
R_10^(4)(r) = 1/16*23^(1/2)*r^4*( -715 +3315*r^2 -4845*r^4 +2261*r^6).
R_10^(6)(r) = 1/8*23^(1/2)*r^6*(255 -646*r^2 +399*r^4).
R_10^(8)(r) = 1/2*23^(1/2)*r^8*( -19 +21*r^2).
R_10^(10)(r) = 23^(1/2)*r^10.
R_11^(1)(r) = 5/256*r*( -3003 +45045*r^2 -218790*r^4 +461890*r^6 -440895*r^8 +156009*r^10).
R_11^(3)(r) = 5/128*r^3*(6435 -48620*r^2 +125970*r^4 -135660*r^6 +52003*r^8).
R_11^(5)(r) = 5/16*r^5*( -1105 +4845*r^2 -6783*r^4 +3059*r^6).
R_11^(7)(r) = 5/8*r^7*(323 -798*r^2 +483*r^4).
R_11^(9)(r) = 5/2*r^9*( -21 +23*r^2).
R_11^(11)(r) = 5*r^11.
R_12^(0)(r) = 3/1024*3^(1/2)*(3003 -90090*r^2 +765765*r^4 -2771340*r^6 +4849845*r^8 -4056234*r^10 
   +1300075*r^12).
R_12^(2)(r) = 3/256*3^(1/2)*r^2*( -9009 +109395*r^2 -461890*r^4 +881790*r^6 -780045*r^8 +260015*r^10).
R_12^(4)(r) = 3/128*3^(1/2)*r^4*(12155 -83980*r^2 +203490*r^4 -208012*r^6 +76475*r^8).
R_12^(6)(r) = 3/16*3^(1/2)*r^6*( -1615 +6783*r^2 -9177*r^4 +4025*r^6).
R_12^(8)(r) = 3/8*3^(1/2)*r^8*(399 -966*r^2 +575*r^4).
R_12^(10)(r) = 3/2*3^(1/2)*r^10*( -23 +25*r^2).
R_12^(12)(r) = 3*3^(1/2)*r^12.
R_13^(1)(r) = 1/1024*29^(1/2)*r*(15015 -306306*r^2 +2078505*r^4 -6466460*r^6 +10140585*r^8 
   -7800450*r^10 +2340135*r^12).
R_13^(3)(r) = 1/256*29^(1/2)*r^3*( -21879 +230945*r^2 -881790*r^4 +1560090*r^6 -1300075*r^8 
   +412965*r^10).
R_13^(5)(r) = 1/128*29^(1/2)*r^5*(20995 -135660*r^2 +312018*r^4 -305900*r^6 +108675*r^8).
R_13^(7)(r) = 1/16*29^(1/2)*r^7*( -2261 +9177*r^2 -12075*r^4 +5175*r^6).
R_13^(9)(r) = 1/8*29^(1/2)*r^9*(483 -1150*r^2 +675*r^4).
R_13^(11)(r) = 1/2*29^(1/2)*r^11*( -25 +27*r^2).
R_13^(13)(r) = 29^(1/2)*r^13.
\end{verbatim}\normalsize

\subsubsection{Inverse Representation}
Inversion of the representations (\ref{eq.Rofr3D}) defines coefficients $f_{j,n,l}$,
\begin{equation}
r^j\equiv \sum_{n=l\mod 2}^j f_{j,n,l} R_n^{(l)}(r),\quad j-l\thickspace \mathrm{even}.
\label{eq.rofR3d}
\end{equation}
The projection technique equivalent to the 2D case based on the orthogonality (\ref{eq.ortho3D}),
(\ref{eq.3dhyp}) and \cite[7.512.2]{GR} yields
\begin{eqnarray}
f_{j,n,l}
&=&
\int_0^1 r^{j+2}R_n^{(l)}(r)dr
=
\frac{\sqrt{2n+3}}{2^{n-l}}
\frac{1}{\binom{n }{ l}}
\sum_{s=0}^{(n-l)/2} \frac{(-1)^s}{j+3+n-2s}
\binom{n}{ s} \binom{l+\frac{n-l}{2}-s }{ l}\binom{2n+1-2s}{ n-l}
\\
&=&
\sqrt{2n+3}
\frac{1}{j+3+l}
\frac{ \left(\frac{j-n}{2}+1\right)_\alpha }
{ \left(\frac{j-l}{2}+q+1\right)_\alpha}
.
\label{eq.fshort}
\end{eqnarray}
Equation  (\ref{eq.rofR3d}) evaluated at $r=1$ using (\ref{eq.R3dat1}) establishes the sum rule
\begin{equation}
\sum_{n=l\mod 2}^j \sqrt{2n+3}f_{j,n,l}
=1
.
\end{equation}
Recurrences derived from (\ref{eq.fshort})
are
\begin{eqnarray}
f_{j+2,n,l}&=&\frac{(j+3+l)(j-l+2)}{(j-n+2)(j+n+5)}f_{j,n,l};
\\
f_{j,n+2,l}&=&\frac{j-n}{j+5+n}\sqrt{\frac{2n+7}{2n+3}}f_{j,n,l};
\\
f_{j,n,l+2}&=&\frac{j+3+l}{j-l}f_{j,n,l}.
\end{eqnarray}
Examples of (\ref{eq.rofR3d}) for small powers $j$ are:

\small \begin{verbatim}
r^0 = 1/3*3^(1/2)*R_0^(0)(r).
r^1 = 1/5*5^(1/2)*R_1^(1)(r).
r^2 = 1/5*3^(1/2)*R_0^(0)(r) +2/35*7^(1/2)*R_2^(0)(r)
  = 1/7*7^(1/2)*R_2^(2)(r).
r^3 = 1/7*5^(1/2)*R_1^(1)(r) +2/21*R_3^(1)(r)
  = 1/3*R_3^(3)(r).
r^4 = 1/7*3^(1/2)*R_0^(0)(r) +4/63*7^(1/2)*R_2^(0)(r) +8/693*11^(1/2)*R_4^(0)(r)
  = 1/9*7^(1/2)*R_2^(2)(r) +2/99*11^(1/2)*R_4^(2)(r)
  = 1/11*11^(1/2)*R_4^(4)(r).
r^5 = 1/9*5^(1/2)*R_1^(1)(r) +4/33*R_3^(1)(r) +8/1287*13^(1/2)*R_5^(1)(r)
  = 3/11*R_3^(3)(r) +2/143*13^(1/2)*R_5^(3)(r)
  = 1/13*13^(1/2)*R_5^(5)(r).
r^6 = 1/9*3^(1/2)*R_0^(0)(r) +2/33*7^(1/2)*R_2^(0)(r) +8/429*11^(1/2)*R_4^(0)(r) 
    +16/6435*15^(1/2)*R_6^(0)(r)
  = 1/11*7^(1/2)*R_2^(2)(r) +4/143*11^(1/2)*R_4^(2)(r) +8/2145*15^(1/2)*R_6^(2)(r)
  = 1/13*11^(1/2)*R_4^(4)(r) +2/195*15^(1/2)*R_6^(4)(r)
  = 1/15*15^(1/2)*R_6^(6)(r).
r^7 = 1/11*5^(1/2)*R_1^(1)(r) +18/143*R_3^(1)(r) +8/715*13^(1/2)*R_5^(1)(r) 
    +16/12155*17^(1/2)*R_7^(1)(r)
  = 3/13*R_3^(3)(r) +4/195*13^(1/2)*R_5^(3)(r) +8/3315*17^(1/2)*R_7^(3)(r)
  = 1/15*13^(1/2)*R_5^(5)(r) +2/255*17^(1/2)*R_7^(5)(r)
  = 1/17*17^(1/2)*R_7^(7)(r).
r^8 = 1/11*3^(1/2)*R_0^(0)(r) +8/143*7^(1/2)*R_2^(0)(r) +16/715*11^(1/2)*R_4^(0)(r) 
    +64/12155*15^(1/2)*R_6^(0)(r) +128/230945*19^(1/2)*R_8^(0)(r)
  = 1/13*7^(1/2)*R_2^(2)(r) +2/65*11^(1/2)*R_4^(2)(r) +8/1105*15^(1/2)*R_6^(2)(r) 
    +16/20995*19^(1/2)*R_8^(2)(r)
  = 1/15*11^(1/2)*R_4^(4)(r) +4/255*15^(1/2)*R_6^(4)(r) +8/4845*19^(1/2)*R_8^(4)(r)
  = 1/17*15^(1/2)*R_6^(6)(r) +2/323*19^(1/2)*R_8^(6)(r)
  = 1/19*19^(1/2)*R_8^(8)(r).
r^9 = 1/13*5^(1/2)*R_1^(1)(r) +8/65*R_3^(1)(r) +16/1105*13^(1/2)*R_5^(1)(r) 
    +64/20995*17^(1/2)*R_7^(1)(r) +128/440895*21^(1/2)*R_9^(1)(r)
  = 1/5*R_3^(3)(r) +2/85*13^(1/2)*R_5^(3)(r) +8/1615*17^(1/2)*R_7^(3)(r) +16/33915*21^(1/2)*R_9^(3)(r)
  = 1/17*13^(1/2)*R_5^(5)(r) +4/323*17^(1/2)*R_7^(5)(r) +8/6783*21^(1/2)*R_9^(5)(r)
  = 1/19*17^(1/2)*R_7^(7)(r) +2/399*21^(1/2)*R_9^(7)(r)
  = 1/21*21^(1/2)*R_9^(9)(r).
r^10 = 1/13*3^(1/2)*R_0^(0)(r) +2/39*7^(1/2)*R_2^(0)(r) +16/663*11^(1/2)*R_4^(0)(r) 
    +32/4199*15^(1/2)*R_6^(0)(r) +128/88179*19^(1/2)*R_8^(0)(r) +256/2028117*23^(1/2)*R_10^(0)(r)
  = 1/15*7^(1/2)*R_2^(2)(r) +8/255*11^(1/2)*R_4^(2)(r) +16/1615*15^(1/2)*R_6^(2)(r) 
    +64/33915*19^(1/2)*R_8^(2)(r) +128/780045*23^(1/2)*R_10^(2)(r)
  = 1/17*11^(1/2)*R_4^(4)(r) +6/323*15^(1/2)*R_6^(4)(r) +8/2261*19^(1/2)*R_8^(4)(r) 
    +16/52003*23^(1/2)*R_10^(4)(r)
  = 1/19*15^(1/2)*R_6^(6)(r) +4/399*19^(1/2)*R_8^(6)(r) +8/9177*23^(1/2)*R_10^(6)(r)
  = 1/21*19^(1/2)*R_8^(8)(r) +2/483*23^(1/2)*R_10^(8)(r)
  = 1/23*23^(1/2)*R_10^(10)(r).
r^11 = 1/15*5^(1/2)*R_1^(1)(r) +2/17*R_3^(1)(r) +16/969*13^(1/2)*R_5^(1)(r) 
    +32/6783*17^(1/2)*R_7^(1)(r) +128/156009*21^(1/2)*R_9^(1)(r) +256/780045*R_11^(1)(r)
  = 3/17*R_3^(3)(r) +8/323*13^(1/2)*R_5^(3)(r) +16/2261*17^(1/2)*R_7^(3)(r) 
    +64/52003*21^(1/2)*R_9^(3)(r) +128/260015*R_11^(3)(r)
  = 1/19*13^(1/2)*R_5^(5)(r) +2/133*17^(1/2)*R_7^(5)(r) +8/3059*21^(1/2)*R_9^(5)(r) 
    +16/15295*R_11^(5)(r)
  = 1/21*17^(1/2)*R_7^(7)(r) +4/483*21^(1/2)*R_9^(7)(r) +8/2415*R_11^(7)(r)
  = 1/23*21^(1/2)*R_9^(9)(r) +2/115*R_11^(9)(r)
  = 1/5*R_11^(11)(r).
r^12 = 1/15*3^(1/2)*R_0^(0)(r) +4/85*7^(1/2)*R_2^(0)(r) +8/323*11^(1/2)*R_4^(0)(r) 
    +64/6783*15^(1/2)*R_6^(0)(r) +128/52003*19^(1/2)*R_8^(0)(r) +512/1300075*23^(1/2)*R_10^(0)(r) 
    +1024/11700675*3^(1/2)*R_12^(0)(r)
  = 1/17*7^(1/2)*R_2^(2)(r) +10/323*11^(1/2)*R_4^(2)(r) +80/6783*15^(1/2)*R_6^(2)(r) 
    +160/52003*19^(1/2)*R_8^(2)(r) +128/260015*23^(1/2)*R_10^(2)(r) +256/2340135*3^(1/2)*R_12^(2)(r)
  = 1/19*11^(1/2)*R_4^(4)(r) +8/399*15^(1/2)*R_6^(4)(r) +16/3059*19^(1/2)*R_8^(4)(r) 
    +64/76475*23^(1/2)*R_10^(4)(r) +128/688275*3^(1/2)*R_12^(4)(r)
  = 1/21*15^(1/2)*R_6^(6)(r) +2/161*19^(1/2)*R_8^(6)(r) +8/4025*23^(1/2)*R_10^(6)(r) 
    +16/36225*3^(1/2)*R_12^(6)(r)
  = 1/23*19^(1/2)*R_8^(8)(r) +4/575*23^(1/2)*R_10^(8)(r) +8/5175*3^(1/2)*R_12^(8)(r)
  = 1/25*23^(1/2)*R_10^(10)(r) +2/225*3^(1/2)*R_12^(10)(r)
  = 1/9*3^(1/2)*R_12^(12)(r).
r^13 = 1/17*5^(1/2)*R_1^(1)(r) +36/323*R_3^(1)(r) +40/2261*13^(1/2)*R_5^(1)(r) 
    +320/52003*17^(1/2)*R_7^(1)(r) +384/260015*21^(1/2)*R_9^(1)(r) +512/468027*R_11^(1)(r) 
    +1024/67863915*29^(1/2)*R_13^(1)(r)
  = 3/19*R_3^(3)(r) +10/399*13^(1/2)*R_5^(3)(r) +80/9177*17^(1/2)*R_7^(3)(r) 
    +32/15295*21^(1/2)*R_9^(3)(r) +128/82593*R_11^(3)(r) +256/11975985*29^(1/2)*R_13^(3)(r)
  = 1/21*13^(1/2)*R_5^(5)(r) +8/483*17^(1/2)*R_7^(5)(r) +16/4025*21^(1/2)*R_9^(5)(r) 
    +64/21735*R_11^(5)(r) +128/3151575*29^(1/2)*R_13^(5)(r)
  = 1/23*17^(1/2)*R_7^(7)(r) +6/575*21^(1/2)*R_9^(7)(r) +8/1035*R_11^(7)(r) 
    +16/150075*29^(1/2)*R_13^(7)(r)
  = 1/25*21^(1/2)*R_9^(9)(r) +4/135*R_11^(9)(r) +8/19575*29^(1/2)*R_13^(9)(r)
  = 5/27*R_11^(11)(r) +2/783*29^(1/2)*R_13^(11)(r)
  = 1/29*29^(1/2)*R_13^(13)(r).
\end{verbatim}\normalsize

This expands $r^j$ in a sum over $R_n^{(l)}$ at constant $l$.
As an alternative, it might be regarded as
solving (\ref{eq.Rofr3D}) as a linear system of equations
with a vector of unknowns from $r^l$ to $r^n$,
\begin{equation}
\left(
\begin{array}{c}
R_l^{(l)} \\
R_{l+2}^{(l)} \\
R_{l+4}^{(l)} \\
\vdots \\
R_n^{(l)} \\
\end{array}
\right)
=
\left( \begin{array}{ccccc}
\sqrt{2l+3} & 0 & 0 & 0 & 0 \\
\cdots & \ddots & 0 & 0 & 0 \\
       &        & \ddots &   0 &  0 \\
\cdots & \cdots & \cdots & \ddots & 0 \\
\cdots & \cdots & \cdots & \cdots & \ddots \\
\end{array}\right)
\cdot 
\left(
\begin{array}{c}
r^l\\
r^{l+2} \\
r^{l+4} \\
\vdots \\
r^n \\
\end{array}
\right).
\end{equation}
The matrix is a lower triangular matrix displaying the coefficients of (\ref{eq.Rofr3D});
the equation is easily solved via forward elimination, for example.
A complementary expansion in a series of $R_n^{(l)}$ with constant $n$
\begin{equation}
r^j\equiv \sum_{l=j\mod 2}^n \hat f_{j,n,l} R_n^{(l)}(r),\quad j-n=0\mod 2,
\label{eq.rofR3dtoo}
\end{equation}
lacks the equivalent orthogonality similar to (\ref{eq.ortho3D}), but
again it can be formulated as a linear system of equations,
\begin{equation}
\left(
\begin{array}{c}
R_n^{(l)} \\
R_n^{(l+2)} \\
R_n^{(l+4)} \\
\vdots \\
R_n^{(n)} \\
\end{array}
\right)
=
\left( \begin{array}{ccccc}
\ddots & \cdots & \cdots & \cdots & \cdots \\
0 & \ddots & \cdots & \cdots & \cdots \\
0      &  0     & \ddots &   &  \\
0 & 0 & 0 & \ddots & \cdots \\
0 &  0 & 0 & 0 & \sqrt{2n+3} \\
\end{array}\right)
\cdot 
\left(
\begin{array}{c}
r^l\\
r^{l+2} \\
r^{l+4} \\
\vdots \\
r^n \\
\end{array}
\right),
\end{equation}
this time with an upper triangular matrix populated by the coefficients of (\ref{eq.Rofr3D}),
solvable with backward elimination. The sum rule according to (\ref{eq.R3dat1}) is
\begin{equation}
1=\sqrt{2n+3}\sum_l \hat f_{j,n,l}
.
\end{equation}
Unlike (\ref{eq.rofR3d}) which turns out to be
helpful in Section \ref{sec.lin3D}, (\ref{eq.rofR3dtoo}) is not used further down, so
we keep the table short:
\small \begin{verbatim}
r^0 / 3^(1/2) = 1/3*R_0^(0).  
r^0 / 7^(1/2) = 5/21*R_2^(2) -2/21*R_2^(0).  
r^0 / 11^(1/2) = 9/55*R_4^(4) -4/33*R_4^(2) +8/165*R_4^(0).  
r^0 / 15^(1/2) = 13/105*R_6^(6) -18/175*R_6^(4) +8/105*R_6^(2) -16/525*R_6^(0).  
r^0 / 19^(1/2) = 17/171*R_8^(8) -104/1197*R_8^(6) +48/665*R_8^(4) -64/1197*R_8^(2) +128/5985*R_8^(0).  
r^0 / 23^(1/2) = 21/253*R_10^(10) -170/2277*R_10^(8) +1040/15939*R_10^(6) -96/1771*R_10^(4) 
    +640/15939*R_10^(2) -256/15939*R_10^(0).  
r^0 / 3*3^(1/2) = 25/351*R_12^(12) -28/429*R_12^(10) +680/11583*R_12^(8) -320/6237*R_12^(6) 
    +128/3003*R_12^(4) -2560/81081*R_12^(2) +1024/81081*R_12^(0).  
r^1 / 5^(1/2) = 1/5*R_1^(1).  
r^1 / 3 = 7/45*R_3^(3) -2/45*R_3^(1).  
r^1 / 13^(1/2) = 11/91*R_5^(5) -4/65*R_5^(3) +8/455*R_5^(1).  
r^1 / 17^(1/2) = 5/51*R_7^(7) -22/357*R_7^(5) +8/255*R_7^(3) -16/1785*R_7^(1).  
r^1 / 21^(1/2) = 19/231*R_9^(9) -40/693*R_9^(7) +16/441*R_9^(5) -64/3465*R_9^(3) +128/24255*R_9^(1).  
r^1 / 5 = 23/325*R_11^(11) -38/715*R_11^(9) +16/429*R_11^(7) -32/1365*R_11^(5) +128/10725*R_11^(3) 
    -256/75075*R_11^(1).  
r^1 / 29^(1/2) = 9/145*R_13^(13) -92/1885*R_13^(11) +152/4147*R_13^(9) -320/12441*R_13^(7) 
    +128/7917*R_13^(5) -512/62205*R_13^(3) +1024/435435*R_13^(1).  
r^2 / 7^(1/2) = 1/7*R_2^(2).  
r^2 / 11^(1/2) = 9/77*R_4^(4) -2/77*R_4^(2).  
r^2 / 15^(1/2) = 13/135*R_6^(6) -4/105*R_6^(4) +8/945*R_6^(2).  
r^2 / 19^(1/2) = 17/209*R_8^(8) -26/627*R_8^(6) +24/1463*R_8^(4) -16/4389*R_8^(2).  
r^2 / 23^(1/2) = 21/299*R_10^(10) -136/3289*R_10^(8) +16/759*R_10^(6) -192/23023*R_10^(4) 
    +128/69069*R_10^(2).  
r^2 / 3*3^(1/2) = 5/81*R_12^(12) -14/351*R_12^(10) +272/11583*R_12^(8) -32/2673*R_12^(6) 
    +128/27027*R_12^(4) -256/243243*R_12^(2).  
r^3 / 3 = 1/9*R_3^(3).  
r^3 / 13^(1/2) = 11/117*R_5^(5) -2/117*R_5^(3).  
r^3 / 17^(1/2) = 15/187*R_7^(7) -4/153*R_7^(5) +8/1683*R_7^(3).  
r^3 / 21^(1/2) = 19/273*R_9^(9) -30/1001*R_9^(7) +8/819*R_9^(5) -16/9009*R_9^(3).  
r^3 / 5 = 23/375*R_11^(11) -152/4875*R_11^(9) +48/3575*R_11^(7) -64/14625*R_11^(5) 
    +128/160875*R_11^(3).  
r^3 / 29^(1/2) = 27/493*R_13^(13) -46/1479*R_13^(11) +304/19227*R_13^(9) -480/70499*R_13^(7) 
    +128/57681*R_13^(5) -256/634491*R_13^(3).  
r^4 / 11^(1/2) = 1/11*R_4^(4).  
r^4 / 15^(1/2) = 13/165*R_6^(6) -2/165*R_6^(4).  
r^4 / 19^(1/2) = 17/247*R_8^(8) -4/209*R_8^(6) +8/2717*R_8^(4).  
r^4 / 23^(1/2) = 7/115*R_10^(10) -34/1495*R_10^(8) +8/1265*R_10^(6) -16/16445*R_10^(4).  
r^4 / 3*3^(1/2) = 25/459*R_12^(12) -56/2295*R_12^(10) +16/1755*R_12^(8) -64/25245*R_12^(6) 
    +128/328185*R_12^(4).  
r^5 / 13^(1/2) = 1/13*R_5^(5).  
r^5 / 17^(1/2) = 15/221*R_7^(7) -2/221*R_7^(5).  
r^5 / 21^(1/2) = 19/315*R_9^(9) -4/273*R_9^(7) +8/4095*R_9^(5).  
r^5 / 5 = 23/425*R_11^(11) -38/2125*R_11^(9) +24/5525*R_11^(7) -16/27625*R_11^(5).  
r^5 / 29^(1/2) = 27/551*R_13^(13) -184/9367*R_13^(11) +16/2465*R_13^(9) -192/121771*R_13^(7) 
    +128/608855*R_13^(5).  
r^6 / 15^(1/2) = 1/15*R_6^(6).  
r^6 / 19^(1/2) = 17/285*R_8^(8) -2/285*R_8^(6).  
r^6 / 23^(1/2) = 21/391*R_10^(10) -4/345*R_10^(8) +8/5865*R_10^(6).  
r^6 / 3*3^(1/2) = 25/513*R_12^(12) -14/969*R_12^(10) +8/2565*R_12^(8) -16/43605*R_12^(6).  
r^7 / 17^(1/2) = 1/17*R_7^(7).  
r^7 / 21^(1/2) = 19/357*R_9^(9) -2/357*R_9^(7).  
r^7 / 5 = 23/475*R_11^(11) -4/425*R_11^(9) +8/8075*R_11^(7).  
r^7 / 29^(1/2) = 9/203*R_13^(13) -46/3857*R_13^(11) +8/3451*R_13^(9) -16/65569*R_13^(7).  
r^8 / 19^(1/2) = 1/19*R_8^(8).  
r^8 / 23^(1/2) = 21/437*R_10^(10) -2/437*R_10^(8).  
r^8 / 3*3^(1/2) = 25/567*R_12^(12) -4/513*R_12^(10) +8/10773*R_12^(8).  
r^9 / 21^(1/2) = 1/21*R_9^(9).  
r^9 / 5 = 23/525*R_11^(11) -2/525*R_11^(9).  
r^9 / 29^(1/2) = 27/667*R_13^(13) -4/609*R_13^(11) +8/14007*R_13^(9).  
r^10 / 23^(1/2) = 1/23*R_10^(10).  
r^10 / 3*3^(1/2) = 25/621*R_12^(12) -2/621*R_12^(10).  
r^11 / 5 = 1/25*R_11^(11).  
r^11 / 29^(1/2) = 27/725*R_13^(13) -2/725*R_13^(11).  
r^12 / 3*3^(1/2) = 1/27*R_12^(12).  
r^13 / 29^(1/2) = 1/29*R_13^(13).  
\end{verbatim}\normalsize

\subsection{Basis in the Unit Sphere} 
Let the 3D Zernike functions
\begin{equation}
Z_{n,l}^{(m)}\equiv R_n^{(l)}(r)Y_l^{(m)}(\varphi,\theta);\quad
0\le \varphi\le 2\pi;\quad 0\le \theta\le \pi;\quad 0\le r\le 1,
\end{equation}
be defined with Edmonds' sign choice of the Spherical Harmonics \cite{Edmonds,Messiah}
\begin{equation}
Y_l^{(m)}=(-1)^m\sqrt{\frac{2l+1}{4\pi}\,\frac{(l-m)!}{(l+m)!}}
P_l^m(\cos\theta)e^{im\varphi},
\quad -l\le m\le l.
\end{equation}
Complex conjugation, denoted by the star, yields \cite[(2.5.6)]{Edmonds}
\begin{equation}
Y_l^{(m)*} = (-1)^m Y_l^{(-m)}.
\label{eq.Ylmcc}
\end{equation}
The latitudes are spanned via Associated Legendre Functions \cite[(2.5.10)]{Edmonds}
\begin{equation}
P_l^{m}(x)= (1-x^2)^{m/2}\frac{d^m}{dx^m}P_l(x),\quad m\ge 0,
\label{eq.Plm}
\end{equation}
which differs by a factor $(-1)^m$ from other sign
conventions \cite{AS,GR}. Extension to negative upper indices is finally defined via
\begin{equation}
P_l^{-m}(x)=(-1)^m\frac{(l-m)!}{(l+m)!}P_l^m(x),\quad m>0.
\end{equation}
The 3D Zernike functions are products of
Vector Harmonics $r^lY_l^{(m)}$ by even polynomials in $r$ of order $n-l$.
They are ortho-normal over the unit sphere,
\begin{equation}
\int_{r<1} Z_{n,l}^{(m)}(r,\varphi,\theta)
Z_{n',l'}^{(m')*}(r,\varphi,\theta) d^3r=\delta_{n,n'}\delta_{m,m'}\delta_{l,l'}
,\quad d^3r=r^2\sin\theta dr\,d\theta\, d\varphi.
\label{eq.orthoZ3D}
\end{equation}
This foundation in spherical coordinates spawns interest in
transformation into (Section \ref{sec.Z2C3D})
or from (Section \ref{sec.C2Z3D}) Cartesian coordinates.

\subsection{Zernike to Cartesian} \label{sec.Z2C3D} 
The Spherical Harmonics are
\cite[(3.2),(3.3)]{MatharIJQC90}
\begin{eqnarray}
Y_l^{(m)}(\theta,\varphi)
&=&
(-1)^m
\sqrt{\frac{2l+1}{4\pi}\,\frac{(l-m)!}{(l+m)!}}
2^l\sin^m\theta\sum_{\nu=0}^{l-m}
\left(\frac{m+\nu+1-l}{2}\right)_l \frac{1}{\nu!(l-m-\nu)!}\cos^\nu\theta
\nonumber
\\
&&
\times
\left(\sum_{j=0,2,4,\ldots}^m+i\sum_{j=1,3,5\ldots}^m\right)
(-1)^{\lfloor j/2\rfloor}
\binom{m}{ j} \cos^{m-j}\varphi \sin^j\varphi
.
\end{eqnarray}
Rendered as Cartesian Coordinates
\begin{equation}
x=r\sin\theta\cos\varphi ,\quad
y=r\sin\theta\cos\varphi ,\quad
z=r\cos\theta ,
\label{eq.xyz}
\end{equation}
the Vector Harmonics are
\cite[(3.7)]{MatharIJQC90}
\begin{eqnarray}
r^l Y_l^{(m)}(\theta,\varphi)
&=&
\left(-\frac{1}{2}\right)^m
\sqrt{\frac{2l+1}{4\pi}(l-m)!(l+m)!}
\sum_{\sigma_1,\sigma_2\ge 0}^{\sigma_1+\sigma_2\le\frac{l-m}{2}}
\!\!
\frac{1}{\sigma_1!\sigma_2!}\left(-\frac{1}{4}\right)^{\sigma_1+\sigma_2}
\!\!
\frac{1}{(m+\sigma_1+\sigma_2)!(l-m-2(\sigma_1+\sigma_2))!}
\nonumber
\\
&&
\times
\left(\sum_{j=0,2,\ldots}^m+i\sum_{j=1,3,5\ldots}^m\right)
(-1)^{\lfloor j/2\rfloor}\binom{m}{ j}
x^{m-j+2\sigma_1}y^{j+2\sigma_2}z^{l-m-2(\sigma_1+\sigma_2)}
.
\end{eqnarray}
This is the 3D version of Eqs.\ (\ref{eq.rjcos}) and (\ref{eq.rjsin}).

Sample outputs are tabulated with real and imaginary part separated
by a comma on each right hand side, and a factor $\sqrt\pi$ moved to the
left hand sides:
\small \begin{verbatim}
Pi^(1/2) r^0 Y_0^(0) = 1/2 , 0.
Pi^(1/2) r^1 Y_1^(0) = 1/2*z*3^(1/2) , 0.
Pi^(1/2) r^1 Y_1^(1) = -1/4*6^(1/2)*x ,  -1/4*6^(1/2)*y.
Pi^(1/2) r^2 Y_2^(0) = -1/4*( -2*z^2 +y^2 +x^2)*5^(1/2) , 0.
Pi^(1/2) r^2 Y_2^(1) = -1/4*30^(1/2)*x*z ,  -1/4*30^(1/2)*z*y.
Pi^(1/2) r^2 Y_2^(2) = 1/8*30^(1/2)*(x -y)*(x +y) , 1/4*30^(1/2)*x*y.
Pi^(1/2) r^3 Y_3^(0) = -1/4*7^(1/2)*( -2*z^2 +3*y^2 +3*x^2)*z , 0.
Pi^(1/2) r^3 Y_3^(1) = 1/8*x*( -4*z^2 +y^2 +x^2)*21^(1/2) , 1/8*21^(1/2)*( -4*z^2 +y^2 +x^2)*y.
Pi^(1/2) r^3 Y_3^(2) = 1/8*210^(1/2)*(x -y)*(x +y)*z , 1/4*210^(1/2)*x*y*z.
Pi^(1/2) r^3 Y_3^(3) = -1/8*x*(x^2 -3*y^2)*35^(1/2) ,  -1/8*35^(1/2)*( -y^2 +3*x^2)*y.
Pi^(1/2) r^4 Y_4^(0) = 3/2*z^4 -9/2*z^2*y^2 +9/16*y^4 -9/2*x^2*z^2 +9/8*x^2*y^2 +9/16*x^4 , 0.
Pi^(1/2) r^4 Y_4^(1) = 3/8*5^(1/2)*( -4*z^2 +3*y^2 +3*x^2)*x*z , 3/8*5^(1/2)*( -4*z^2 +3*y^2 
    +3*x^2)*y*z.
Pi^(1/2) r^4 Y_4^(2) = -3/16*10^(1/2)*(x^2 -6*z^2 +y^2)*(x +y)*(x -y) ,  -3/8*10^(1/2)*x*(x^2 -6*z^2 
    +y^2)*y.
Pi^(1/2) r^4 Y_4^(3) = -3/8*35^(1/2)*x*(x^2 -3*y^2)*z ,  -3/8*35^(1/2)*( -y^2 +3*x^2)*y*z.
Pi^(1/2) r^4 Y_4^(4) = 3/32*70^(1/2)*( -y^2 +2*x*y +x^2)*( -y^2 -2*x*y +x^2) , 3/8*70^(1/2)*(x -y)*(x 
    +y)*x*y.
Pi^(1/2) r^5 Y_5^(0) = 1/16*11^(1/2)*(8*z^4 -40*x^2*z^2 -40*z^2*y^2 +30*x^2*y^2 +15*x^4 +15*y^4)*z , 0.
Pi^(1/2) r^5 Y_5^(1) = -1/32*x*(8*z^4 -12*z^2*y^2 +y^4 -12*x^2*z^2 +2*x^2*y^2 +x^4)*330^(1/2) ,  
    -1/32*330^(1/2)*(8*z^4 -12*z^2*y^2 +y^4 -12*x^2*z^2 +2*x^2*y^2 +x^4)*y.
Pi^(1/2) r^5 Y_5^(2) = -1/16*2310^(1/2)*( -2*z^2 +y^2 +x^2)*(x -y)*(x +y)*z ,  -1/8*2310^(1/2)*x*( 
    -2*z^2 +y^2 +x^2)*y*z.
Pi^(1/2) r^5 Y_5^(3) = 1/32*385^(1/2)*(x^2 -8*z^2 +y^2)*(x^2 -3*y^2)*x , 1/32*385^(1/2)*(x^2 -8*z^2 
    +y^2)*( -y^2 +3*x^2)*y.
Pi^(1/2) r^5 Y_5^(4) = 3/32*770^(1/2)*( -y^2 -2*x*y +x^2)*( -y^2 +2*x*y +x^2)*z , 3/8*770^(1/2)*(x 
    -y)*(x +y)*x*y*z.
Pi^(1/2) r^5 Y_5^(5) = -3/32*x*(x^4 -10*x^2*y^2 +5*y^4)*77^(1/2) ,  -3/32*77^(1/2)*(y^4 -10*x^2*y^2 
    +5*x^4)*y.
Pi^(1/2) r^6 Y_6^(0) = -1/32*( -16*z^6 +120*z^4*y^2 -90*z^2*y^4 +5*y^6 +120*z^4*x^2 -180*x^2*y^2*z^2 
    +15*x^2*y^4 -90*x^4*z^2 +15*x^4*y^2 +5*x^6)*13^(1/2) , 0.
Pi^(1/2) r^6 Y_6^(1) = -1/32*546^(1/2)*(8*z^4 -20*x^2*z^2 -20*z^2*y^2 +10*x^2*y^2 +5*x^4 +5*y^4)*x*z , 
     -1/32*546^(1/2)*(8*z^4 -20*x^2*z^2 -20*z^2*y^2 +10*x^2*y^2 +5*x^4 +5*y^4)*y*z.
Pi^(1/2) r^6 Y_6^(2) = 1/64*1365^(1/2)*(x^4 -16*x^2*z^2 +2*x^2*y^2 +y^4 +16*z^4 -16*z^2*y^2)*( -y 
    +x)*(x +y) , 1/32*1365^(1/2)*x*(x^4 -16*x^2*z^2 +2*x^2*y^2 +y^4 +16*z^4 -16*z^2*y^2)*y.
Pi^(1/2) r^6 Y_6^(3) = 1/32*1365^(1/2)*(3*x^2 -8*z^2 +3*y^2)*(x^2 -3*y^2)*x*z , 1/32*1365^(1/2)*(3*x^2 
    -8*z^2 +3*y^2)*( -y^2 +3*x^2)*y*z.
Pi^(1/2) r^6 Y_6^(4) = -3/64*182^(1/2)*(x^2 +y^2 -10*z^2)*( -y^2 -2*x*y +x^2)*( -y^2 +2*x*y +x^2) ,  
    -3/16*182^(1/2)*(x^2 +y^2 -10*z^2)*(y +x)*( -y +x)*x*y.
Pi^(1/2) r^6 Y_6^(5) = -3/32*1001^(1/2)*x*(x^4 -10*x^2*y^2 +5*y^4)*z ,  -3/32*1001^(1/2)*(y^4 
    -10*x^2*y^2 +5*x^4)*y*z.
Pi^(1/2) r^6 Y_6^(6) = 1/64*3003^(1/2)*( -y +x)*(y +x)*(x^2 +4*x*y +y^2)*(y^2 -4*x*y +x^2) , 
    1/32*3003^(1/2)*(x^2 -3*y^2)*( -y^2 +3*x^2)*x*y.
\end{verbatim}\normalsize

More general Cartesian representations of $r^nY_l^{(m)}$, for even $n-l>0$,
follow from there by multiplication with
the trinomials
\begin{equation}
r^{n-l}=(x^2+y^2+z^2)^{(n-l)/2}
=
\sum_{\sigma_1,\sigma_2\ge 0}^{\sigma_1+\sigma_2\le (n-l)/2}
\frac{[(n-l)/2]!}{\sigma_1!\sigma_2!(\frac{n-l}{2}-\sigma_1-\sigma_2)!}x^{2\sigma_1}y^{2\sigma_2}z^{n-l-2(\sigma_1+\sigma_2)}
.
\end{equation}

Gathering terms with the aid of the tables in Section \ref{sec.3dR} generates
Cartesian expansions for $Z_{n,l}^{(m)}$:
\small \begin{verbatim}
Z_0,0^(0) = 1/2*3^(1/2)/Pi^(1/2) , 0.
Z_1,1^(0) = 1/2*z/Pi^(1/2)*15^(1/2) , 0.
Z_1,1^(1) = -1/4*30^(1/2)/Pi^(1/2)*x ,  -1/4*30^(1/2)/Pi^(1/2)*y.
Z_2,0^(0) = 1/4*7^(1/2)*( -3 +5*x^2 +5*y^2 +5*z^2)/Pi^(1/2) , 0.
Z_2,2^(0) = -1/4*( -2*z^2 +y^2 +x^2)*35^(1/2)/Pi^(1/2) , 0.
Z_2,2^(1) = -1/4*210^(1/2)/Pi^(1/2)*x*z ,  -1/4*210^(1/2)/Pi^(1/2)*y*z.
Z_2,2^(2) = 1/8*210^(1/2)*( -y +x)*(y +x)/Pi^(1/2) , 1/4*210^(1/2)/Pi^(1/2)*x*y.
Z_3,1^(0) = 3/4*z*3^(1/2)*( -5 +7*x^2 +7*y^2 +7*z^2)/Pi^(1/2) , 0.
Z_3,1^(1) = -3/8*6^(1/2)*x*( -5 +7*x^2 +7*y^2 +7*z^2)/Pi^(1/2) ,  -3/8*6^(1/2)*y*( -5 +7*x^2 +7*y^2 
    +7*z^2)/Pi^(1/2).
Z_3,3^(0) = -3/4*z*( -2*z^2 +3*y^2 +3*x^2)*7^(1/2)/Pi^(1/2) , 0.
Z_3,3^(1) = 3/8*21^(1/2)*x*( -4*z^2 +y^2 +x^2)/Pi^(1/2) , 3/8*21^(1/2)*y*( -4*z^2 +y^2 +x^2)/Pi^(1/2).
Z_3,3^(2) = 3/8*210^(1/2)*z*( -y +x)*(y +x)/Pi^(1/2) , 3/4*210^(1/2)/Pi^(1/2)*x*y*z.
Z_3,3^(3) = -3/8*35^(1/2)*x*(x^2 -3*y^2)/Pi^(1/2) ,  -3/8*35^(1/2)*y*(3*x^2 -y^2)/Pi^(1/2).
Z_4,0^(0) = 1/16*11^(1/2)*(15 -70*x^2 -70*y^2 -70*z^2 +63*x^4 +126*x^2*y^2 +126*x^2*z^2 +63*y^4 
    +126*z^2*y^2 +63*z^4)/Pi^(1/2) , 0.
Z_4,2^(0) = -1/8*( -2*z^2 +y^2 +x^2)*55^(1/2)*( -7 +9*x^2 +9*y^2 +9*z^2)/Pi^(1/2) , 0.
Z_4,2^(1) = -1/8*330^(1/2)*x*z*( -7 +9*x^2 +9*y^2 +9*z^2)/Pi^(1/2) ,  -1/8*330^(1/2)*y*z*( -7 +9*x^2 
    +9*y^2 +9*z^2)/Pi^(1/2).
Z_4,2^(2) = 1/16*330^(1/2)*(x -y)*(x +y)*( -7 +9*x^2 +9*y^2 +9*z^2)/Pi^(1/2) , 1/8*330^(1/2)*x*y*( -7 
    +9*x^2 +9*y^2 +9*z^2)/Pi^(1/2).
Z_4,4^(0) = 3/16*11^(1/2)*(8*z^4 -24*z^2*y^2 +3*y^4 -24*x^2*z^2 +6*x^2*y^2 +3*x^4)/Pi^(1/2) , 0.
Z_4,4^(1) = 3/8*55^(1/2)*x*z*( -4*z^2 +3*y^2 +3*x^2)/Pi^(1/2) , 3/8*55^(1/2)*y*z*( -4*z^2 +3*y^2 
    +3*x^2)/Pi^(1/2).
Z_4,4^(2) = -3/16*110^(1/2)*(x -y)*(x +y)*(x^2 -6*z^2 +y^2)/Pi^(1/2) ,  -3/8*110^(1/2)*x*y*(x^2 -6*z^2 
    +y^2)/Pi^(1/2).
Z_4,4^(3) = -3/8*385^(1/2)*x*z*(x^2 -3*y^2)/Pi^(1/2) ,  -3/8*385^(1/2)*y*z*(3*x^2 -y^2)/Pi^(1/2).
Z_4,4^(4) = 3/32*770^(1/2)*( -y^2 +2*x*y +x^2)*( -y^2 -2*x*y +x^2)/Pi^(1/2) , 3/8*770^(1/2)*x*y*(x 
    -y)*(x +y)/Pi^(1/2).
Z_5,1^(0) = 1/16*z*39^(1/2)*(35 -126*x^2 -126*y^2 -126*z^2 +99*x^4 +198*x^2*y^2 +198*x^2*z^2 +99*y^4 
    +198*z^2*y^2 +99*z^4)/Pi^(1/2) , 0.
Z_5,1^(1) = -1/32*78^(1/2)*x*(35 -126*x^2 -126*y^2 -126*z^2 +99*x^4 +198*x^2*y^2 +198*x^2*z^2 +99*y^4 
    +198*z^2*y^2 +99*z^4)/Pi^(1/2) ,  -1/32*78^(1/2)*y*(35 -126*x^2 -126*y^2 -126*z^2 +99*x^4 +198*x^2*y^2 
    +198*x^2*z^2 +99*y^4 +198*z^2*y^2 +99*z^4)/Pi^(1/2).
Z_5,3^(0) = -1/8*z*( -2*z^2 +3*y^2 +3*x^2)*91^(1/2)*( -9 +11*x^2 +11*y^2 +11*z^2)/Pi^(1/2) , 0.
Z_5,3^(1) = 1/16*273^(1/2)*x*( -4*z^2 +y^2 +x^2)*( -9 +11*x^2 +11*y^2 +11*z^2)/Pi^(1/2) , 
    1/16*273^(1/2)*y*( -4*z^2 +y^2 +x^2)*( -9 +11*x^2 +11*y^2 +11*z^2)/Pi^(1/2).
Z_5,3^(2) = 1/16*2730^(1/2)*z*(x -y)*(x +y)*( -9 +11*x^2 +11*y^2 +11*z^2)/Pi^(1/2) , 
    1/8*2730^(1/2)*x*y*z*( -9 +11*x^2 +11*y^2 +11*z^2)/Pi^(1/2).
Z_5,3^(3) = -1/16*455^(1/2)*x*(x^2 -3*y^2)*( -9 +11*x^2 +11*y^2 +11*z^2)/Pi^(1/2) ,  
    -1/16*455^(1/2)*y*(3*x^2 -y^2)*( -9 +11*x^2 +11*y^2 +11*z^2)/Pi^(1/2).
Z_5,5^(0) = 1/16*z*(8*z^4 -40*z^2*y^2 +15*y^4 -40*x^2*z^2 +30*x^2*y^2 +15*x^4)*143^(1/2)/Pi^(1/2) , 0.
Z_5,5^(1) = -1/32*4290^(1/2)*x*(8*z^4 -12*z^2*y^2 +y^4 -12*x^2*z^2 +2*x^2*y^2 +x^4)/Pi^(1/2) ,  
    -1/32*4290^(1/2)*y*(8*z^4 -12*z^2*y^2 +y^4 -12*x^2*z^2 +2*x^2*y^2 +x^4)/Pi^(1/2).
Z_5,5^(2) = -1/16*30030^(1/2)*z*(x -y)*(x +y)*( -2*z^2 +y^2 +x^2)/Pi^(1/2) ,  -1/8*30030^(1/2)*x*y*z*( 
    -2*z^2 +y^2 +x^2)/Pi^(1/2).
Z_5,5^(3) = 1/32*5005^(1/2)*x*(x^2 -3*y^2)*(x^2 -8*z^2 +y^2)/Pi^(1/2) , 1/32*5005^(1/2)*y*(3*x^2 
    -y^2)*(x^2 -8*z^2 +y^2)/Pi^(1/2).
Z_5,5^(4) = 3/32*10010^(1/2)*z*( -y^2 +2*x*y +x^2)*( -y^2 -2*x*y +x^2)/Pi^(1/2) , 
    3/8*10010^(1/2)*x*y*z*(x -y)*(x +y)/Pi^(1/2).
Z_5,5^(5) = -3/32*1001^(1/2)*x*(x^4 -10*x^2*y^2 +5*y^4)/Pi^(1/2) ,  -3/32*1001^(1/2)*y*(5*x^4 
    -10*x^2*y^2 +y^4)/Pi^(1/2).
Z_6,0^(0) = 1/32*15^(1/2)*( -35 +315*x^2 +315*y^2 +315*z^2 -693*x^4 -1386*x^2*y^2 -1386*x^2*z^2 
    -693*y^4 -1386*z^2*y^2 -693*z^4 +429*x^6 +1287*x^4*y^2 +1287*x^4*z^2 +1287*x^2*y^4 +2574*x^2*y^2*z^2 
    +1287*z^4*x^2 +429*y^6 +1287*z^2*y^4 +1287*z^4*y^2 +429*z^6)/Pi^(1/2) , 0.
Z_6,2^(0) = -5/32*3^(1/2)*( -2*z^2 +y^2 +x^2)*(63 -198*x^2 -198*y^2 -198*z^2 +143*x^4 +286*x^2*y^2 
    +286*x^2*z^2 +143*y^4 +286*z^2*y^2 +143*z^4)/Pi^(1/2) , 0.
Z_6,2^(1) = -15/32*2^(1/2)*x*z*(63 -198*x^2 -198*y^2 -198*z^2 +143*x^4 +286*x^2*y^2 +286*x^2*z^2 
    +143*y^4 +286*z^2*y^2 +143*z^4)/Pi^(1/2) ,  -15/32*2^(1/2)*y*z*(63 -198*x^2 -198*y^2 -198*z^2 +143*x^4 
    +286*x^2*y^2 +286*x^2*z^2 +143*y^4 +286*z^2*y^2 +143*z^4)/Pi^(1/2).
Z_6,2^(2) = 15/64*2^(1/2)*(x -y)*(x +y)*(63 -198*x^2 -198*y^2 -198*z^2 +143*x^4 +286*x^2*y^2 
    +286*x^2*z^2 +143*y^4 +286*z^2*y^2 +143*z^4)/Pi^(1/2) , 15/32*2^(1/2)*x*y*(63 -198*x^2 -198*y^2 
    -198*z^2 +143*x^4 +286*x^2*y^2 +286*x^2*z^2 +143*y^4 +286*z^2*y^2 +143*z^4)/Pi^(1/2).
Z_6,4^(0) = 3/32*15^(1/2)*(8*z^4 -24*z^2*y^2 +3*y^4 -24*x^2*z^2 +6*x^2*y^2 +3*x^4)*( -11 +13*x^2 
    +13*y^2 +13*z^2)/Pi^(1/2) , 0.
Z_6,4^(1) = 15/16*3^(1/2)*x*z*( -4*z^2 +3*y^2 +3*x^2)*( -11 +13*x^2 +13*y^2 +13*z^2)/Pi^(1/2) , 
    15/16*3^(1/2)*y*z*( -4*z^2 +3*y^2 +3*x^2)*( -11 +13*x^2 +13*y^2 +13*z^2)/Pi^(1/2).
Z_6,4^(2) = -15/32*6^(1/2)*(x -y)*(x +y)*(x^2 -6*z^2 +y^2)*( -11 +13*x^2 +13*y^2 +13*z^2)/Pi^(1/2) ,  
    -15/16*6^(1/2)*x*y*(x^2 -6*z^2 +y^2)*( -11 +13*x^2 +13*y^2 +13*z^2)/Pi^(1/2).
Z_6,4^(3) = -15/16*21^(1/2)*x*z*(x^2 -3*y^2)*( -11 +13*x^2 +13*y^2 +13*z^2)/Pi^(1/2) ,  
    -15/16*21^(1/2)*y*z*(3*x^2 -y^2)*( -11 +13*x^2 +13*y^2 +13*z^2)/Pi^(1/2).
Z_6,4^(4) = 15/64*42^(1/2)*( -y^2 -2*x*y +x^2)*( -y^2 +2*x*y +x^2)*( -11 +13*x^2 +13*y^2 
    +13*z^2)/Pi^(1/2) , 15/16*42^(1/2)*x*y*(x -y)*(x +y)*( -11 +13*x^2 +13*y^2 +13*z^2)/Pi^(1/2).
Z_6,6^(0) = -1/32*( -16*z^6 +120*z^4*y^2 -90*z^2*y^4 +5*y^6 +120*z^4*x^2 -180*x^2*y^2*z^2 +15*x^2*y^4 
    -90*x^4*z^2 +15*x^4*y^2 +5*x^6)*195^(1/2)/Pi^(1/2) , 0.
Z_6,6^(1) = -3/32*910^(1/2)*x*z*(8*z^4 -20*z^2*y^2 +5*y^4 -20*x^2*z^2 +10*x^2*y^2 +5*x^4)/Pi^(1/2) ,  
    -3/32*910^(1/2)*y*z*(8*z^4 -20*z^2*y^2 +5*y^4 -20*x^2*z^2 +10*x^2*y^2 +5*x^4)/Pi^(1/2).
Z_6,6^(2) = 15/64*91^(1/2)*(x -y)*(x +y)*(x^4 -16*x^2*z^2 +2*x^2*y^2 +y^4 +16*z^4 
    -16*z^2*y^2)/Pi^(1/2) , 15/32*91^(1/2)*x*y*(x^4 -16*x^2*z^2 +2*x^2*y^2 +y^4 +16*z^4 
    -16*z^2*y^2)/Pi^(1/2).
Z_6,6^(3) = 15/32*91^(1/2)*x*z*(x^2 -3*y^2)*(3*x^2 -8*z^2 +3*y^2)/Pi^(1/2) , 15/32*91^(1/2)*y*z*(3*x^2 
    -y^2)*(3*x^2 -8*z^2 +3*y^2)/Pi^(1/2).
Z_6,6^(4) = -3/64*2730^(1/2)*( -y^2 +2*x*y +x^2)*( -y^2 -2*x*y +x^2)*(x^2 +y^2 -10*z^2)/Pi^(1/2) ,  
    -3/16*2730^(1/2)*x*y*(x -y)*(x +y)*(x^2 +y^2 -10*z^2)/Pi^(1/2).
Z_6,6^(5) = -3/32*15015^(1/2)*x*z*(x^4 -10*x^2*y^2 +5*y^4)/Pi^(1/2) ,  -3/32*15015^(1/2)*y*z*(5*x^4 
    -10*x^2*y^2 +y^4)/Pi^(1/2).
Z_6,6^(6) = 3/64*5005^(1/2)*(x -y)*(x +y)*(x^2 +4*x*y +y^2)*(y^2 -4*x*y +x^2)/Pi^(1/2) , 
    3/32*5005^(1/2)*x*y*(x^2 -3*y^2)*(3*x^2 -y^2)/Pi^(1/2).
Z_7,1^(0) = 1/32*z*51^(1/2)*( -105 +693*x^2 +693*y^2 +693*z^2 -1287*x^4 -2574*x^2*y^2 -2574*x^2*z^2 
    -1287*y^4 -2574*z^2*y^2 -1287*z^4 +715*x^6 +2145*x^4*y^2 +2145*x^4*z^2 +2145*x^2*y^4 +4290*x^2*y^2*z^2 
    +2145*z^4*x^2 +715*y^6 +2145*z^2*y^4 +2145*z^4*y^2 +715*z^6)/Pi^(1/2) , 0.
Z_7,1^(1) = -1/64*102^(1/2)*x*( -105 +693*x^2 +693*y^2 +693*z^2 -1287*x^4 -2574*x^2*y^2 -2574*x^2*z^2 
    -1287*y^4 -2574*z^2*y^2 -1287*z^4 +715*x^6 +2145*x^4*y^2 +2145*x^4*z^2 +2145*x^2*y^4 +4290*x^2*y^2*z^2 
    +2145*z^4*x^2 +715*y^6 +2145*z^2*y^4 +2145*z^4*y^2 +715*z^6)/Pi^(1/2) ,  -1/64*102^(1/2)*y*( -105 
    +693*x^2 +693*y^2 +693*z^2 -1287*x^4 -2574*x^2*y^2 -2574*x^2*z^2 -1287*y^4 -2574*z^2*y^2 -1287*z^4 
    +715*x^6 +2145*x^4*y^2 +2145*x^4*z^2 +2145*x^2*y^4 +4290*x^2*y^2*z^2 +2145*z^4*x^2 +715*y^6 
    +2145*z^2*y^4 +2145*z^4*y^2 +715*z^6)/Pi^(1/2).
Z_7,3^(0) = -1/32*z*( -2*z^2 +3*y^2 +3*x^2)*119^(1/2)*(99 -286*x^2 -286*y^2 -286*z^2 +195*x^4 
    +390*x^2*y^2 +390*x^2*z^2 +195*y^4 +390*z^2*y^2 +195*z^4)/Pi^(1/2) , 0.
Z_7,3^(1) = 1/64*357^(1/2)*x*( -4*z^2 +y^2 +x^2)*(99 -286*x^2 -286*y^2 -286*z^2 +195*x^4 +390*x^2*y^2 
    +390*x^2*z^2 +195*y^4 +390*z^2*y^2 +195*z^4)/Pi^(1/2) , 1/64*357^(1/2)*y*( -4*z^2 +y^2 +x^2)*(99 
    -286*x^2 -286*y^2 -286*z^2 +195*x^4 +390*x^2*y^2 +390*x^2*z^2 +195*y^4 +390*z^2*y^2 +195*z^4)/Pi^(1/2).
Z_7,3^(2) = 1/64*3570^(1/2)*z*(x -y)*(x +y)*(99 -286*x^2 -286*y^2 -286*z^2 +195*x^4 +390*x^2*y^2 
    +390*x^2*z^2 +195*y^4 +390*z^2*y^2 +195*z^4)/Pi^(1/2) , 1/32*3570^(1/2)*x*y*z*(99 -286*x^2 -286*y^2 
    -286*z^2 +195*x^4 +390*x^2*y^2 +390*x^2*z^2 +195*y^4 +390*z^2*y^2 +195*z^4)/Pi^(1/2).
Z_7,3^(3) = -1/64*595^(1/2)*x*(x^2 -3*y^2)*(99 -286*x^2 -286*y^2 -286*z^2 +195*x^4 +390*x^2*y^2 
    +390*x^2*z^2 +195*y^4 +390*z^2*y^2 +195*z^4)/Pi^(1/2) ,  -1/64*595^(1/2)*y*(3*x^2 -y^2)*(99 -286*x^2 
    -286*y^2 -286*z^2 +195*x^4 +390*x^2*y^2 +390*x^2*z^2 +195*y^4 +390*z^2*y^2 +195*z^4)/Pi^(1/2).
Z_7,5^(0) = 1/32*z*(8*z^4 -40*z^2*y^2 +15*y^4 -40*x^2*z^2 +30*x^2*y^2 +15*x^4)*187^(1/2)*( -13 +15*x^2 
    +15*y^2 +15*z^2)/Pi^(1/2) , 0.
Z_7,5^(1) = -1/64*5610^(1/2)*x*(8*z^4 -12*z^2*y^2 +y^4 -12*x^2*z^2 +2*x^2*y^2 +x^4)*( -13 +15*x^2 
    +15*y^2 +15*z^2)/Pi^(1/2) ,  -1/64*5610^(1/2)*y*(8*z^4 -12*z^2*y^2 +y^4 -12*x^2*z^2 +2*x^2*y^2 +x^4)*( 
    -13 +15*x^2 +15*y^2 +15*z^2)/Pi^(1/2).
Z_7,5^(2) = -1/32*39270^(1/2)*z*(x -y)*(x +y)*( -2*z^2 +y^2 +x^2)*( -13 +15*x^2 +15*y^2 
    +15*z^2)/Pi^(1/2) ,  -1/16*39270^(1/2)*x*y*z*( -2*z^2 +y^2 +x^2)*( -13 +15*x^2 +15*y^2 
    +15*z^2)/Pi^(1/2).
Z_7,5^(3) = 1/64*6545^(1/2)*x*(x^2 -3*y^2)*(x^2 -8*z^2 +y^2)*( -13 +15*x^2 +15*y^2 +15*z^2)/Pi^(1/2) , 
    1/64*6545^(1/2)*y*(3*x^2 -y^2)*(x^2 -8*z^2 +y^2)*( -13 +15*x^2 +15*y^2 +15*z^2)/Pi^(1/2).
Z_7,5^(4) = 3/64*13090^(1/2)*z*( -y^2 -2*x*y +x^2)*( -y^2 +2*x*y +x^2)*( -13 +15*x^2 +15*y^2 
    +15*z^2)/Pi^(1/2) , 3/16*13090^(1/2)*x*y*z*(x -y)*(x +y)*( -13 +15*x^2 +15*y^2 +15*z^2)/Pi^(1/2).
Z_7,5^(5) = -3/64*1309^(1/2)*x*(x^4 -10*x^2*y^2 +5*y^4)*( -13 +15*x^2 +15*y^2 +15*z^2)/Pi^(1/2) ,  
    -3/64*1309^(1/2)*y*(5*x^4 -10*x^2*y^2 +y^4)*( -13 +15*x^2 +15*y^2 +15*z^2)/Pi^(1/2).
Z_7,7^(0) = -1/32*z*( -16*z^6 +168*z^4*y^2 -210*z^2*y^4 +35*y^6 +168*z^4*x^2 -420*x^2*y^2*z^2 
    +105*x^2*y^4 -210*x^4*z^2 +105*x^4*y^2 +35*x^6)*255^(1/2)/Pi^(1/2) , 0.
Z_7,7^(1) = 1/128*3570^(1/2)*x*( -64*z^6 +240*z^4*y^2 -120*z^2*y^4 +5*y^6 +240*z^4*x^2 
    -240*x^2*y^2*z^2 +15*x^2*y^4 -120*x^4*z^2 +15*x^4*y^2 +5*x^6)/Pi^(1/2) , 1/128*3570^(1/2)*y*( -64*z^6 
    +240*z^4*y^2 -120*z^2*y^4 +5*y^6 +240*z^4*x^2 -240*x^2*y^2*z^2 +15*x^2*y^4 -120*x^4*z^2 +15*x^4*y^2 
    +5*x^6)/Pi^(1/2).
Z_7,7^(2) = 3/64*595^(1/2)*z*(x -y)*(x +y)*(15*x^4 -80*x^2*z^2 +30*x^2*y^2 +15*y^4 +48*z^4 
    -80*z^2*y^2)/Pi^(1/2) , 3/32*595^(1/2)*x*y*z*(15*x^4 -80*x^2*z^2 +30*x^2*y^2 +15*y^4 +48*z^4 
    -80*z^2*y^2)/Pi^(1/2).
Z_7,7^(3) = -3/128*1190^(1/2)*x*(x^2 -3*y^2)*(3*x^4 -60*x^2*z^2 +6*x^2*y^2 +80*z^4 -60*z^2*y^2 
    +3*y^4)/Pi^(1/2) ,  -3/128*1190^(1/2)*y*(3*x^2 -y^2)*(3*x^4 -60*x^2*z^2 +6*x^2*y^2 +80*z^4 -60*z^2*y^2 
    +3*y^4)/Pi^(1/2).
Z_7,7^(4) = -3/64*13090^(1/2)*z*( -y^2 +2*x*y +x^2)*( -y^2 -2*x*y +x^2)*(3*x^2 +3*y^2 
    -10*z^2)/Pi^(1/2) ,  -3/16*13090^(1/2)*x*y*z*(x -y)*(x +y)*(3*x^2 +3*y^2 -10*z^2)/Pi^(1/2).
Z_7,7^(5) = 3/128*13090^(1/2)*x*(x^4 -10*x^2*y^2 +5*y^4)*(x^2 +y^2 -12*z^2)/Pi^(1/2) , 
    3/128*13090^(1/2)*y*(5*x^4 -10*x^2*y^2 +y^4)*(x^2 +y^2 -12*z^2)/Pi^(1/2).
Z_7,7^(6) = 3/64*85085^(1/2)*z*(x -y)*(x +y)*(x^2 +4*x*y +y^2)*(y^2 -4*x*y +x^2)/Pi^(1/2) , 
    3/32*85085^(1/2)*x*y*z*(x^2 -3*y^2)*(3*x^2 -y^2)/Pi^(1/2).
Z_7,7^(7) = -3/128*24310^(1/2)*x*(x^6 -21*x^4*y^2 +35*x^2*y^4 -7*y^6)/Pi^(1/2) ,  
    -3/128*24310^(1/2)*y*(7*x^6 -35*x^4*y^2 +21*x^2*y^4 -y^6)/Pi^(1/2).
\end{verbatim}\normalsize
Terms with negative upper index $m$ of $Z$ have not been listed and follow from (\ref{eq.Ylmcc}):
If $m$ is even, the imaginary part of $Z$ changes sign; if $m$ is odd, the real part changes sign.

\subsection{Cartesian to Zernike} \label{sec.C2Z3D}

The inverse problem to Section \ref{sec.Z2C3D} is finding the $u$-coefficients
in the ansatz
\begin{equation}
x^p y^q z^t=\sum_{\begin{array}{c}-p-q\le m \le p+q\\
          l\le n\le p+q+t\end{array}}u_{p,q,t,n,l,m} R_n^{(l)}(r)Y_l^{(m)}(\theta,\varphi)
,
\label{eq.udef}
\end{equation}
given $p$, $q$ and $t$.
In practise we use (\ref{eq.orthoZ3D}) to construct
\begin{equation}
u_{p,q,t,n,l,m} = \int_{r<1} r^{p+q+t} \sin^{p+q}\theta\cos^t\theta \cos^p\varphi\sin^q\varphi
   R_n^{(l)}(r)Y_l^{(m)*}(\theta,\varphi) d^3r
\end{equation}
by projection onto each individual $Z_{n,l}^{(m)}$.
After dropping a common exponential factor in the wave function bases,
Section 3.3 of my earlier work \cite{MatharIJQC90} states,
\begin{equation}
u_{p,q,t,n,l,m}= (-1)^m\sqrt{\frac{2l+1}{4\pi}\,\frac{(l-m)!}{(l+m)!}}
I_r(p+q+t,n,l) I_\theta(p+q,t,l,m) I_\varphi(p,q,m)^*,
\end{equation}
with two factors defined as
\begin{equation}
I_r(j,n,l)
= \int_0^1 r^{j+2} R_n^{(l)}(r) dr
=\left\{
\begin{array}{ll}
0, & j-l \, \mathrm{odd }\,\mathrm{ or}\, n-l\,\mathrm{ odd};\\
f_{j,n,l}, & j+l\, \mathrm{even}.
\end{array}
\right.
\end{equation}
and
\begin{eqnarray}
I_\varphi(p,q,m)
&=& \int_0^{2\pi} e^{im\varphi}\cos^p\varphi\sin^q\varphi d\varphi
\\
&=&
\left\{
\begin{array}{ll}
0, & p+q-m\,\mathrm{odd}; \\
\dfrac{\pi}{2^{p+q-1}}i^q \sum\limits_{\sigma=\max(0,(p-q-m)/2)}^{\min(p,(p+q-m)/2)}
\dbinom{ p }{ \sigma}
\dbinom{ q }{ \frac{p+q-m}{2}-\sigma }
(-1)^{\sigma-(p+q-m)/2}, & p+q-m\,\mathrm{even}.
\end{array}
\right.
\end{eqnarray}
An obvious recurrence is
\begin{equation}
I_\varphi(p,q+2,m)=I_\varphi(p,q,m)-I_\varphi(p+2,q,m).
\end{equation}
The third factor is only evaluated for even $k-m$---otherwise $I_\varphi$ equals zero
which turns $u$ to zero and $I_\theta$ is not of interest---
\begin{eqnarray}
&&
I_\theta(k,t,l,m)
= \int_0^{\pi} \sin^{k+1}\theta \cos^t\theta P_l^m(\cos\theta) d\theta
\\
&&=
\left\{
\begin{array}{ll}
0, & l-m+t\,\mathrm{odd} ;\\
\dfrac{2^{l+1}(-1)^{(m-|m|)/2}}{(l-m)!}\sum\limits_{\nu=0}^{(l-|m|)/2}
\dfrac{ \left(\frac{1}{2}-\nu\right)_l \dbinom{l-|m| }{ 2\nu} }
{(1+t+l-|m|-2\nu)\dbinom{(1+t+l+k)/2-\nu}{(k+|m|)/2}}
      ,& l-m+t\,\mathrm{even}.
\end{array}
\right.
\end{eqnarray}
An actual table of (\ref{eq.udef}) looks as follows:
\small \begin{verbatim}
x / Pi^(1/2) = 1/15*30^(1/2)*Z_1,1^(-1) -1/15*30^(1/2)*Z_1,1^(1).
y / Pi^(1/2) = 1/15*i*30^(1/2)*Z_1,1^(-1) +1/15*i*30^(1/2)*Z_1,1^(1).
z / Pi^(1/2) = 2/15*15^(1/2)*Z_1,1^(0).
x^2 / Pi^(1/2) = 2/15*3^(1/2)*Z_0,0^(0) +4/105*7^(1/2)*Z_2,0^(0) +1/105*210^(1/2)*Z_2,2^(-2) 
    -2/105*35^(1/2)*Z_2,2^(0) +1/105*210^(1/2)*Z_2,2^(2).
x y / Pi^(1/2) = 1/105*i*210^(1/2)*Z_2,2^(-2) -1/105*i*210^(1/2)*Z_2,2^(2).
x z / Pi^(1/2) = 1/105*210^(1/2)*Z_2,2^(-1) -1/105*210^(1/2)*Z_2,2^(1).
y^2 / Pi^(1/2) = 2/15*3^(1/2)*Z_0,0^(0) +4/105*7^(1/2)*Z_2,0^(0) -1/105*210^(1/2)*Z_2,2^(-2) 
    -2/105*35^(1/2)*Z_2,2^(0) -1/105*210^(1/2)*Z_2,2^(2).
y z / Pi^(1/2) = 1/105*i*210^(1/2)*Z_2,2^(-1) +1/105*i*210^(1/2)*Z_2,2^(1).
z^2 / Pi^(1/2) = 2/15*3^(1/2)*Z_0,0^(0) +4/105*7^(1/2)*Z_2,0^(0) +4/105*35^(1/2)*Z_2,2^(0).
x^3 / Pi^(1/2) = 1/35*30^(1/2)*Z_1,1^(-1) -1/35*30^(1/2)*Z_1,1^(1) +2/105*6^(1/2)*Z_3,1^(-1) 
    -2/105*6^(1/2)*Z_3,1^(1) +1/105*35^(1/2)*Z_3,3^(-3) -1/105*21^(1/2)*Z_3,3^(-1) 
    +1/105*21^(1/2)*Z_3,3^(1) -1/105*35^(1/2)*Z_3,3^(3).
x^2 y / Pi^(1/2) = 1/105*i*30^(1/2)*Z_1,1^(-1) +1/105*i*30^(1/2)*Z_1,1^(1) +2/315*i*6^(1/2)*Z_3,1^(-1) 
    +2/315*i*6^(1/2)*Z_3,1^(1) +1/105*i*35^(1/2)*Z_3,3^(-3) -1/315*i*21^(1/2)*Z_3,3^(-1) 
    -1/315*i*21^(1/2)*Z_3,3^(1) +1/105*i*35^(1/2)*Z_3,3^(3).
x^2 z / Pi^(1/2) = 2/105*15^(1/2)*Z_1,1^(0) +4/315*3^(1/2)*Z_3,1^(0) +1/315*210^(1/2)*Z_3,3^(-2) 
    -2/105*7^(1/2)*Z_3,3^(0) +1/315*210^(1/2)*Z_3,3^(2).
x y^2 / Pi^(1/2) = 1/105*30^(1/2)*Z_1,1^(-1) -1/105*30^(1/2)*Z_1,1^(1) +2/315*6^(1/2)*Z_3,1^(-1) 
    -2/315*6^(1/2)*Z_3,1^(1) -1/105*35^(1/2)*Z_3,3^(-3) -1/315*21^(1/2)*Z_3,3^(-1) 
    +1/315*21^(1/2)*Z_3,3^(1) +1/105*35^(1/2)*Z_3,3^(3).
x y z / Pi^(1/2) = 1/315*i*210^(1/2)*Z_3,3^(-2) -1/315*i*210^(1/2)*Z_3,3^(2).
x z^2 / Pi^(1/2) = 1/105*30^(1/2)*Z_1,1^(-1) -1/105*30^(1/2)*Z_1,1^(1) +2/315*6^(1/2)*Z_3,1^(-1) 
    -2/315*6^(1/2)*Z_3,1^(1) +4/315*21^(1/2)*Z_3,3^(-1) -4/315*21^(1/2)*Z_3,3^(1).
y^3 / Pi^(1/2) = 1/35*i*30^(1/2)*Z_1,1^(-1) +1/35*i*30^(1/2)*Z_1,1^(1) +2/105*i*6^(1/2)*Z_3,1^(-1) 
    +2/105*i*6^(1/2)*Z_3,1^(1) -1/105*i*35^(1/2)*Z_3,3^(-3) -1/105*i*21^(1/2)*Z_3,3^(-1) 
    -1/105*i*21^(1/2)*Z_3,3^(1) -1/105*i*35^(1/2)*Z_3,3^(3).
y^2 z / Pi^(1/2) = 2/105*15^(1/2)*Z_1,1^(0) +4/315*3^(1/2)*Z_3,1^(0) -1/315*210^(1/2)*Z_3,3^(-2) 
    -2/105*7^(1/2)*Z_3,3^(0) -1/315*210^(1/2)*Z_3,3^(2).
y z^2 / Pi^(1/2) = 1/105*i*30^(1/2)*Z_1,1^(-1) +1/105*i*30^(1/2)*Z_1,1^(1) +2/315*i*6^(1/2)*Z_3,1^(-1) 
    +2/315*i*6^(1/2)*Z_3,1^(1) +4/315*i*21^(1/2)*Z_3,3^(-1) +4/315*i*21^(1/2)*Z_3,3^(1).
z^3 / Pi^(1/2) = 2/35*15^(1/2)*Z_1,1^(0) +4/105*3^(1/2)*Z_3,1^(0) +4/105*7^(1/2)*Z_3,3^(0).
x^4 / Pi^(1/2) = 2/35*3^(1/2)*Z_0,0^(0) +8/315*7^(1/2)*Z_2,0^(0) +2/315*210^(1/2)*Z_2,2^(-2) 
    -4/315*35^(1/2)*Z_2,2^(0) +2/315*210^(1/2)*Z_2,2^(2) +16/3465*11^(1/2)*Z_4,0^(0) 
    +4/3465*330^(1/2)*Z_4,2^(-2) -8/3465*55^(1/2)*Z_4,2^(0) +4/3465*330^(1/2)*Z_4,2^(2) 
    +1/1155*770^(1/2)*Z_4,4^(-4) -2/1155*110^(1/2)*Z_4,4^(-2) +2/385*11^(1/2)*Z_4,4^(0) 
    -2/1155*110^(1/2)*Z_4,4^(2) +1/1155*770^(1/2)*Z_4,4^(4).
x^3 y / Pi^(1/2) = 1/315*i*210^(1/2)*Z_2,2^(-2) -1/315*i*210^(1/2)*Z_2,2^(2) 
    +2/3465*i*330^(1/2)*Z_4,2^(-2) -2/3465*i*330^(1/2)*Z_4,2^(2) +1/1155*i*770^(1/2)*Z_4,4^(-4) 
    -1/1155*i*110^(1/2)*Z_4,4^(-2) +1/1155*i*110^(1/2)*Z_4,4^(2) -1/1155*i*770^(1/2)*Z_4,4^(4).
x^3 z / Pi^(1/2) = 1/315*210^(1/2)*Z_2,2^(-1) -1/315*210^(1/2)*Z_2,2^(1) +2/3465*330^(1/2)*Z_4,2^(-1) 
    -2/3465*330^(1/2)*Z_4,2^(1) +1/1155*385^(1/2)*Z_4,4^(-3) -1/385*55^(1/2)*Z_4,4^(-1) 
    +1/385*55^(1/2)*Z_4,4^(1) -1/1155*385^(1/2)*Z_4,4^(3).
x^2 y^2 / Pi^(1/2) = 2/105*3^(1/2)*Z_0,0^(0) +8/945*7^(1/2)*Z_2,0^(0) -4/945*35^(1/2)*Z_2,2^(0) 
    +16/10395*11^(1/2)*Z_4,0^(0) -8/10395*55^(1/2)*Z_4,2^(0) -1/1155*770^(1/2)*Z_4,4^(-4) 
    +2/1155*11^(1/2)*Z_4,4^(0) -1/1155*770^(1/2)*Z_4,4^(4).
x^2 y z / Pi^(1/2) = 1/945*i*210^(1/2)*Z_2,2^(-1) +1/945*i*210^(1/2)*Z_2,2^(1) 
    +2/10395*i*330^(1/2)*Z_4,2^(-1) +2/10395*i*330^(1/2)*Z_4,2^(1) +1/1155*i*385^(1/2)*Z_4,4^(-3) 
    -1/1155*i*55^(1/2)*Z_4,4^(-1) -1/1155*i*55^(1/2)*Z_4,4^(1) +1/1155*i*385^(1/2)*Z_4,4^(3).
x^2 z^2 / Pi^(1/2) = 2/105*3^(1/2)*Z_0,0^(0) +8/945*7^(1/2)*Z_2,0^(0) +1/945*210^(1/2)*Z_2,2^(-2) 
    +2/945*35^(1/2)*Z_2,2^(0) +1/945*210^(1/2)*Z_2,2^(2) +16/10395*11^(1/2)*Z_4,0^(0) 
    +2/10395*330^(1/2)*Z_4,2^(-2) +4/10395*55^(1/2)*Z_4,2^(0) +2/10395*330^(1/2)*Z_4,2^(2) 
    +2/1155*110^(1/2)*Z_4,4^(-2) -8/1155*11^(1/2)*Z_4,4^(0) +2/1155*110^(1/2)*Z_4,4^(2).
x y^3 / Pi^(1/2) = 1/315*i*210^(1/2)*Z_2,2^(-2) -1/315*i*210^(1/2)*Z_2,2^(2) 
    +2/3465*i*330^(1/2)*Z_4,2^(-2) -2/3465*i*330^(1/2)*Z_4,2^(2) -1/1155*i*770^(1/2)*Z_4,4^(-4) 
    -1/1155*i*110^(1/2)*Z_4,4^(-2) +1/1155*i*110^(1/2)*Z_4,4^(2) +1/1155*i*770^(1/2)*Z_4,4^(4).
x y^2 z / Pi^(1/2) = 1/945*210^(1/2)*Z_2,2^(-1) -1/945*210^(1/2)*Z_2,2^(1) 
    +2/10395*330^(1/2)*Z_4,2^(-1) -2/10395*330^(1/2)*Z_4,2^(1) -1/1155*385^(1/2)*Z_4,4^(-3) 
    -1/1155*55^(1/2)*Z_4,4^(-1) +1/1155*55^(1/2)*Z_4,4^(1) +1/1155*385^(1/2)*Z_4,4^(3).
x y z^2 / Pi^(1/2) = 1/945*i*210^(1/2)*Z_2,2^(-2) -1/945*i*210^(1/2)*Z_2,2^(2) 
    +2/10395*i*330^(1/2)*Z_4,2^(-2) -2/10395*i*330^(1/2)*Z_4,2^(2) +2/1155*i*110^(1/2)*Z_4,4^(-2) 
    -2/1155*i*110^(1/2)*Z_4,4^(2).
x z^3 / Pi^(1/2) = 1/315*210^(1/2)*Z_2,2^(-1) -1/315*210^(1/2)*Z_2,2^(1) +2/3465*330^(1/2)*Z_4,2^(-1) 
    -2/3465*330^(1/2)*Z_4,2^(1) +4/1155*55^(1/2)*Z_4,4^(-1) -4/1155*55^(1/2)*Z_4,4^(1).
y^4 / Pi^(1/2) = 2/35*3^(1/2)*Z_0,0^(0) +8/315*7^(1/2)*Z_2,0^(0) -2/315*210^(1/2)*Z_2,2^(-2) 
    -4/315*35^(1/2)*Z_2,2^(0) -2/315*210^(1/2)*Z_2,2^(2) +16/3465*11^(1/2)*Z_4,0^(0) 
    -4/3465*330^(1/2)*Z_4,2^(-2) -8/3465*55^(1/2)*Z_4,2^(0) -4/3465*330^(1/2)*Z_4,2^(2) 
    +1/1155*770^(1/2)*Z_4,4^(-4) +2/1155*110^(1/2)*Z_4,4^(-2) +2/385*11^(1/2)*Z_4,4^(0) 
    +2/1155*110^(1/2)*Z_4,4^(2) +1/1155*770^(1/2)*Z_4,4^(4).
y^3 z / Pi^(1/2) = 1/315*i*210^(1/2)*Z_2,2^(-1) +1/315*i*210^(1/2)*Z_2,2^(1) 
    +2/3465*i*330^(1/2)*Z_4,2^(-1) +2/3465*i*330^(1/2)*Z_4,2^(1) -1/1155*i*385^(1/2)*Z_4,4^(-3) 
    -1/385*i*55^(1/2)*Z_4,4^(-1) -1/385*i*55^(1/2)*Z_4,4^(1) -1/1155*i*385^(1/2)*Z_4,4^(3).
y^2 z^2 / Pi^(1/2) = 2/105*3^(1/2)*Z_0,0^(0) +8/945*7^(1/2)*Z_2,0^(0) -1/945*210^(1/2)*Z_2,2^(-2) 
    +2/945*35^(1/2)*Z_2,2^(0) -1/945*210^(1/2)*Z_2,2^(2) +16/10395*11^(1/2)*Z_4,0^(0) 
    -2/10395*330^(1/2)*Z_4,2^(-2) +4/10395*55^(1/2)*Z_4,2^(0) -2/10395*330^(1/2)*Z_4,2^(2) 
    -2/1155*110^(1/2)*Z_4,4^(-2) -8/1155*11^(1/2)*Z_4,4^(0) -2/1155*110^(1/2)*Z_4,4^(2).
y z^3 / Pi^(1/2) = 1/315*i*210^(1/2)*Z_2,2^(-1) +1/315*i*210^(1/2)*Z_2,2^(1) 
    +2/3465*i*330^(1/2)*Z_4,2^(-1) +2/3465*i*330^(1/2)*Z_4,2^(1) +4/1155*i*55^(1/2)*Z_4,4^(-1) 
    +4/1155*i*55^(1/2)*Z_4,4^(1).
z^4 / Pi^(1/2) = 2/35*3^(1/2)*Z_0,0^(0) +8/315*7^(1/2)*Z_2,0^(0) +8/315*35^(1/2)*Z_2,2^(0) 
    +16/3465*11^(1/2)*Z_4,0^(0) +16/3465*55^(1/2)*Z_4,2^(0) +16/1155*11^(1/2)*Z_4,4^(0).
x^5 / Pi^(1/2) = 1/63*30^(1/2)*Z_1,1^(-1) -1/63*30^(1/2)*Z_1,1^(1) +4/231*6^(1/2)*Z_3,1^(-1) 
    -4/231*6^(1/2)*Z_3,1^(1) +2/231*35^(1/2)*Z_3,3^(-3) -2/231*21^(1/2)*Z_3,3^(-1) 
    +2/231*21^(1/2)*Z_3,3^(1) -2/231*35^(1/2)*Z_3,3^(3) +8/9009*78^(1/2)*Z_5,1^(-1) 
    -8/9009*78^(1/2)*Z_5,1^(1) +4/9009*455^(1/2)*Z_5,3^(-3) -4/9009*273^(1/2)*Z_5,3^(-1) 
    +4/9009*273^(1/2)*Z_5,3^(1) -4/9009*455^(1/2)*Z_5,3^(3) +1/3003*1001^(1/2)*Z_5,5^(-5) 
    -1/9009*5005^(1/2)*Z_5,5^(-3) +1/9009*4290^(1/2)*Z_5,5^(-1) -1/9009*4290^(1/2)*Z_5,5^(1) 
    +1/9009*5005^(1/2)*Z_5,5^(3) -1/3003*1001^(1/2)*Z_5,5^(5).
x^4 y / Pi^(1/2) = 1/315*i*30^(1/2)*Z_1,1^(-1) +1/315*i*30^(1/2)*Z_1,1^(1) 
    +4/1155*i*6^(1/2)*Z_3,1^(-1) +4/1155*i*6^(1/2)*Z_3,1^(1) +2/385*i*35^(1/2)*Z_3,3^(-3) 
    -2/1155*i*21^(1/2)*Z_3,3^(-1) -2/1155*i*21^(1/2)*Z_3,3^(1) +2/385*i*35^(1/2)*Z_3,3^(3) 
    +8/45045*i*78^(1/2)*Z_5,1^(-1) +8/45045*i*78^(1/2)*Z_5,1^(1) +4/15015*i*455^(1/2)*Z_5,3^(-3) 
    -4/45045*i*273^(1/2)*Z_5,3^(-1) -4/45045*i*273^(1/2)*Z_5,3^(1) +4/15015*i*455^(1/2)*Z_5,3^(3) 
    +1/3003*i*1001^(1/2)*Z_5,5^(-5) -1/15015*i*5005^(1/2)*Z_5,5^(-3) +1/45045*i*4290^(1/2)*Z_5,5^(-1) 
    +1/45045*i*4290^(1/2)*Z_5,5^(1) -1/15015*i*5005^(1/2)*Z_5,5^(3) +1/3003*i*1001^(1/2)*Z_5,5^(5).
x^4 z / Pi^(1/2) = 2/315*15^(1/2)*Z_1,1^(0) +8/1155*3^(1/2)*Z_3,1^(0) +2/1155*210^(1/2)*Z_3,3^(-2) 
    -4/385*7^(1/2)*Z_3,3^(0) +2/1155*210^(1/2)*Z_3,3^(2) +16/45045*39^(1/2)*Z_5,1^(0) 
    +4/45045*2730^(1/2)*Z_5,3^(-2) -8/15015*91^(1/2)*Z_5,3^(0) +4/45045*2730^(1/2)*Z_5,3^(2) 
    +1/15015*10010^(1/2)*Z_5,5^(-4) -2/45045*30030^(1/2)*Z_5,5^(-2) +2/3003*143^(1/2)*Z_5,5^(0) 
    -2/45045*30030^(1/2)*Z_5,5^(2) +1/15015*10010^(1/2)*Z_5,5^(4).
x^3 y^2 / Pi^(1/2) = 1/315*30^(1/2)*Z_1,1^(-1) -1/315*30^(1/2)*Z_1,1^(1) +4/1155*6^(1/2)*Z_3,1^(-1) 
    -4/1155*6^(1/2)*Z_3,1^(1) -2/1155*35^(1/2)*Z_3,3^(-3) -2/1155*21^(1/2)*Z_3,3^(-1) 
    +2/1155*21^(1/2)*Z_3,3^(1) +2/1155*35^(1/2)*Z_3,3^(3) +8/45045*78^(1/2)*Z_5,1^(-1) 
    -8/45045*78^(1/2)*Z_5,1^(1) -4/45045*455^(1/2)*Z_5,3^(-3) -4/45045*273^(1/2)*Z_5,3^(-1) 
    +4/45045*273^(1/2)*Z_5,3^(1) +4/45045*455^(1/2)*Z_5,3^(3) -1/3003*1001^(1/2)*Z_5,5^(-5) 
    +1/45045*5005^(1/2)*Z_5,5^(-3) +1/45045*4290^(1/2)*Z_5,5^(-1) -1/45045*4290^(1/2)*Z_5,5^(1) 
    -1/45045*5005^(1/2)*Z_5,5^(3) +1/3003*1001^(1/2)*Z_5,5^(5).
x^3 y z / Pi^(1/2) = 1/1155*i*210^(1/2)*Z_3,3^(-2) -1/1155*i*210^(1/2)*Z_3,3^(2) 
    +2/45045*i*2730^(1/2)*Z_5,3^(-2) -2/45045*i*2730^(1/2)*Z_5,3^(2) +1/15015*i*10010^(1/2)*Z_5,5^(-4) 
    -1/45045*i*30030^(1/2)*Z_5,5^(-2) +1/45045*i*30030^(1/2)*Z_5,5^(2) -1/15015*i*10010^(1/2)*Z_5,5^(4).
x^3 z^2 / Pi^(1/2) = 1/315*30^(1/2)*Z_1,1^(-1) -1/315*30^(1/2)*Z_1,1^(1) +4/1155*6^(1/2)*Z_3,1^(-1) 
    -4/1155*6^(1/2)*Z_3,1^(1) +1/1155*35^(1/2)*Z_3,3^(-3) +1/385*21^(1/2)*Z_3,3^(-1) 
    -1/385*21^(1/2)*Z_3,3^(1) -1/1155*35^(1/2)*Z_3,3^(3) +8/45045*78^(1/2)*Z_5,1^(-1) 
    -8/45045*78^(1/2)*Z_5,1^(1) +2/45045*455^(1/2)*Z_5,3^(-3) +2/15015*273^(1/2)*Z_5,3^(-1) 
    -2/15015*273^(1/2)*Z_5,3^(1) -2/45045*455^(1/2)*Z_5,3^(3) +4/45045*5005^(1/2)*Z_5,5^(-3) 
    -2/15015*4290^(1/2)*Z_5,5^(-1) +2/15015*4290^(1/2)*Z_5,5^(1) -4/45045*5005^(1/2)*Z_5,5^(3).
x^2 y^3 / Pi^(1/2) = 1/315*i*30^(1/2)*Z_1,1^(-1) +1/315*i*30^(1/2)*Z_1,1^(1) 
    +4/1155*i*6^(1/2)*Z_3,1^(-1) +4/1155*i*6^(1/2)*Z_3,1^(1) +2/1155*i*35^(1/2)*Z_3,3^(-3) 
    -2/1155*i*21^(1/2)*Z_3,3^(-1) -2/1155*i*21^(1/2)*Z_3,3^(1) +2/1155*i*35^(1/2)*Z_3,3^(3) 
    +8/45045*i*78^(1/2)*Z_5,1^(-1) +8/45045*i*78^(1/2)*Z_5,1^(1) +4/45045*i*455^(1/2)*Z_5,3^(-3) 
    -4/45045*i*273^(1/2)*Z_5,3^(-1) -4/45045*i*273^(1/2)*Z_5,3^(1) +4/45045*i*455^(1/2)*Z_5,3^(3) 
    -1/3003*i*1001^(1/2)*Z_5,5^(-5) -1/45045*i*5005^(1/2)*Z_5,5^(-3) +1/45045*i*4290^(1/2)*Z_5,5^(-1) 
    +1/45045*i*4290^(1/2)*Z_5,5^(1) -1/45045*i*5005^(1/2)*Z_5,5^(3) -1/3003*i*1001^(1/2)*Z_5,5^(5).
x^2 y^2 z / Pi^(1/2) = 2/945*15^(1/2)*Z_1,1^(0) +8/3465*3^(1/2)*Z_3,1^(0) -4/1155*7^(1/2)*Z_3,3^(0) 
    +16/135135*39^(1/2)*Z_5,1^(0) -8/45045*91^(1/2)*Z_5,3^(0) -1/15015*10010^(1/2)*Z_5,5^(-4) 
    +2/9009*143^(1/2)*Z_5,5^(0) -1/15015*10010^(1/2)*Z_5,5^(4).
x^2 y z^2 / Pi^(1/2) = 1/945*i*30^(1/2)*Z_1,1^(-1) +1/945*i*30^(1/2)*Z_1,1^(1) 
    +4/3465*i*6^(1/2)*Z_3,1^(-1) +4/3465*i*6^(1/2)*Z_3,1^(1) +1/1155*i*35^(1/2)*Z_3,3^(-3) 
    +1/1155*i*21^(1/2)*Z_3,3^(-1) +1/1155*i*21^(1/2)*Z_3,3^(1) +1/1155*i*35^(1/2)*Z_3,3^(3) 
    +8/135135*i*78^(1/2)*Z_5,1^(-1) +8/135135*i*78^(1/2)*Z_5,1^(1) +2/45045*i*455^(1/2)*Z_5,3^(-3) 
    +2/45045*i*273^(1/2)*Z_5,3^(-1) +2/45045*i*273^(1/2)*Z_5,3^(1) +2/45045*i*455^(1/2)*Z_5,3^(3) 
    +4/45045*i*5005^(1/2)*Z_5,5^(-3) -2/45045*i*4290^(1/2)*Z_5,5^(-1) -2/45045*i*4290^(1/2)*Z_5,5^(1) 
    +4/45045*i*5005^(1/2)*Z_5,5^(3).
x^2 z^3 / Pi^(1/2) = 2/315*15^(1/2)*Z_1,1^(0) +8/1155*3^(1/2)*Z_3,1^(0) +1/1155*210^(1/2)*Z_3,3^(-2) 
    -2/1155*7^(1/2)*Z_3,3^(0) +1/1155*210^(1/2)*Z_3,3^(2) +16/45045*39^(1/2)*Z_5,1^(0) 
    +2/45045*2730^(1/2)*Z_5,3^(-2) -4/45045*91^(1/2)*Z_5,3^(0) +2/45045*2730^(1/2)*Z_5,3^(2) 
    +2/45045*30030^(1/2)*Z_5,5^(-2) -8/9009*143^(1/2)*Z_5,5^(0) +2/45045*30030^(1/2)*Z_5,5^(2).
x y^4 / Pi^(1/2) = 1/315*30^(1/2)*Z_1,1^(-1) -1/315*30^(1/2)*Z_1,1^(1) +4/1155*6^(1/2)*Z_3,1^(-1) 
    -4/1155*6^(1/2)*Z_3,1^(1) -2/385*35^(1/2)*Z_3,3^(-3) -2/1155*21^(1/2)*Z_3,3^(-1) 
    +2/1155*21^(1/2)*Z_3,3^(1) +2/385*35^(1/2)*Z_3,3^(3) +8/45045*78^(1/2)*Z_5,1^(-1) 
    -8/45045*78^(1/2)*Z_5,1^(1) -4/15015*455^(1/2)*Z_5,3^(-3) -4/45045*273^(1/2)*Z_5,3^(-1) 
    +4/45045*273^(1/2)*Z_5,3^(1) +4/15015*455^(1/2)*Z_5,3^(3) +1/3003*1001^(1/2)*Z_5,5^(-5) 
    +1/15015*5005^(1/2)*Z_5,5^(-3) +1/45045*4290^(1/2)*Z_5,5^(-1) -1/45045*4290^(1/2)*Z_5,5^(1) 
    -1/15015*5005^(1/2)*Z_5,5^(3) -1/3003*1001^(1/2)*Z_5,5^(5).
x y^3 z / Pi^(1/2) = 1/1155*i*210^(1/2)*Z_3,3^(-2) -1/1155*i*210^(1/2)*Z_3,3^(2) 
    +2/45045*i*2730^(1/2)*Z_5,3^(-2) -2/45045*i*2730^(1/2)*Z_5,3^(2) -1/15015*i*10010^(1/2)*Z_5,5^(-4) 
    -1/45045*i*30030^(1/2)*Z_5,5^(-2) +1/45045*i*30030^(1/2)*Z_5,5^(2) +1/15015*i*10010^(1/2)*Z_5,5^(4).
x y^2 z^2 / Pi^(1/2) = 1/945*30^(1/2)*Z_1,1^(-1) -1/945*30^(1/2)*Z_1,1^(1) +4/3465*6^(1/2)*Z_3,1^(-1) 
    -4/3465*6^(1/2)*Z_3,1^(1) -1/1155*35^(1/2)*Z_3,3^(-3) +1/1155*21^(1/2)*Z_3,3^(-1) 
    -1/1155*21^(1/2)*Z_3,3^(1) +1/1155*35^(1/2)*Z_3,3^(3) +8/135135*78^(1/2)*Z_5,1^(-1) 
    -8/135135*78^(1/2)*Z_5,1^(1) -2/45045*455^(1/2)*Z_5,3^(-3) +2/45045*273^(1/2)*Z_5,3^(-1) 
    -2/45045*273^(1/2)*Z_5,3^(1) +2/45045*455^(1/2)*Z_5,3^(3) -4/45045*5005^(1/2)*Z_5,5^(-3) 
    -2/45045*4290^(1/2)*Z_5,5^(-1) +2/45045*4290^(1/2)*Z_5,5^(1) +4/45045*5005^(1/2)*Z_5,5^(3).
x y z^3 / Pi^(1/2) = 1/1155*i*210^(1/2)*Z_3,3^(-2) -1/1155*i*210^(1/2)*Z_3,3^(2) 
    +2/45045*i*2730^(1/2)*Z_5,3^(-2) -2/45045*i*2730^(1/2)*Z_5,3^(2) +2/45045*i*30030^(1/2)*Z_5,5^(-2) 
    -2/45045*i*30030^(1/2)*Z_5,5^(2).
x z^4 / Pi^(1/2) = 1/315*30^(1/2)*Z_1,1^(-1) -1/315*30^(1/2)*Z_1,1^(1) +4/1155*6^(1/2)*Z_3,1^(-1) 
    -4/1155*6^(1/2)*Z_3,1^(1) +8/1155*21^(1/2)*Z_3,3^(-1) -8/1155*21^(1/2)*Z_3,3^(1) 
    +8/45045*78^(1/2)*Z_5,1^(-1) -8/45045*78^(1/2)*Z_5,1^(1) +16/45045*273^(1/2)*Z_5,3^(-1) 
    -16/45045*273^(1/2)*Z_5,3^(1) +8/45045*4290^(1/2)*Z_5,5^(-1) -8/45045*4290^(1/2)*Z_5,5^(1).
y^5 / Pi^(1/2) = 1/63*i*30^(1/2)*Z_1,1^(-1) +1/63*i*30^(1/2)*Z_1,1^(1) +4/231*i*6^(1/2)*Z_3,1^(-1) 
    +4/231*i*6^(1/2)*Z_3,1^(1) -2/231*i*35^(1/2)*Z_3,3^(-3) -2/231*i*21^(1/2)*Z_3,3^(-1) 
    -2/231*i*21^(1/2)*Z_3,3^(1) -2/231*i*35^(1/2)*Z_3,3^(3) +8/9009*i*78^(1/2)*Z_5,1^(-1) 
    +8/9009*i*78^(1/2)*Z_5,1^(1) -4/9009*i*455^(1/2)*Z_5,3^(-3) -4/9009*i*273^(1/2)*Z_5,3^(-1) 
    -4/9009*i*273^(1/2)*Z_5,3^(1) -4/9009*i*455^(1/2)*Z_5,3^(3) +1/3003*i*1001^(1/2)*Z_5,5^(-5) 
    +1/9009*i*5005^(1/2)*Z_5,5^(-3) +1/9009*i*4290^(1/2)*Z_5,5^(-1) +1/9009*i*4290^(1/2)*Z_5,5^(1) 
    +1/9009*i*5005^(1/2)*Z_5,5^(3) +1/3003*i*1001^(1/2)*Z_5,5^(5).
y^4 z / Pi^(1/2) = 2/315*15^(1/2)*Z_1,1^(0) +8/1155*3^(1/2)*Z_3,1^(0) -2/1155*210^(1/2)*Z_3,3^(-2) 
    -4/385*7^(1/2)*Z_3,3^(0) -2/1155*210^(1/2)*Z_3,3^(2) +16/45045*39^(1/2)*Z_5,1^(0) 
    -4/45045*2730^(1/2)*Z_5,3^(-2) -8/15015*91^(1/2)*Z_5,3^(0) -4/45045*2730^(1/2)*Z_5,3^(2) 
    +1/15015*10010^(1/2)*Z_5,5^(-4) +2/45045*30030^(1/2)*Z_5,5^(-2) +2/3003*143^(1/2)*Z_5,5^(0) 
    +2/45045*30030^(1/2)*Z_5,5^(2) +1/15015*10010^(1/2)*Z_5,5^(4).
y^3 z^2 / Pi^(1/2) = 1/315*i*30^(1/2)*Z_1,1^(-1) +1/315*i*30^(1/2)*Z_1,1^(1) 
    +4/1155*i*6^(1/2)*Z_3,1^(-1) +4/1155*i*6^(1/2)*Z_3,1^(1) -1/1155*i*35^(1/2)*Z_3,3^(-3) 
    +1/385*i*21^(1/2)*Z_3,3^(-1) +1/385*i*21^(1/2)*Z_3,3^(1) -1/1155*i*35^(1/2)*Z_3,3^(3) 
    +8/45045*i*78^(1/2)*Z_5,1^(-1) +8/45045*i*78^(1/2)*Z_5,1^(1) -2/45045*i*455^(1/2)*Z_5,3^(-3) 
    +2/15015*i*273^(1/2)*Z_5,3^(-1) +2/15015*i*273^(1/2)*Z_5,3^(1) -2/45045*i*455^(1/2)*Z_5,3^(3) 
    -4/45045*i*5005^(1/2)*Z_5,5^(-3) -2/15015*i*4290^(1/2)*Z_5,5^(-1) -2/15015*i*4290^(1/2)*Z_5,5^(1) 
    -4/45045*i*5005^(1/2)*Z_5,5^(3).
y^2 z^3 / Pi^(1/2) = 2/315*15^(1/2)*Z_1,1^(0) +8/1155*3^(1/2)*Z_3,1^(0) -1/1155*210^(1/2)*Z_3,3^(-2) 
    -2/1155*7^(1/2)*Z_3,3^(0) -1/1155*210^(1/2)*Z_3,3^(2) +16/45045*39^(1/2)*Z_5,1^(0) 
    -2/45045*2730^(1/2)*Z_5,3^(-2) -4/45045*91^(1/2)*Z_5,3^(0) -2/45045*2730^(1/2)*Z_5,3^(2) 
    -2/45045*30030^(1/2)*Z_5,5^(-2) -8/9009*143^(1/2)*Z_5,5^(0) -2/45045*30030^(1/2)*Z_5,5^(2).
y z^4 / Pi^(1/2) = 1/315*i*30^(1/2)*Z_1,1^(-1) +1/315*i*30^(1/2)*Z_1,1^(1) 
    +4/1155*i*6^(1/2)*Z_3,1^(-1) +4/1155*i*6^(1/2)*Z_3,1^(1) +8/1155*i*21^(1/2)*Z_3,3^(-1) 
    +8/1155*i*21^(1/2)*Z_3,3^(1) +8/45045*i*78^(1/2)*Z_5,1^(-1) +8/45045*i*78^(1/2)*Z_5,1^(1) 
    +16/45045*i*273^(1/2)*Z_5,3^(-1) +16/45045*i*273^(1/2)*Z_5,3^(1) +8/45045*i*4290^(1/2)*Z_5,5^(-1) 
    +8/45045*i*4290^(1/2)*Z_5,5^(1).
z^5 / Pi^(1/2) = 2/63*15^(1/2)*Z_1,1^(0) +8/231*3^(1/2)*Z_3,1^(0) +8/231*7^(1/2)*Z_3,3^(0) 
    +16/9009*39^(1/2)*Z_5,1^(0) +16/9009*91^(1/2)*Z_5,3^(0) +16/9009*143^(1/2)*Z_5,5^(0).
\end{verbatim}\normalsize
Above, \texttt{i} is the imaginary unit which appears on the right hand sides if $q$ is odd.

\subsection{Product Expansion (Linearization Coefficients) } \label{sec.lin3D} 
The product of the angular variables are the established
expansions with Wigner $3j$ coefficients \cite[(4.6.5)]{Edmonds}
\begin{equation}
Y_{l_1}^{(m_1)} Y_{l_2}^{(m_2)}=
\sum_{l=|l_1-l_2|}^{l_1+l_2}\sum_{m=-l}^l \sqrt{\frac{(2l_1+1)(2l_2+1)(2l+1)}{4\pi}}
\left(
\begin{array}{ccc}
l_1 & l_2 & l \\
m_1 & m_2 & m \\
\end{array}
\right)
\left(
\begin{array}{ccc}
l_1 & l_2 & l \\
0 & 0 & 0 \\
\end{array}
\right)
Y_l^{(m)*}
,
\label{eq.Ycoupl}
\end{equation}
where the sum over $m$
is nonzero only at $m_1+m_2+m=0$.
The relation to Clebsch-Gordan coefficients
\begin{equation}
\left(
\begin{array}{ccc}
j_1 & j_2 & j_3 \\
m_1 & m_2 & m_3 \\
\end{array}
\right)
=
(-1)^{j_1-j_2-m_3}\frac{1}{\sqrt{2j_3+1}}
(j_1m_1j_2m_2\mid j_1j_2j_3-m_3)
\end{equation}
opens a route to computation via \cite[(3.6.11)]{Edmonds}\cite[(7.5)]{Fano}\cite[27.9.1]{AS}
\begin{eqnarray}
(j_1m_1j_2m_2\mid j_1j_2jm)
&=&
\delta(m_1+m_2,m)
\sqrt{\frac{(2j+1)(j_1+j_2-j)!(j_1-j_2+j)!(-j_1+j_2+j)!}{(j_1+j_2+j+1)!}}
\nonumber
\\
&&
\times
\sqrt{(j_1+m_1)!(j_1-m_1)!(j_2+m_2)!(j_2-m_2)!(j+m)!(j-m)!}
\\
&&
\times
\sum_z
\frac{(-1)^z}{z!(j_1+j_2-j-z)!(j_1-m_1-z)!(j_2+m_2-z)!(j-j_2+m_1+z)!(j-j_1-m_2+z)!}
,
\nonumber
\end{eqnarray}
where the sum is over all $z$ that keep the arguments of the factorials in the denominator non-negative.
The familiar table of (\ref{eq.Ycoupl}) starts in conjunction
with (\ref{eq.Ylmcc}) as:
\small \begin{verbatim}
Pi^(1/2) Y_0^(0) Y_0^(0) = 2^(1/2)*Y_0^(0).
Pi^(1/2) Y_1^(-1) Y_0^(0) = 1/3*2^(1/2)*Y_1^(-1).
Pi^(1/2) Y_1^(-1) Y_1^(-1) = 1/30*6^(1/2)*Y_2^(-2).
Pi^(1/2) Y_1^(0) Y_0^(0) = 2/3*2^(1/2)*Y_1^(0).
Pi^(1/2) Y_1^(0) Y_1^(-1) = 1/15*6^(1/2)*Y_2^(-1).
Pi^(1/2) Y_1^(0) Y_1^(0) = 2/3*2^(1/2)*Y_0^(0) +4/15*6^(1/2)*Y_2^(0).
Pi^(1/2) Y_1^(1) Y_0^(0) = 1/3*2^(1/2)*Y_1^(1).
Pi^(1/2) Y_1^(1) Y_1^(-1) = -1/3*2^(1/2)*Y_0^(0) +1/30*6^(1/2)*Y_2^(0).
Pi^(1/2) Y_1^(1) Y_1^(0) = 1/15*6^(1/2)*Y_2^(1).
Pi^(1/2) Y_1^(1) Y_1^(1) = 1/30*6^(1/2)*Y_2^(2).
Pi^(1/2) Y_2^(-2) Y_0^(0) = 1/40*2^(1/2)*Y_2^(-2).
Pi^(1/2) Y_2^(-2) Y_1^(-1) = 1/420*3^(1/2)*Y_3^(-3).
Pi^(1/2) Y_2^(-2) Y_1^(0) = 1/210*3^(1/2)*Y_3^(-2).
Pi^(1/2) Y_2^(-2) Y_1^(1) = -1/30*6^(1/2)*Y_1^(-1) +1/420*3^(1/2)*Y_3^(-1).
Pi^(1/2) Y_2^(-2) Y_2^(-2) = 1/20160*10^(1/2)*Y_4^(-4).
Pi^(1/2) Y_2^(-1) Y_0^(0) = 1/10*2^(1/2)*Y_2^(-1).
Pi^(1/2) Y_2^(-1) Y_1^(-1) = 1/105*3^(1/2)*Y_3^(-2).
Pi^(1/2) Y_2^(-1) Y_1^(0) = 1/15*6^(1/2)*Y_1^(-1) +2/105*3^(1/2)*Y_3^(-1).
Pi^(1/2) Y_2^(-1) Y_1^(1) = -1/15*6^(1/2)*Y_1^(0) +1/105*3^(1/2)*Y_3^(0).
Pi^(1/2) Y_2^(-1) Y_2^(-2) = 1/5040*10^(1/2)*Y_4^(-3).
Pi^(1/2) Y_2^(-1) Y_2^(-1) = 1/140*6^(1/2)*Y_2^(-2) +1/1260*10^(1/2)*Y_4^(-2).
Pi^(1/2) Y_2^(0) Y_0^(0) = 3/5*2^(1/2)*Y_2^(0).
Pi^(1/2) Y_2^(0) Y_1^(-1) = -1/30*6^(1/2)*Y_1^(-1) +1/70*3^(1/2)*Y_3^(-1).
Pi^(1/2) Y_2^(0) Y_1^(0) = 4/15*6^(1/2)*Y_1^(0) +12/35*3^(1/2)*Y_3^(0).
Pi^(1/2) Y_2^(0) Y_1^(1) = -1/30*6^(1/2)*Y_1^(1) +1/70*3^(1/2)*Y_3^(1).
Pi^(1/2) Y_2^(0) Y_2^(-2) = -1/280*6^(1/2)*Y_2^(-2) +1/3360*10^(1/2)*Y_4^(-2).
Pi^(1/2) Y_2^(0) Y_2^(-1) = 1/140*6^(1/2)*Y_2^(-1) +1/840*10^(1/2)*Y_4^(-1).
Pi^(1/2) Y_2^(0) Y_2^(0) = 3/5*2^(1/2)*Y_0^(0) +6/35*6^(1/2)*Y_2^(0) +6/35*10^(1/2)*Y_4^(0).
Pi^(1/2) Y_2^(1) Y_0^(0) = 1/10*2^(1/2)*Y_2^(1).
Pi^(1/2) Y_2^(1) Y_1^(-1) = -1/15*6^(1/2)*Y_1^(0) +1/105*3^(1/2)*Y_3^(0).
Pi^(1/2) Y_2^(1) Y_1^(0) = 1/15*6^(1/2)*Y_1^(1) +2/105*3^(1/2)*Y_3^(1).
Pi^(1/2) Y_2^(1) Y_1^(1) = 1/105*3^(1/2)*Y_3^(2).
Pi^(1/2) Y_2^(1) Y_2^(-2) = -1/140*6^(1/2)*Y_2^(-1) +1/5040*10^(1/2)*Y_4^(-1).
Pi^(1/2) Y_2^(1) Y_2^(-1) = -1/10*2^(1/2)*Y_0^(0) -1/140*6^(1/2)*Y_2^(0) +1/1260*10^(1/2)*Y_4^(0).
Pi^(1/2) Y_2^(1) Y_2^(0) = 1/140*6^(1/2)*Y_2^(1) +1/840*10^(1/2)*Y_4^(1).
Pi^(1/2) Y_2^(1) Y_2^(1) = 1/140*6^(1/2)*Y_2^(2) +1/1260*10^(1/2)*Y_4^(2).
Pi^(1/2) Y_2^(2) Y_0^(0) = 1/40*2^(1/2)*Y_2^(2).
Pi^(1/2) Y_2^(2) Y_1^(-1) = -1/30*6^(1/2)*Y_1^(1) +1/420*3^(1/2)*Y_3^(1).
Pi^(1/2) Y_2^(2) Y_1^(0) = 1/210*3^(1/2)*Y_3^(2).
Pi^(1/2) Y_2^(2) Y_1^(1) = 1/420*3^(1/2)*Y_3^(3).
Pi^(1/2) Y_2^(2) Y_2^(-2) = 1/40*2^(1/2)*Y_0^(0) -1/280*6^(1/2)*Y_2^(0) +1/20160*10^(1/2)*Y_4^(0).
Pi^(1/2) Y_2^(2) Y_2^(-1) = -1/140*6^(1/2)*Y_2^(1) +1/5040*10^(1/2)*Y_4^(1).
Pi^(1/2) Y_2^(2) Y_2^(0) = -1/280*6^(1/2)*Y_2^(2) +1/3360*10^(1/2)*Y_4^(2).
Pi^(1/2) Y_2^(2) Y_2^(1) = 1/5040*10^(1/2)*Y_4^(3).
Pi^(1/2) Y_2^(2) Y_2^(2) = 1/20160*10^(1/2)*Y_4^(4).
\end{verbatim}\normalsize

The products of the radial
polynomials are essentially products of Jacobi polynomials and are
expanded according to Srivastava's equation \cite[(21)]{SrivastavaJPA21}.
Since we are concerned about products of two Zernike functions,
$R_{n_1}^{(l_1)}Y_{l_1}^{(m_1)} R_{n_2}^{(l_2)}Y_{l_2}^{(m_2)}$,
with coupling of the angular variables to products in the range $|l_1-l_2|\le l \le l_1+l_2$
via (\ref{eq.Ycoupl}),
the request is to expand
$R_{n_1}^{(l_1)}R_{n_2}^{(l_2)}$
as $R_n^{(l)}$ with $l$ fixed within that interval.
The 3D equivalent to (\ref{eq.gdef}) and (\ref{eq.gtripls}) is
\begin{eqnarray}
R_{n_1}^{(l_1)}(r)R_{n_2}^{(l_2)}(r)
&\equiv& \sum_{n_3=l_3}^{n_1+n_2} k_{n_1,l_1,n_2,l_2,n_3,l_3} R_{n_3}^{(l_3)}(r);
\label{eq.kdef}
\\
k_{n_1,l_1,n_2,l_2,n_3,l_3}&=&\int_0^1 r^2
R_{n_1}^{(l_1)}(r)R_{n_2}^{(l_2)}(r)R_{n_3}^{(l_3)}(r) dr,
\end{eqnarray}
with a $k$-sum rule immediate from taking $r=1$ with (\ref{eq.R3dat1}).
In practise, given the product of the
radial polynomials as a polynomial in $r$, double insertion of (\ref{eq.rofR3d}) leads to 
products of the following format:
\small \begin{verbatim}
R_1^(1)(r)*R_1^(1)(r) = 3^(1/2)*R_0^(0)(r) +2/7*7^(1/2)*R_2^(0)(r)
   = 5/7*7^(1/2)*R_2^(2)(r).
R_1^(1)(r)*R_2^(0)(r) = 2/7*7^(1/2)*R_1^(1)(r) +5/21*35^(1/2)*R_3^(1)(r).
R_1^(1)(r)*R_2^(2)(r) = 5/7*7^(1/2)*R_1^(1)(r) +2/21*35^(1/2)*R_3^(1)(r)
   = 1/3*35^(1/2)*R_3^(3)(r).
R_1^(1)(r)*R_3^(1)(r) = 5/21*35^(1/2)*R_2^(0)(r) +4/33*55^(1/2)*R_4^(0)(r)
   = 2/21*35^(1/2)*R_2^(2)(r) +7/33*55^(1/2)*R_4^(2)(r).
R_1^(1)(r)*R_3^(3)(r) = 1/3*35^(1/2)*R_2^(2)(r) +2/33*55^(1/2)*R_4^(2)(r)
   = 3/11*55^(1/2)*R_4^(4)(r).
R_2^(0)(r)*R_2^(0)(r) = 3^(1/2)*R_0^(0)(r) -2/9*7^(1/2)*R_2^(0)(r) +50/99*11^(1/2)*R_4^(0)(r).
R_2^(0)(r)*R_2^(2)(r) = 4/9*7^(1/2)*R_2^(2)(r) +35/99*11^(1/2)*R_4^(2)(r).
R_2^(2)(r)*R_2^(2)(r) = 3^(1/2)*R_0^(0)(r) +4/9*7^(1/2)*R_2^(0)(r) +8/99*11^(1/2)*R_4^(0)(r)
   = 7/9*7^(1/2)*R_2^(2)(r) +14/99*11^(1/2)*R_4^(2)(r)
   = 7/11*11^(1/2)*R_4^(4)(r).
R_1^(1)(r)*R_4^(0)(r) = 4/33*55^(1/2)*R_3^(1)(r) +7/143*715^(1/2)*R_5^(1)(r).
R_1^(1)(r)*R_4^(2)(r) = 7/33*55^(1/2)*R_3^(1)(r) +4/143*715^(1/2)*R_5^(1)(r)
   = 2/33*55^(1/2)*R_3^(3)(r) +9/143*715^(1/2)*R_5^(3)(r).
R_1^(1)(r)*R_4^(4)(r) = 3/11*55^(1/2)*R_3^(3)(r) +2/143*715^(1/2)*R_5^(3)(r)
   = 1/13*715^(1/2)*R_5^(5)(r).
R_2^(0)(r)*R_3^(1)(r) = 5/21*35^(1/2)*R_1^(1)(r) -8/77*7^(1/2)*R_3^(1)(r) +70/429*91^(1/2)*R_5^(1)(r).
R_2^(0)(r)*R_3^(3)(r) = 6/11*7^(1/2)*R_3^(3)(r) +15/143*91^(1/2)*R_5^(3)(r).
R_2^(2)(r)*R_3^(1)(r) = 2/21*35^(1/2)*R_1^(1)(r) +43/77*7^(1/2)*R_3^(1)(r) +28/429*91^(1/2)*R_5^(1)(r)
   = 4/11*7^(1/2)*R_3^(3)(r) +21/143*91^(1/2)*R_5^(3)(r).
R_2^(2)(r)*R_3^(3)(r) = 1/3*35^(1/2)*R_1^(1)(r) +4/11*7^(1/2)*R_3^(1)(r) +8/429*91^(1/2)*R_5^(1)(r)
   = 9/11*7^(1/2)*R_3^(3)(r) +6/143*91^(1/2)*R_5^(3)(r)
   = 3/13*91^(1/2)*R_5^(5)(r).
R_1^(1)(r)*R_5^(1)(r) = 7/143*715^(1/2)*R_4^(0)(r) +2/65*975^(1/2)*R_6^(0)(r)
   = 4/143*715^(1/2)*R_4^(2)(r) +3/65*975^(1/2)*R_6^(2)(r).
R_1^(1)(r)*R_5^(3)(r) = 9/143*715^(1/2)*R_4^(2)(r) +4/195*975^(1/2)*R_6^(2)(r)
   = 2/143*715^(1/2)*R_4^(4)(r) +11/195*975^(1/2)*R_6^(4)(r).
R_1^(1)(r)*R_5^(5)(r) = 1/13*715^(1/2)*R_4^(4)(r) +2/195*975^(1/2)*R_6^(4)(r)
   = 1/15*975^(1/2)*R_6^(6)(r).
R_2^(0)(r)*R_4^(0)(r) = 50/99*11^(1/2)*R_2^(0)(r) -28/117*7^(1/2)*R_4^(0)(r) 
    +7/143*1155^(1/2)*R_6^(0)(r).
R_2^(0)(r)*R_4^(2)(r) = 35/99*11^(1/2)*R_2^(2)(r) +2/117*7^(1/2)*R_4^(2)(r) 
    +6/143*1155^(1/2)*R_6^(2)(r).
R_2^(0)(r)*R_4^(4)(r) = 8/13*7^(1/2)*R_4^(4)(r) +1/39*1155^(1/2)*R_6^(4)(r).
R_2^(2)(r)*R_4^(0)(r) = 8/99*11^(1/2)*R_2^(2)(r) +56/117*7^(1/2)*R_4^(2)(r) 
    +21/715*1155^(1/2)*R_6^(2)(r).
R_2^(2)(r)*R_4^(2)(r) = 35/99*11^(1/2)*R_2^(0)(r) +56/117*7^(1/2)*R_4^(0)(r) 
    +8/715*1155^(1/2)*R_6^(0)(r)
   = 14/99*11^(1/2)*R_2^(2)(r) +71/117*7^(1/2)*R_4^(2)(r) +12/715*1155^(1/2)*R_6^(2)(r)
   = 4/13*7^(1/2)*R_4^(4)(r) +3/65*1155^(1/2)*R_6^(4)(r).
R_2^(2)(r)*R_4^(4)(r) = 7/11*11^(1/2)*R_2^(2)(r) +4/13*7^(1/2)*R_4^(2)(r) +8/2145*1155^(1/2)*R_6^(2)(r)
   = 11/13*7^(1/2)*R_4^(4)(r) +2/195*1155^(1/2)*R_6^(4)(r)
   = 1/15*1155^(1/2)*R_6^(6)(r).
R_3^(1)(r)*R_3^(1)(r) = 3^(1/2)*R_0^(0)(r) -8/77*7^(1/2)*R_2^(0)(r) +34/143*11^(1/2)*R_4^(0)(r) 
    +196/715*15^(1/2)*R_6^(0)(r)
   = 43/77*7^(1/2)*R_2^(2)(r) -14/143*11^(1/2)*R_4^(2)(r) +294/715*15^(1/2)*R_6^(2)(r).
R_3^(1)(r)*R_3^(3)(r) = 4/11*7^(1/2)*R_2^(2)(r) +61/143*11^(1/2)*R_4^(2)(r) +84/715*15^(1/2)*R_6^(2)(r)
   = 54/143*11^(1/2)*R_4^(4)(r) +21/65*15^(1/2)*R_6^(4)(r).
R_3^(3)(r)*R_3^(3)(r) = 3^(1/2)*R_0^(0)(r) +6/11*7^(1/2)*R_2^(0)(r) +24/143*11^(1/2)*R_4^(0)(r) 
    +16/715*15^(1/2)*R_6^(0)(r)
   = 9/11*7^(1/2)*R_2^(2)(r) +36/143*11^(1/2)*R_4^(2)(r) +24/715*15^(1/2)*R_6^(2)(r)
   = 9/13*11^(1/2)*R_4^(4)(r) +6/65*15^(1/2)*R_6^(4)(r)
   = 3/5*15^(1/2)*R_6^(6)(r).
R_1^(1)(r)*R_6^(0)(r) = 2/65*975^(1/2)*R_5^(1)(r) +3/85*1275^(1/2)*R_7^(1)(r).
R_1^(1)(r)*R_6^(2)(r) = 3/65*975^(1/2)*R_5^(1)(r) +2/85*1275^(1/2)*R_7^(1)(r)
   = 4/195*975^(1/2)*R_5^(3)(r) +11/255*1275^(1/2)*R_7^(3)(r).
R_1^(1)(r)*R_6^(4)(r) = 11/195*975^(1/2)*R_5^(3)(r) +4/255*1275^(1/2)*R_7^(3)(r)
   = 2/195*975^(1/2)*R_5^(5)(r) +13/255*1275^(1/2)*R_7^(5)(r).
R_1^(1)(r)*R_6^(6)(r) = 1/15*975^(1/2)*R_5^(5)(r) +2/255*1275^(1/2)*R_7^(5)(r)
   = 1/17*1275^(1/2)*R_7^(7)(r).
R_2^(0)(r)*R_5^(1)(r) = 70/429*91^(1/2)*R_3^(1)(r) -2/11*7^(1/2)*R_5^(1)(r) 
    +9/221*1547^(1/2)*R_7^(1)(r).
R_2^(0)(r)*R_5^(3)(r) = 15/143*91^(1/2)*R_3^(3)(r) +4/33*7^(1/2)*R_5^(3)(r) 
    +22/663*1547^(1/2)*R_7^(3)(r).
R_2^(0)(r)*R_5^(5)(r) = 2/3*7^(1/2)*R_5^(5)(r) +1/51*1547^(1/2)*R_7^(5)(r).
R_2^(2)(r)*R_5^(1)(r) = 28/429*91^(1/2)*R_3^(1)(r) +29/55*7^(1/2)*R_5^(1)(r) 
    +18/1105*1547^(1/2)*R_7^(1)(r)
   = 8/429*91^(1/2)*R_3^(3)(r) +24/55*7^(1/2)*R_5^(3)(r) +33/1105*1547^(1/2)*R_7^(3)(r).
R_2^(2)(r)*R_5^(3)(r) = 21/143*91^(1/2)*R_3^(1)(r) +24/55*7^(1/2)*R_5^(1)(r) 
    +8/1105*1547^(1/2)*R_7^(1)(r)
   = 6/143*91^(1/2)*R_3^(3)(r) +107/165*7^(1/2)*R_5^(3)(r) +44/3315*1547^(1/2)*R_7^(3)(r)
   = 4/15*7^(1/2)*R_5^(5)(r) +11/255*1547^(1/2)*R_7^(5)(r).
R_2^(2)(r)*R_5^(5)(r) = 3/13*91^(1/2)*R_3^(3)(r) +4/15*7^(1/2)*R_5^(3)(r) +8/3315*1547^(1/2)*R_7^(3)(r)
   = 13/15*7^(1/2)*R_5^(5)(r) +2/255*1547^(1/2)*R_7^(5)(r)
   = 1/17*1547^(1/2)*R_7^(7)(r).
R_3^(1)(r)*R_4^(0)(r) = 4/33*55^(1/2)*R_1^(1)(r) +34/143*11^(1/2)*R_3^(1)(r) 
    -28/2145*143^(1/2)*R_5^(1)(r) +1323/12155*187^(1/2)*R_7^(1)(r).
R_3^(1)(r)*R_4^(2)(r) = 7/33*55^(1/2)*R_1^(1)(r) -14/143*11^(1/2)*R_3^(1)(r) 
    +194/2145*143^(1/2)*R_5^(1)(r) +756/12155*187^(1/2)*R_7^(1)(r)
   = 61/143*11^(1/2)*R_3^(3)(r) -12/715*143^(1/2)*R_5^(3)(r) +126/1105*187^(1/2)*R_7^(3)(r).
R_3^(1)(r)*R_4^(4)(r) = 54/143*11^(1/2)*R_3^(3)(r) +79/715*143^(1/2)*R_5^(3)(r) 
    +28/1105*187^(1/2)*R_7^(3)(r)
   = 8/65*143^(1/2)*R_5^(5)(r) +7/85*187^(1/2)*R_7^(5)(r).
R_3^(3)(r)*R_4^(0)(r) = 24/143*11^(1/2)*R_3^(3)(r) +84/715*143^(1/2)*R_5^(3)(r) 
    +63/1105*187^(1/2)*R_7^(3)(r).
R_3^(3)(r)*R_4^(2)(r) = 2/33*55^(1/2)*R_1^(1)(r) +61/143*11^(1/2)*R_3^(1)(r) 
    +184/2145*143^(1/2)*R_5^(1)(r) +216/12155*187^(1/2)*R_7^(1)(r)
   = 36/143*11^(1/2)*R_3^(3)(r) +93/715*143^(1/2)*R_5^(3)(r) +36/1105*187^(1/2)*R_7^(3)(r)
   = 6/65*143^(1/2)*R_5^(5)(r) +9/85*187^(1/2)*R_7^(5)(r).
R_3^(3)(r)*R_4^(4)(r) = 3/11*55^(1/2)*R_1^(1)(r) +54/143*11^(1/2)*R_3^(1)(r) 
    +24/715*143^(1/2)*R_5^(1)(r) +48/12155*187^(1/2)*R_7^(1)(r)
   = 9/13*11^(1/2)*R_3^(3)(r) +4/65*143^(1/2)*R_5^(3)(r) +8/1105*187^(1/2)*R_7^(3)(r)
   = 1/5*143^(1/2)*R_5^(5)(r) +2/85*187^(1/2)*R_7^(5)(r)
   = 3/17*187^(1/2)*R_7^(7)(r).
\end{verbatim}\normalsize
Products of two entries of the previous two tables expand products
$Z_{n_1,l_1}^{(m_1)} Z_{n_2,l_2}^{(m_2)}$ of two
3D Zernike functions into a sum over 3D Zernike functions by picking
lines with $l$-parameters which match.

\section{Summary} 
The 2D and 3D Zernike functions are orthogonal basis sets defined
in the unit circle and unit sphere, where the specific notation  introduced
by Noll is the most common standard in 2D, and a more rational (but
square-root loaded) standard seems to emerge in 3D\@.
We have demonstrated the transformation of the basis functions between
the radial-angular and the centered Cartesian systems for both dimensions.

The applications are numerical field simulations, where the foundation
is backed by an isotropy in 2D or 3D space, but followup calculations
employ vector field operators which have simpler representations
in global Cartesian than in circular or spherical local coordinates.

\acknowledgments{
This work is supported by the NWO VICI grant 639.043.201
to A. Quirrenbach,
``Optical Interferometry: A new Method for Studies of Extrasolar Planets.''
}

\appendix

\section{Notations } \label{sec.nots} 
\begin{tabular}{ll}
$(z)_n$        & Pochhammer's Symbol, $(z)_0\equiv 1$, $(z)_n=z(z+1)(z+2)\cdots(z+n-1)=\Gamma(z+n)/\Gamma(z)$ \cite[6.1.22]{AS}\\
$_2F_1(.,.;.;.)$        & Gaussian Hypergeometric Function \cite[\S 15]{AS}\\
$_.F_.(.,.;.;.)$        & generalized Hypergeometric Function \cite{SlaterHyp}\\
$(....|....)$        & Clebsch-Gordan coefficient \cite[27.9]{AS}\\
$\left(\begin{array}{ccc}.&.&.\\.&.&.\end{array}\right)$       & Wigner's $3j$ coefficient, Section \ref{sec.3D}\\
$\binom{ . }{ .}$       & binomial coefficient \cite[3.1.2]{AS}\\
$\ldots ^*$       & complex conjugate\\
$\lfloor\ldots\rfloor$       & floor function, the largest integer not larger than the embraced argument\\
$j!$       & factorial $j\cdot(j-1)\cdot(j-2)\cdots 1$ \\
$j!!$       & double factorial $j\cdot(j-2)\cdot(j-4)\cdots = \prod_{s=0}^{\lfloor(j-1)/2\rfloor} (j-2s)$ \\
\texttt{f\symbol{94}(1/2)}       & square root of $f$ \\
$a$        & index excess (\ref{eq.R2Dhyp})\\
$\alpha$        & index excess (\ref{eq.Jac3D})\\
$b$        & index mean (\ref{eq.R2Dhyp}) \\
$\varphi$, \texttt{phi}        & azimuth angle\\
$f$, $\hat f$        & linearization coefficients (\ref{eq.rofR3d}), (\ref{eq.rofR3dtoo}) \\
$g$        & linearization coefficient (\ref{eq.gdef}) \\
$h$        & expansion coefficient (\ref{eq.rofR}) \\
$i$, \texttt{i}        & imaginary unit\\
$k$        & expansion coefficient (\ref{eq.kdef}) \\
$P_.^.(.)$        & Associated Legendre Polynomial (\ref{eq.Plm}) \cite[\S 8]{AS}\\
$P_.^{(.,.)}(.)$        & Jacobi Polynomial \cite[\S 22]{AS}\\
\texttt{Pi}        & $\pi$ \\
$q$        & 3D radial polynomial group parameter (\ref{eq.Jac3D})\\
$R_n^m(.)$, \texttt{R\_.\symbol{94}.}        & Zernike radial polynomial, circular, Section \ref{sec.2D}\\
$R_n^{(l)}(.)$, \texttt{R\_.\symbol{94}(.)}        & Zernike radial polynomial, spherical, Section \ref{sec.3D}\\
$\theta$        & polar angle, Section \ref{sec.3D}\\
$u$        & expansion coefficient (\ref{eq.udef})\\
$x,y,z$        & Cartesian coordinates (\ref{eq.xy}), (\ref{eq.xyz})\\
$Y_{.}^{(.)}$, \texttt{Y\_.\symbol{94}(.)}       & Spherical Harmonics, Section \ref{sec.3D}\\
$Z_.$, \texttt{Z\_.}       & Zernike function, circular, Section \ref{sec.2D}\\
$Z_{.,.}^{(.)}$, \texttt{Z\_.,.\symbol{94}(.)}        & Zernike function, spherical, Section \ref{sec.3D}\\
\end{tabular}

\bibliographystyle{apsrmp}
\bibliography{all}

\end{document}